\documentclass[pof,superscriptaddress]{revtex4-1}
\usepackage[dvips]{graphicx}
\usepackage{amssymb,amsfonts,amsmath}
\usepackage{color}
\usepackage[caption=false]{subfig}

\newcommand{\bs}{{\bf {s}}}
\newcommand{\br}{{\bf {r}}}
\newcommand{\bv}{{\bf {v}}}

\newcommand{\etal}{{\it et al.~}}

\begin{document}


\title{Coherent laminar and turbulent motion of toroidal vortex bundles}



\author{D.H.~Wacks}
\affiliation{School of Mechanical and Systems Engineering,
Newcastle University, Newcastle upon Tyne NE1 7RU, United Kingdom}

\author{A.~W.~Baggaley}
\affiliation{School of Mathematics and Statistics, 
University of Glasgow, Glasgow G12 8QW, Scotland, United Kingdom}

\author{C.~F.~Barenghi}
\affiliation{Joint Quantum Centre (JQC) Durham-Newcastle, and
School of Mathematics and Statistics, 
Newcastle University, Newcastle upon Tyne, NE1 7RU, United Kingdom}


\date{\today}

\begin{abstract}
Motivated by experiments performed in superfluid helium, we study
numerically the motion of toroidal bundles of vortex filaments in an
inviscid fluid. We find that the evolution of these large-scale 
vortex structures involves the generalised leapfrogging of the constituent
vortex rings. Despite three dimensional perturbations in the form of
Kelvin waves and vortex reconnections, toroidal vortex 
bundles retain their coherence over a relatively large distance
(compared to their size), in agreement with experimental
observations.
\end{abstract}

\pacs{\\
47.32.C- (Vortex dynamics) \\
47.32.cf (Vortex reconnection and rings)\\
67.25.dk (Vortices and turbulence in superfluid helium 4)
}
\maketitle 

\section{Introduction}
\label{sec:intro}

This work is motivated by a series of experiments 
\cite{Borner1981,Borner1983,Borner1985} performed in the 1980's
in which Borner and collaborators forced
superfluid helium ($^4$He) through the circular orifice of a cylindrical tube 
(diameter $D=0.8~\rm cm$) by the single stroke of a piston,
at temperatures between $1.3$ and $2.15~\rm K$  above absolute
zero. Using acoustic methods, Borner \etal 
determined size, position, velocity and superfluid circulation
of the localised vortex structure which was ejected from the tube's orifice.  
Their finding that velocity and circulation 
remained reproducibly constant during the evolution 
suggested
that the vortex structure maintained its identity and travelled coherently.

To appreciate the significance of Borner's result, we recall \cite{Donnelly}
that the flow of superfluid helium is potential (in a quantum fluid,
the velocity is proportional to the gradient of the phase of the
quantum mechanical wave function).  Superfluid vorticity exists only 
in the form of thin vortex lines of atomic thickness (the radius of the
vortex core is $a \approx 10^{-8}~\rm cm$) and
fixed circulation $\kappa=9.97 \times 10^{-4}~\rm cm^2/s$ 
proportional to Planck's constant.
The superfluid literature
is familiar \cite{BD2009} with individual vortex rings,
ranging in diameter from approximately one micron (for rings
generated by a high voltage tip \cite{Golov}) to half a millimetre
(as recently visualized by Bewley and Sreenivasan \cite{Bewley-rings}).
However, the circulation $\Gamma$ observed by Borner was much larger
than the quantum of circulation $\kappa$:
they reported values in the range from $\Gamma=2.3$ to $4.85~\rm cm^2/s$
depending on piston's stroke, velocity and temperature.
According to Borner, the cylinder-piston arrangement generates
a macroscopic, self-propelling, toroidal vortex structure - some kind of 
large-scale vortex ring. The natural interpretation is that such
structure actually consists of a toroidal bundle of 
$N = \Gamma/\kappa \approx  10^3$
individual vortex rings of the same polarity travelling in the same
direction at close distance from each other, hence interacting strongly
but remaining coherent during the observed evolution over a distance
of about $6~\rm cm$. 

Borner's interpretation was strengthened by further experiments
performed by Murakami and collaborators \cite{Murakami}. 
Using a cylinder-piston arrangement
of the same size as Borner's, Murakami \etal succeeded in visualizing Borner's
large-scale vortex ring and tracking its motion
by means of frozen hydrogen-deuterium 
particles trapped in the cores of individual vortex rings.
Stamm and collaborators \cite{Stamm1,Stamm2} also generated large-scale
vortex rings in liquid helium, but using a different technique (second-sound
heat pulses), which resulted in vortex structures less stable than Borner's
and Murakami's.

Unfortunately the experiments of Borner, Murakami and Stamm were not followed up
and became almost forgotten. Today the situation has changed. 
Current experimental, numerical and theoretical
work is concerned with similarities
between  superfluid turbulence and ordinary turbulence.
In particular, 
the energy spectrum \cite{Salort2010,Salort2012,LNS,Sasa2011,BLR}
which is observed in turbulent superfluid helium obeys the same
Kolmogorov scaling of ordinary turbulence. It has been argued
\cite{Vinen,SS2012} that this property arises from the partial 
polarization of superfluid vortex lines.
The existence of bundles of vortices in turbulent helium is also debated 
\cite{Baggaley-structures,Sherwin2012}, in analogy with the ``worms" which are
responsible for the intermittency of ordinary turbulence \cite{Frisch}.  
Issues raised by Borner's and Murakami's experiments,
such as the interaction of vortex lines, their polarization,
coherence and stability, have thus become important. 
Moreover, Borner's cylinder-piston arrangement is a convenient configuration
to study polarized vortex lines under controlled experimental conditions.

Thus motivated, we revisit Borner's and Murakami's
experiments. Our work is not to model the experiments precisely, 
or to perform linear stability calculations 
(it is not even clear what should be the state to perturb).
We simply want to gain  
qualitative insight into the three-dimensional motion and stability 
of a number of coaxial, thin cored vortex rings arranged in the shape of 
a toroidal bundle, as envisaged by Borner. 
Will such a vortex configuration travel a significant
distance (larger than its diameter) in a coherent way, before
becoming unstable, or will vortex interactions destroy the structure
straight away ?

Borner's experiment has another interesting fluid dynamics aspect.
It is well-known that two coaxial vortex rings of the
same polarity, when close to each other, execute a peculiar leapfrogging
motion \cite{Oshima,Riley,Satti,Niemi}: 
the ring behind shrinks, speeds up, goes inside the ring which was ahead,
overtakes it, grows in size, slows down, and then the process repeats
in a periodic fashion. 
For a toroidal bundle of many ($N \gg 2$) vortex rings to remain coherent, 
some kind of generalized leapfrogging motion is likely to take place. 

Before we finish this introduction, we recall \cite{Donnelly}
 the classical expressions
for the velocity $v$ and the energy (per unit density)
$E$ of a vortex ring of radius $R$:

\begin{equation}
v=\frac{\Gamma}{4 \pi R} \left( \ln{(8R/a)}-\alpha \right),
\label{eq:v}
\end{equation}

\begin{equation}
E=\frac{1}{2} \Gamma^2 R \left( \ln{(8R/a)}-\beta \right),
\label{eq:E}
\end{equation}

\noindent
where $a$ is the core radius and 
$\Gamma$ is the circulation; the quantities $\alpha$ and $\beta$ 
depend on the assumed vortex core structure \cite{Donnelly}. 
For hollow core we have $\alpha=1/2$ and $\beta=3/2$;
for core  rotating at uniform angular velocity as a solid--body
(Rankine vortex)
we have $\alpha=1/4$ and $\beta=7/4$ \cite{note-core}.
For a single quantum vortex ring, the quantities $\Gamma$ and $a$ 
must be replaced by the quantum of circulation
$\kappa$ and the superfluid vortex core radius
$a_0=1.3 \times 10^{-8}~\rm cm$ (proportional to the superfluid healing
length). For Borner's large-scale vortex ring
structure, which we model as a torus, $R$ will be the major radius 
of the torus and $a$ the minor radius.

\section{Method}
\label{sec:method}

For the sake of simplicity, we  ignore the thermal excitations 
which make up helium's normal fluid \cite{Donnelly}.
This approximation is justified by the
fact that, at the lowest temperature explored by Borner, the normal fluid 
fraction is only 4.5 percent of the total density \cite{DB1998}. 
In this low temperature limit, 
friction is negligible \cite{DB1998}
and vortex lines move with the superfluid.
We are thus in the limit of classical Euler dynamics.
We represent vortex lines as three-dimensional, closed
space curves $\bs=\bs(\xi,t)$
of infinitesimal thickness and circulation $\kappa$ moving in an
inviscid, incompressible Euler fluid. The curves are
parametrised by arc length $\xi$ and evolve in time $t$
in an infinite computational domain according to the 
Biot-Savart law \cite{Saffman}

\begin{equation}
\frac{d \bs}{dt}=
-\frac{\kappa}{4 \pi} \oint_{\cal L} \frac{(\bs-\br) }
{\vert \bs - \br \vert^3}
\times {\bf d}\br,
\label{eq:BS}
\end{equation}

\noindent
where the line integral extends over the entire vortex 
configuration $\cal L$. 
Vortex reconnections are forbidden by Euler
dynamics, but are possible for quantum vortices 
\cite{Bewley-reconnections,Zuccher,Paoletti}.
Therefore we supplement Eq.~\ref{eq:BS} with
an algorithmic reconnection procedure which changes the topology
of the vortex configuration when the distance between two filaments
is less than a prescribed cutoff value.
In the next sections, when describing results, we shall state explicitly
whether vortex reconnections have taken place or not; in this way
we shall distinguish the aspects of our problem 
which refer to classical Eulerian fluid dynamics
from the aspects which (since reconnections have taken place)
are relevant only to superfluid liquid helium.

The numerical techniques which we use to discretize the
vortex lines, compute their evolution and
de-singularize the Biot-Savart integral (using the superfluid vortex
core radius $a_0$)
are described in our previous papers
\cite{Baggaley-cascade,Baggaley-fluctuations}. Here it suffices to
recall that the relative distance between discretization points along
filaments is held approximately
between $\Delta \xi$ and $\Delta \xi/2$
where $\Delta \xi$ represents the prescribed numerical resolution.
To make direct comparison with related vortex filament calculations
performed by Schwarz \cite{Schwarz} and Tsubota \& Adachi \cite{Adachi}
we run our simulations with vortex core cutoff corresponding to
uniformly rotating solid--body core, and find good agreement \cite{Sherwin2012}.
The reconnection algorithm,
triggered when the vortex separation is closer than $\Delta \xi/2$,
was described by Baggaley \cite{Baggaley-reconnections} who
compared it to other algorithms used in the liquid helium literature. 
A more microscopic model based on the Gross-Pitaevskii equation for
a Bose-Einstein condensate suggests that at each reconnection
a small amount of kinetic energy is turned into sound 
\cite{Leadbeater,Zuccher}. 
To account for this effect, at each reconnection the algorithm
reduces the total vortex length (used as a proxy for energy),
whereas, in the absence of reconnections,
according to Eulerian dynamics the energy is conserved during the evolution.

Our model is based on the assumption
that, apart from isolated reconnection events, 
the vortex filaments are usually far from each other compared to
the vortex core thickness $a_0$. This assumption is justified by
the following argument:
let $R$ and $a$ be respectively
the major and minor radius of the toroidal structure generated by
the piston in Borner's experiment;
using Eq.~\ref{eq:v} and the observed
values of $\Gamma$ and $v$, we estimate the minor radius of the torus, $a$,
and find that the typical distance between vortex lines, 
$\ell \approx n_0^{-1/2} \approx 0.003~\rm cm$ (where
$n_0=N/(\pi a^2)$ is the number of vortex lines per unit cross section
of the torus) is indeed many orders of magnitude bigger than $a_0$.

\section{Initial condition}
\label{sec:init}

\begin{figure}
\centering
\subfloat[] {
  \includegraphics[scale=0.4]{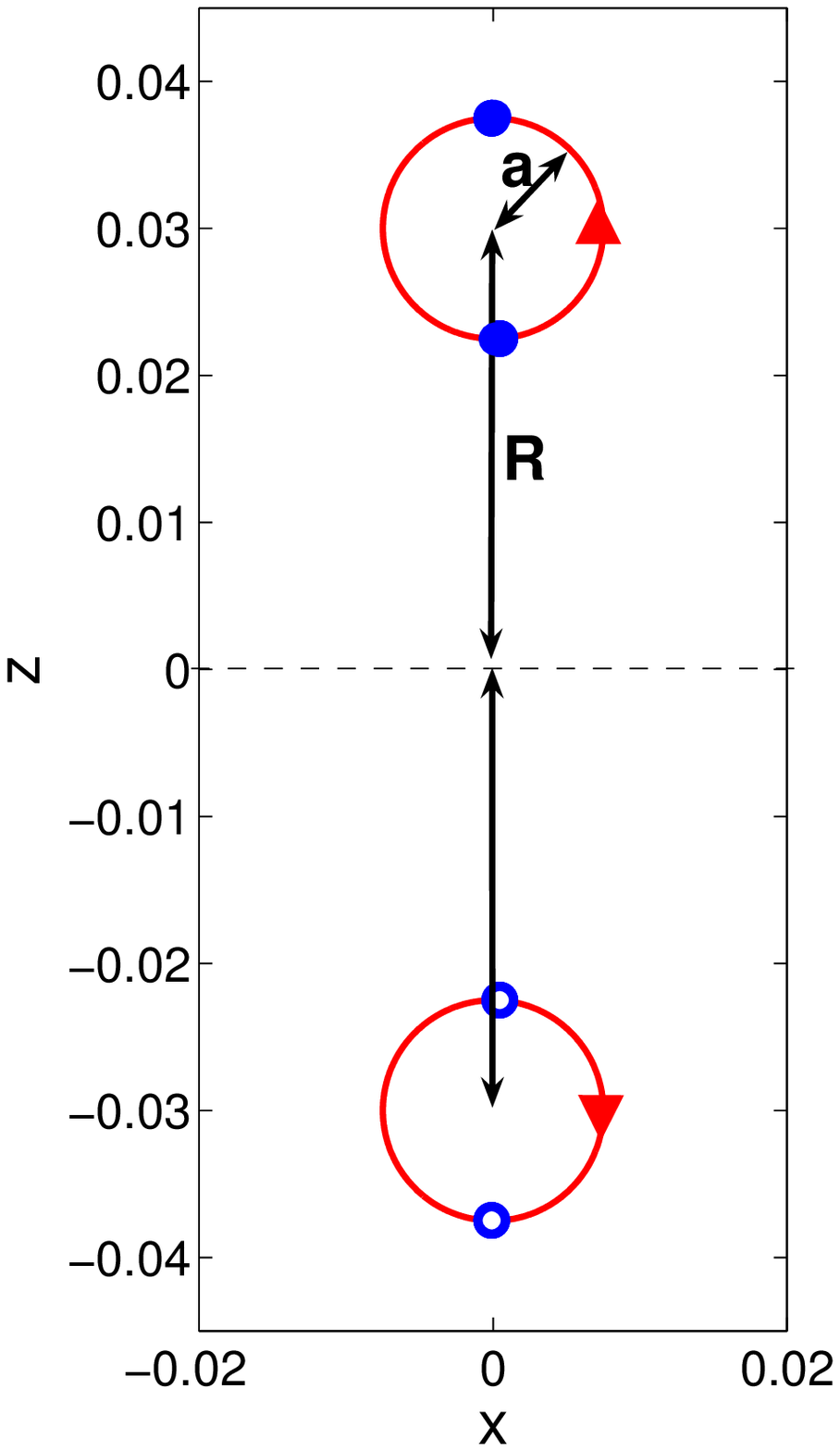}
  \label{fig:sub1}
}
\subfloat[] {
  \includegraphics[scale=0.4]{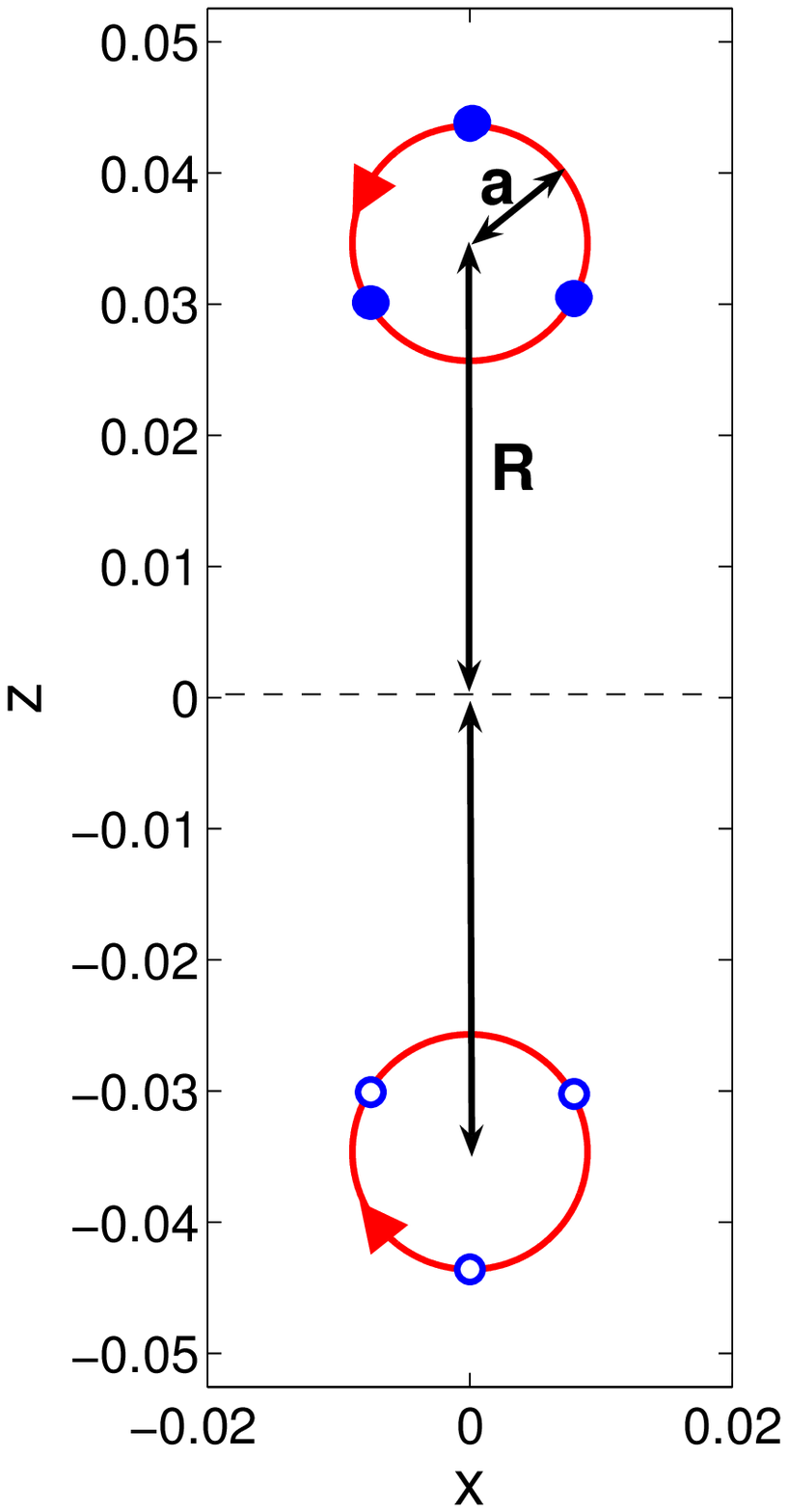}
  \label{fig:sub2}
}\\
\subfloat[] {
  \includegraphics[scale=0.4]{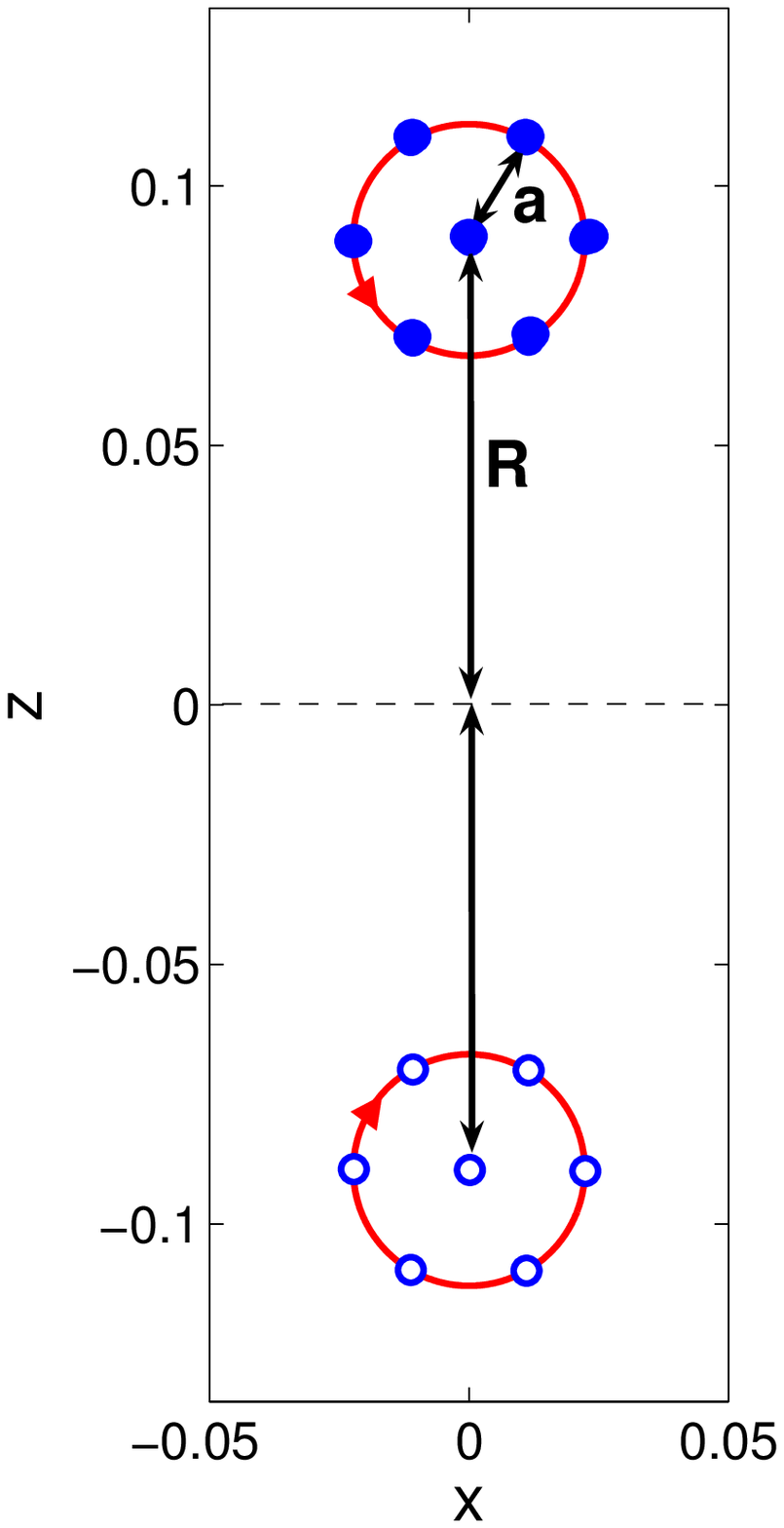}
  \label{fig:sub3}
}
\subfloat[] {
  \includegraphics[scale=0.4]{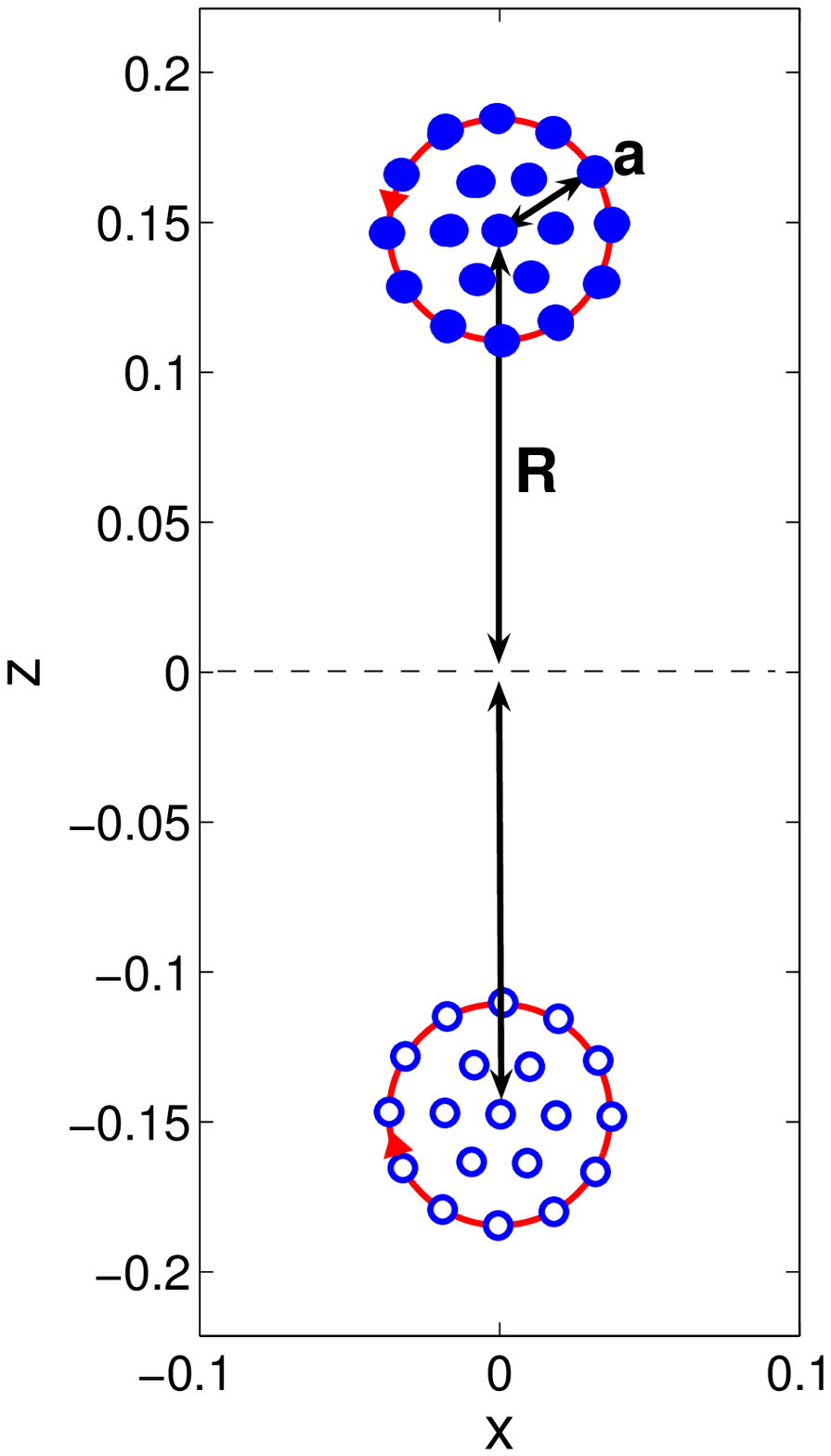}
  \label{fig:sub4}
}
\caption{Initial cross-sectional configurations for
(a): $N=2$, $R=0.03~\rm cm$, $a=\ell/2=0.0075~\rm cm$;
(b): $N=3$, $R=0.035~\rm cm$, $a=\ell/\sqrt{3}=0.00866~\rm cm$;
(c): $N=7$, $R=0.0896~\rm cm$, $a=0.0223~\rm cm$;
(d): $N=19$, $R=0.1476~\rm cm$, $a=0.0369~\rm cm$.
In all cases $R/a=4$. Solid (hollow) blue dots show where the vortex
rings cut the xz plane with positive anticlockwise (negative clockwise)
circulation; the red circles are for visualization only.
}
\label{fig:1}
\end{figure}

We do not have enough experimental
information about the formation of Borner's large-scale
helium vortex ring as it rolls out of the orifice. Moreover, the vortex
filament method is not suitable for describing the process of vortex
nucleation - the Gross-Pitaevskii nonlinear Schroedinger Equation 
should be used instead.
Even neglecting the formation process near the orifice, 
we do not know if there is a steady or periodic vortex configuration 
which travels away from the orifice, whose linear or nonlinear
stability can be studied -- perhaps there is not, and the
problem is essentially an initial value one.
For the sake of simplicity,
our initial condition consists of $N$ coaxial
vortex rings lying in the yz-plane within a torus of major radius 
$R$ and minor radius $a$, oriented to travel in the x-direction..
For $N \ge 7$ the rings are arranged in a hexagonal pattern on the cross
section of the torus such that each successive hexagonal layer is at
distance $\ell$ further way from the centre than the previous layer. 
This configuration is the most energetically favourable for a 
vortex lattice in a rotating cylinder, as described
by Donnelly \cite{Donnelly}. The number of initial vortex rings is thus
$N=3n(n-1)+1=7,19,37,61,91,\cdots$ for $n>1$ where $n$ is the layer's number. 
$N=2$ is a special case, as the two rings
are at the same time co-axial and concentric, lying on the same yz-plane
at distance $\ell=2 a$ from each other. $N=3$ is another special case:
the cross section has the shape of an equilateral triangle 
and $a=\ell/\sqrt{3}$. In general, for $n \geq 2$, $a=(n-1)\ell$.
Examples of initial conditions with $N=2,3,7$ and $19$ rings are shown 
in Fig.~\ref{fig:1}. 

Borner and collaborators measured experimentally the large and
small radii of the large-scale rings, $R$ and $a$, 
as well as the circulation $\Gamma$.
They reported that $R \approx 0.4~\rm cm$ or slightly larger,
consistently with the observation \cite{Didden,Shariff} that, in an 
ordinary fluid, the ring which is generated by the piston-cylinder 
arrangement is somewhat larger than the orifice, and that $R/a \approx 4$.
The estimated values of the intervortex distance $\ell$
vary from approximately $10^{-2}$ to $10^{-4}~\rm cm$.

\section{Computational constraints and tests}
\label{sec:constraints}

Our main computational constraint is the evaluation of  
Biot-Savart integrals, which grows with $N_p^2$, where
$N_p$ is the number of discretization points along the vortices. 
$N_p$ is determined by
the parameter $\Delta \xi$ (the distance between discretization
points) and the total vortex length (which depends on the number $N$ of
vortex rings and their radius).
Clearly $\Delta \xi$ must be smaller
than the distance between vortex rings, which depends on $N$ and $a$.
Furthermore, to ensure the numerical stability of the evolution,
the time step $\Delta t$ 
must be small enough to resolve
the shortest, fastest-rotating Kelvin waves (helical perturbations along
the vortex filaments) which have wavelength
of the order of $\Delta \xi$. Therefore a decrease of $\Delta \xi$ requires
a decrease of $\Delta t$. This means that, 
whereas the cost of a single time-step scales as $N_p^2$,
the cost of computing the evolution up to a given time actually scales
as $N_p^3$.

We do not have the computing power to compute
the long-term evolution of vortex bundles with the very large number of 
vortex rings of Borner's experiments ($N \sim 10^3$),
but we shall see that the values of $N_p$ which we can afford give us
qualitative information
about the motion of small vortex bundles, and provide 
us with insight into the conditions of the experiments.

\begin{figure}
\centering
\includegraphics[scale=0.7]{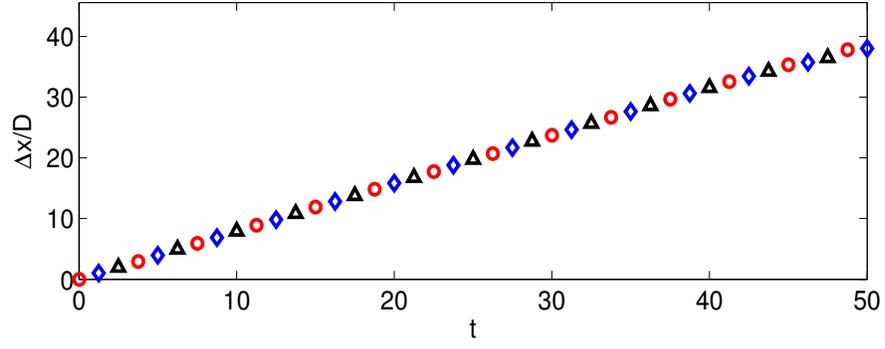} 

\vskip 3cm

\caption{
Ordinary leapfrogging of $N=2$ vortex rings. 
Distance travelled $\Delta x$ in the
x-direction in units of initial diameter $D=2R$ vs time $t~\rm (s)$
at $R=0.03~\rm cm$, $a=0.0075~\rm cm$ ($R/a=4$) for
$\Delta \xi=0.00149~\rm cm$ (red circles), 
$\Delta \xi=0.001~\rm cm$ (blue diamonds), and
$\Delta \xi=0.0005~\rm cm$ (black triangle), where the parameter $\Delta \xi$ 
determines the spatial numerical resolution. The time step
is respectively $\Delta t=5 \times 10^{-5}~\rm s$, $2.5 \times 10^{-5}~\rm s$
and $0.625 \times 10^{-5}~\rm s$.
}
\label{fig:2}
\end{figure}

To test the computer code, we verify that a single ($N=1$)
vortex ring travels
at constant energy $E$ with steady translational velocity $v$.
The velocity is computed from the time to travel a given distance
(for $N>1$ we track the motion of the centre of vorticity instead).
For example, for a single vortex ring of radius $R=0.0896~\rm cm$
with $\Delta \xi=0.00015~\rm cm$,
we find that $v=0.01573~\rm cm/s$ differs from the value predicted by
Eq.~(\ref{eq:v}) with the Rankine core model, $v=0.01556~\rm cm/s$,
by approximately $1\%$.
The energy (per unit density) is computed using the
expression

\begin{equation}
E=\frac{1}{2} \int_V \bv^2 dV=\kappa \oint_{\cal L} \bv \cdot \bs \times \bs' d\xi,
\label{eq:E-Saffman}
\end{equation}
\noindent
where we have assumed that the vorticity is concentrated along 
the vortex filaments and that the velocity goes to zero at infinity.
For a single vortex ring, we find that energy is conserved 
to within $10^{-12} \%$
for typical time scales and length scales investigated. 
The case of $N=2$ leapfrogging vortex rings is more challenging because 
energy conservation is tested under the action of
a re-meshing algorithm (the number of discretization points on a ring
changes as the ring grows and shrinks in size).
We find that energy is conserved to within $0.5\%$.
Fig.~\ref{fig:2}
shows stable leapfrogging of two concentric coaxial rings
for a distance of about 40 diameters.
The figure also demonstrates that the motion does not depend on the
numerical resolution parameter $\Delta \xi$. 
The speed of translation of the vortex bundles is tested by
changing the resolution $\Delta \xi$. For example, consider the
bundle of $N=3$ vortex rings with $R=0.06~\rm cm$, $\ell=0.015~\rm cm$,
$a=\ell/\sqrt{3}$, $R/a=6.92$. Using resolution $\Delta \xi=0.00149~\rm cm$
($\ell/\Delta\xi=10$) we find that (with time step
$\Delta t=5 \times 10^{-5}~\rm s$) it travels the distance $\Delta x/D=10.24$
(compared to its initial diameter) in $t=40~\rm s$, hence the average
speed is $v=0.031~\rm cm/s$. If we reduce the resolution to
$\Delta \xi=0.00125~\rm cm$ (with $\Delta t= 3 \times 10^{-5}~\rm s$)
and $\Delta \xi=0.00075~\rm cm$ (with $\Delta t=1.5 \times 10^{-5}~\rm s$)
we find that in both cases it reaches $\Delta x/D=11.525$ in $t=45~\rm s$,
which still corresponds to $v=0.031~\rm cm/s$.

Finally, we test energy conservation during the evolution of larger
vortex bundles ($N=3, 7, 19$). For the length scales and time scales
investigated, we find that the relative error $(E(t)-E(0))/E(0)$
is always well below $1 \%$, provided that no vortex reconnections
have taken place.

\section{Results}
\label{sec:results}

\subsection{Coherent motion of toroidal vortex bundles}
\label{subsec:coherent}

\begin{figure}
\begin{center}
\includegraphics[scale=0.65]{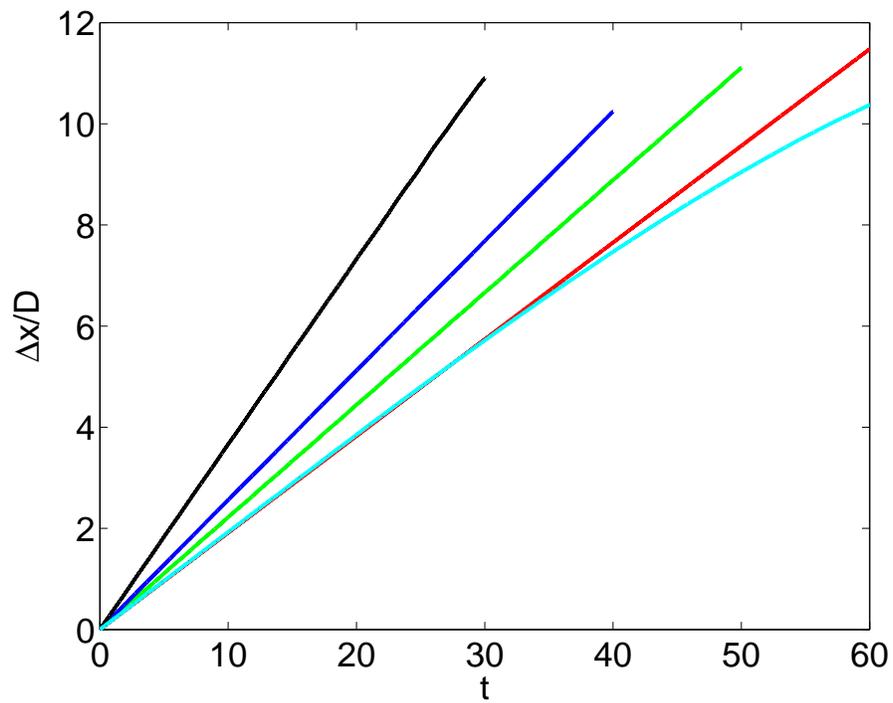} 
\end{center}

\vskip 3cm

\caption{
Distance travelled $\Delta x$ in units of initial diameter $D$ vs time 
$t$ (s) for vortex bundles consisting of $N=1$ (red, fourth curve from left), 
$N=2$ (green, third curve from left), $N=3$ (blue, second curve from left), 
$N=7$ ( black, first curve at left) and $N=19$ (cyan, first curve at right)
vortex rings.
}
\label{fig:3}
\end{figure}

We now proceed to investigate  the motion of 
bundles of vortex rings with $N=1,3,7,19$. 
The parameters of the simulations are
shown in Table~\ref{tab:1}. In  all runs we set up the initial condition
so that $\ell=0.015~\rm cm$ and $R=0.06~\rm cm$. For $N=19$, since
$a=2 \ell$, $R=0.06~\rm cm$ would have meant that $a=0.003~\rm cm$, hence
$R/a=2$, and the leapfrogging motion, at least in two-dimensions, may
not be stable \cite{Acheson}. Therefore, for $N=19$ only,
we set $R=0.12~\rm cm$ and $R/a=4$.
In all cases we find that the vortex bundles travel at 
approximately constant speed without disintegrating,
and reach the distance $10 D$ 
(observed by Borner) after approximately $50$, $40$, $35$ 
and $25~\rm s$ respectively.
Fig.~\ref{fig:3} shows the distance travelled, $\Delta x$,
in units of the initial diameter $D$ vs time $t$ for toroidal bundles
consisting of $N=1,2,3,7$ and $19$ vortex rings; it is apparent that
all bundles travel at essentially constant velocity (the largest bundle
slows down slightly).

Table~\ref{tab:2} gives quantitative information about each simulation:
average curvature $\bar c$, translational velocity $v$ of centre of
vorticity, 
distance travelled $\Delta x$ in units of the initial diameter $D$, and
vortex length $\Lambda$. 
Changes in length and curvature reflect the appearance of Kelvin waves
along the filaments. 
The labels ``start", ``finish" and ``reconnection"
denote the times at which we start and stop a simulation, and the time
at which the first vortex reconnection (if any) has taken place.
As mentioned in Section~\ref{sec:method},
classical Eulerian dynamics breaks down at the first reconnection; 
beyond the first reconnection, our results are still valid 
for superfluid liquid.

\subsection{Velocity and energy}

\begin{figure}
\centering
\subfloat[] {
  \includegraphics[scale=0.4]{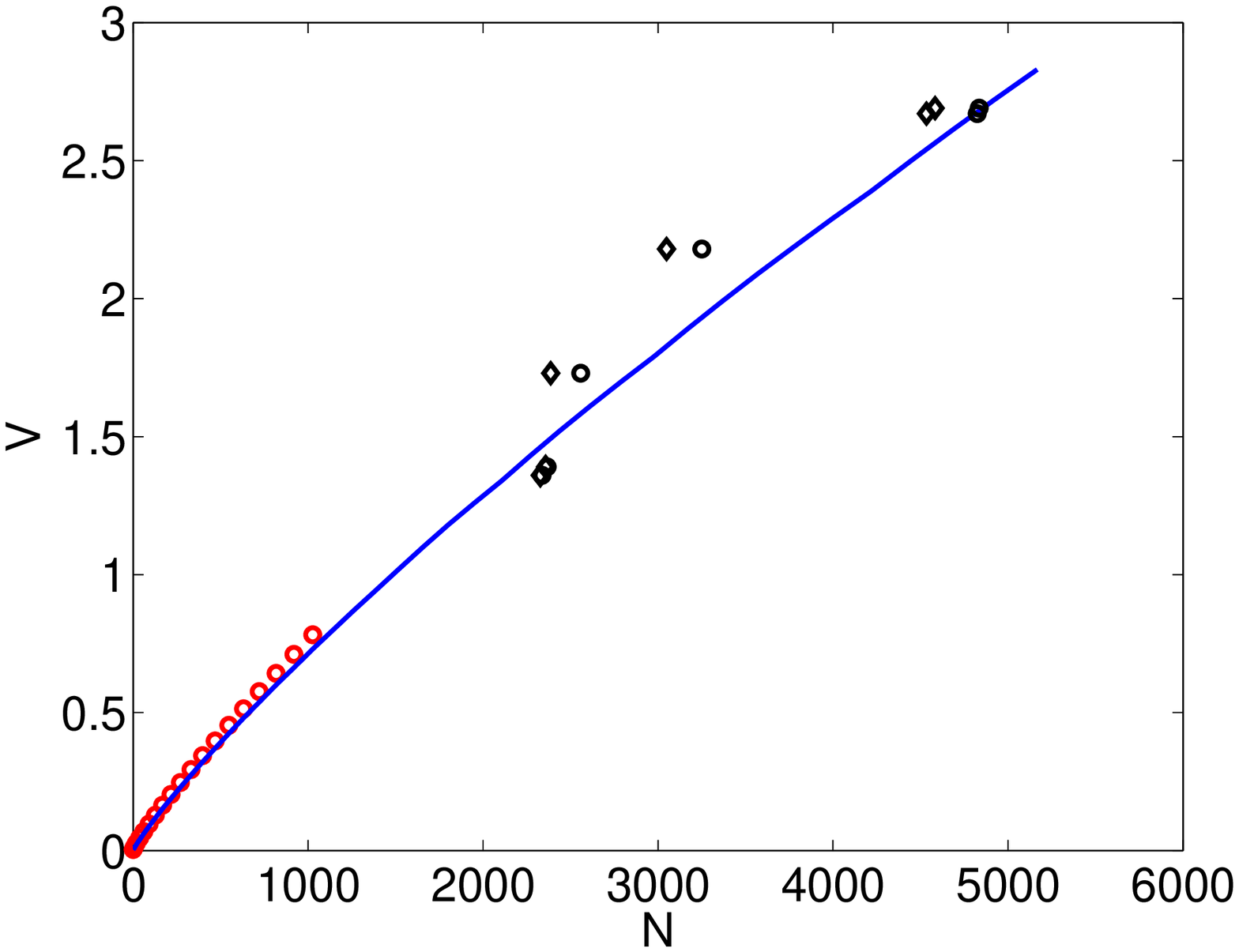}
}
\subfloat[] {
  \includegraphics[scale=0.4]{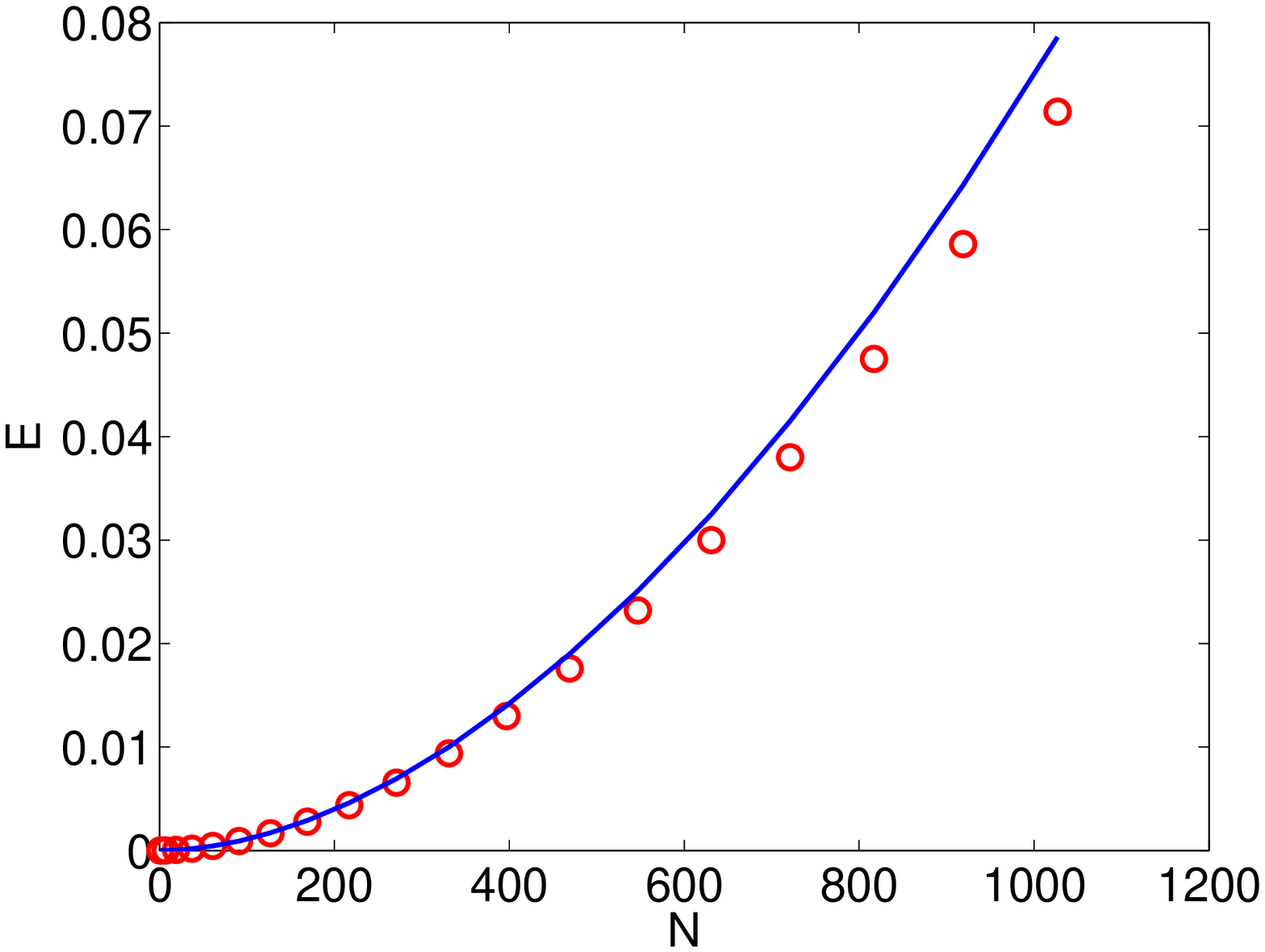}
}

\vskip 3cm

\caption{(a): Translational velocity, $v$, and (b): energy, $E$, of vortex bundles with
$N$ rings vs number of rings $N$.
Red circles: numerical simulations for $N \leq 1027$ 
(parameters used: $R=0.4~\rm cm$, $\ell=0.003~\rm cm$, $a=(n-1) \ell$ 
where $n$ is the number of hexagonal layers in the initial condition, 
$\Delta t=10^{-6}~\rm s$ and $\Delta \xi \approx 0.0335~\rm cm$, 
corresponding to about 100 discretization points per ring). Black squares
and black diamonds: Borner's experimental values at distance $x=1.81~\rm cm$ and
$x=4.65~\rm cm$ respectively from the orifice. Blue lines: models for velocity
$v'$ and energy $E'$ from Eq.~\ref{eq:v'} and
\ref{eq:E'} respectively.
}
\label{fig:4}
\end{figure}

As explained in Section~\ref{sec:constraints},
the available computing power limits our simulations
to a number of vortex rings (up to $N=19$ in the previous subsection) which is
much less than in Borner's experiment. To bridge this gap,
we have performed shorter numerical simulations of only few time steps in order
to determine the initial instantaneous velocity of vortex bundles 
with experimentally realistic values of $N$, up to $N=1027$, 
the first ${\rm O}(10^3)$ hexagonally
centred number. The limitation of this comparison is that the vortex bundle 
does not have time to distort its initially circular shape and acquire the 
slightly elliptical cross section which we observe in long-term simulations.
Fig.~\ref{fig:4}(a) shows that the translational velocity $v$ determined 
numerically in this way for large vortex bundles (red circles) 
is in good agreement with
Borner's experiments (black squares and diamonds).

In the limit $R \gg a$, simple models of the translational velocity and
energy of a vortex bundle of $N$ vortex
rings is obtained by generalizing Eq.~\ref{eq:v} and ~\ref{eq:E}:

\begin{equation}
v'=\frac{N \kappa}{4 \pi R} \left( \ln{(8R/a)}-\alpha \right),
\label{eq:v'}
\end{equation}

\begin{equation}
E'=\frac{1}{2} N^2 \kappa^2 R \left( \ln{(8R/a)}-\beta \right),
\label{eq:E'}
\end{equation}

\noindent
where $a$ is now the minor radius of the torus 
and $N \kappa$ has replaced $\kappa$. 
The blue line in Fig.~\ref{fig:4}(a,b), calculated for the hollow core model,
shows that this model is indeed a good approximation.

\subsection{Generalized leapfrogging}

\begin{figure}
\centering
  \includegraphics[scale=0.35]{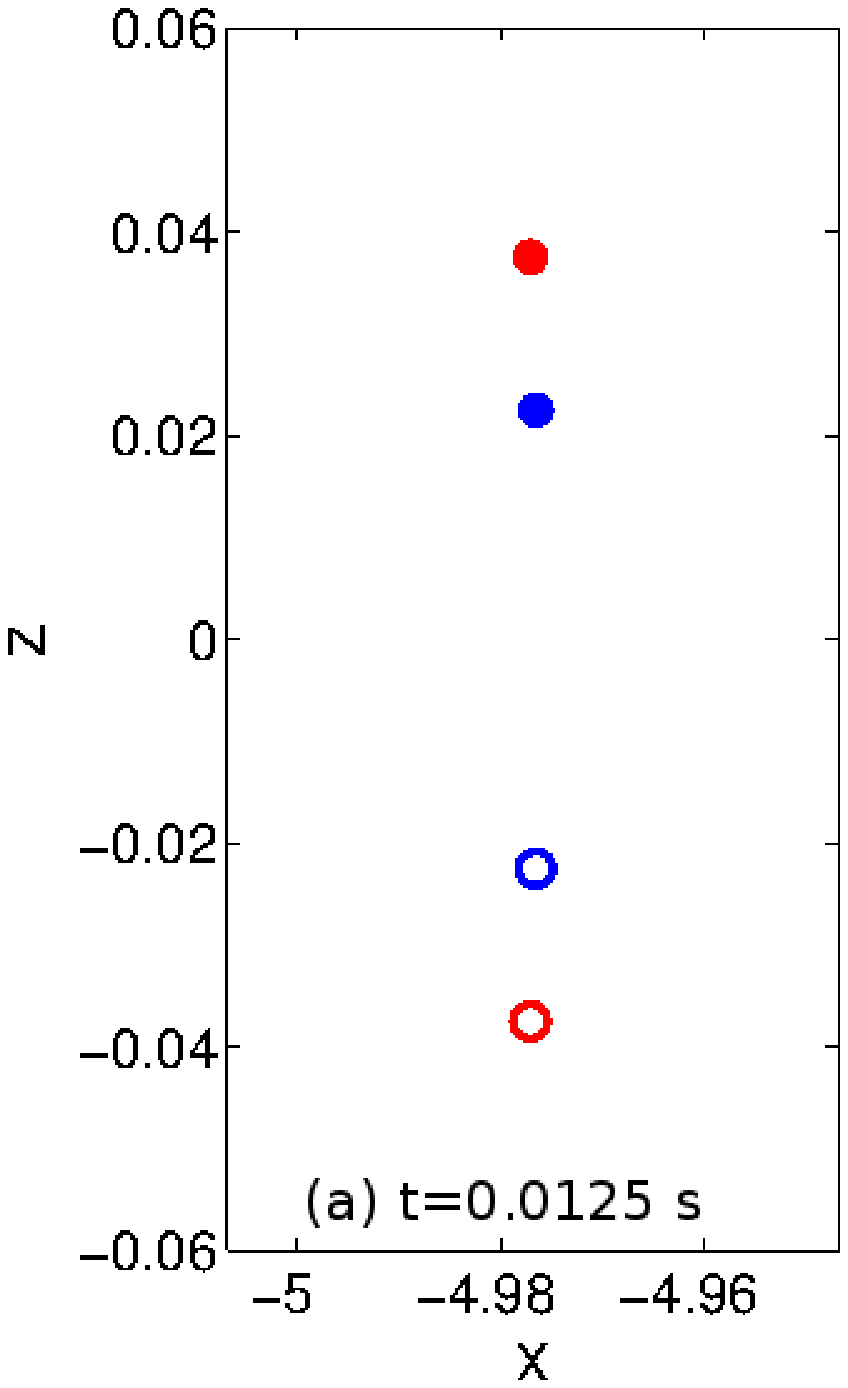}
  \includegraphics[scale=0.35]{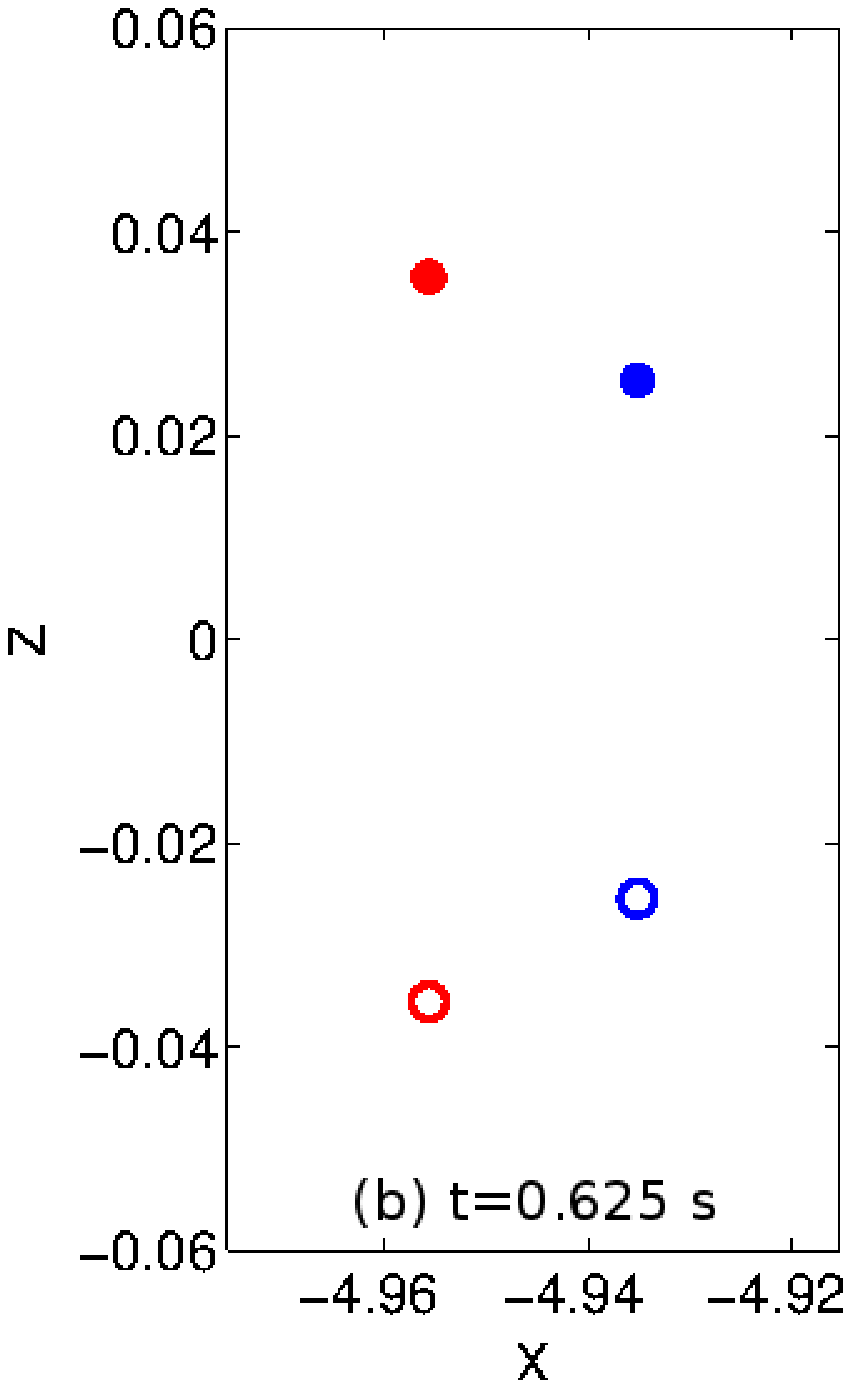}
  \includegraphics[scale=0.35]{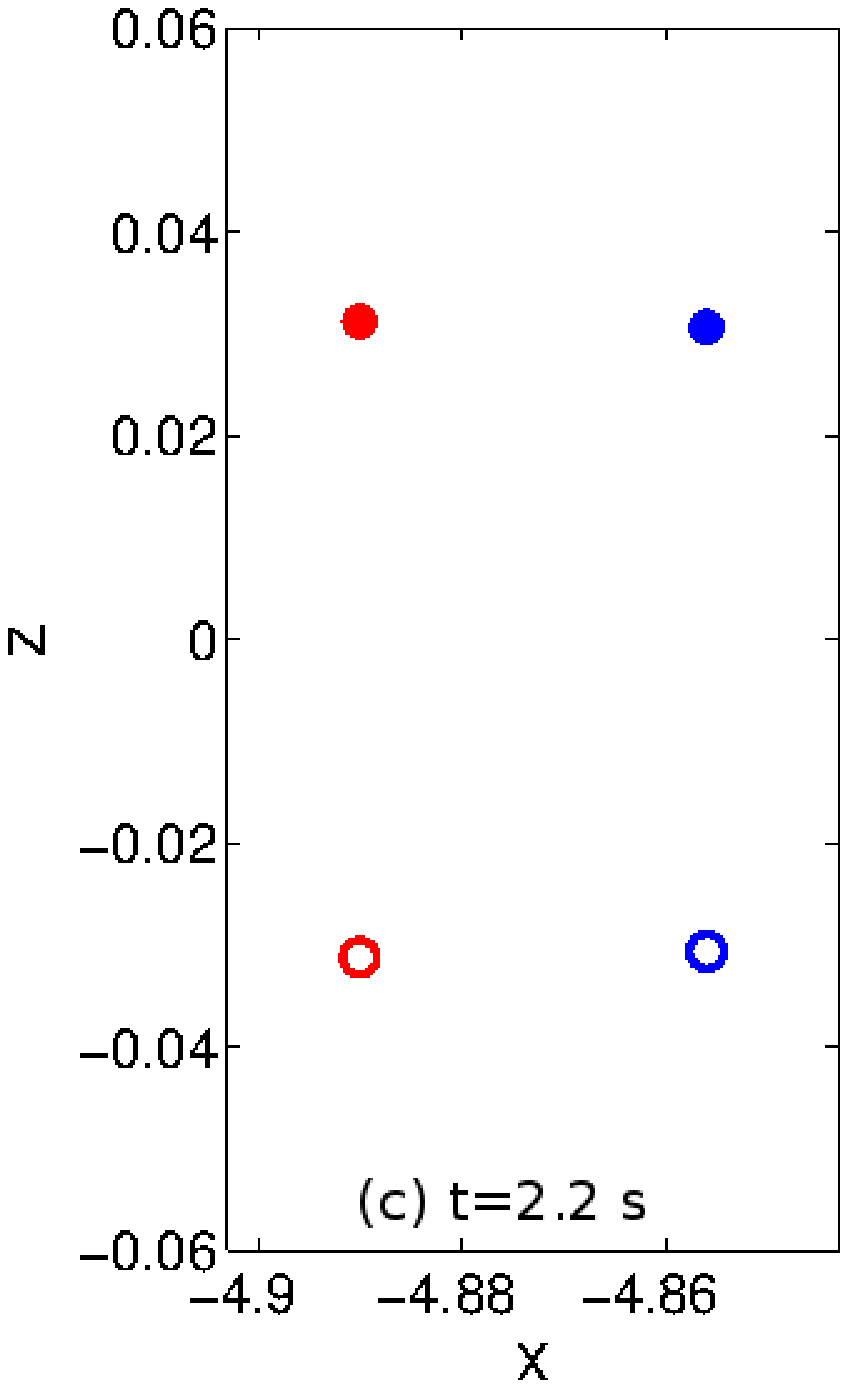} \\
\vskip 1cm
  \includegraphics[scale=0.35]{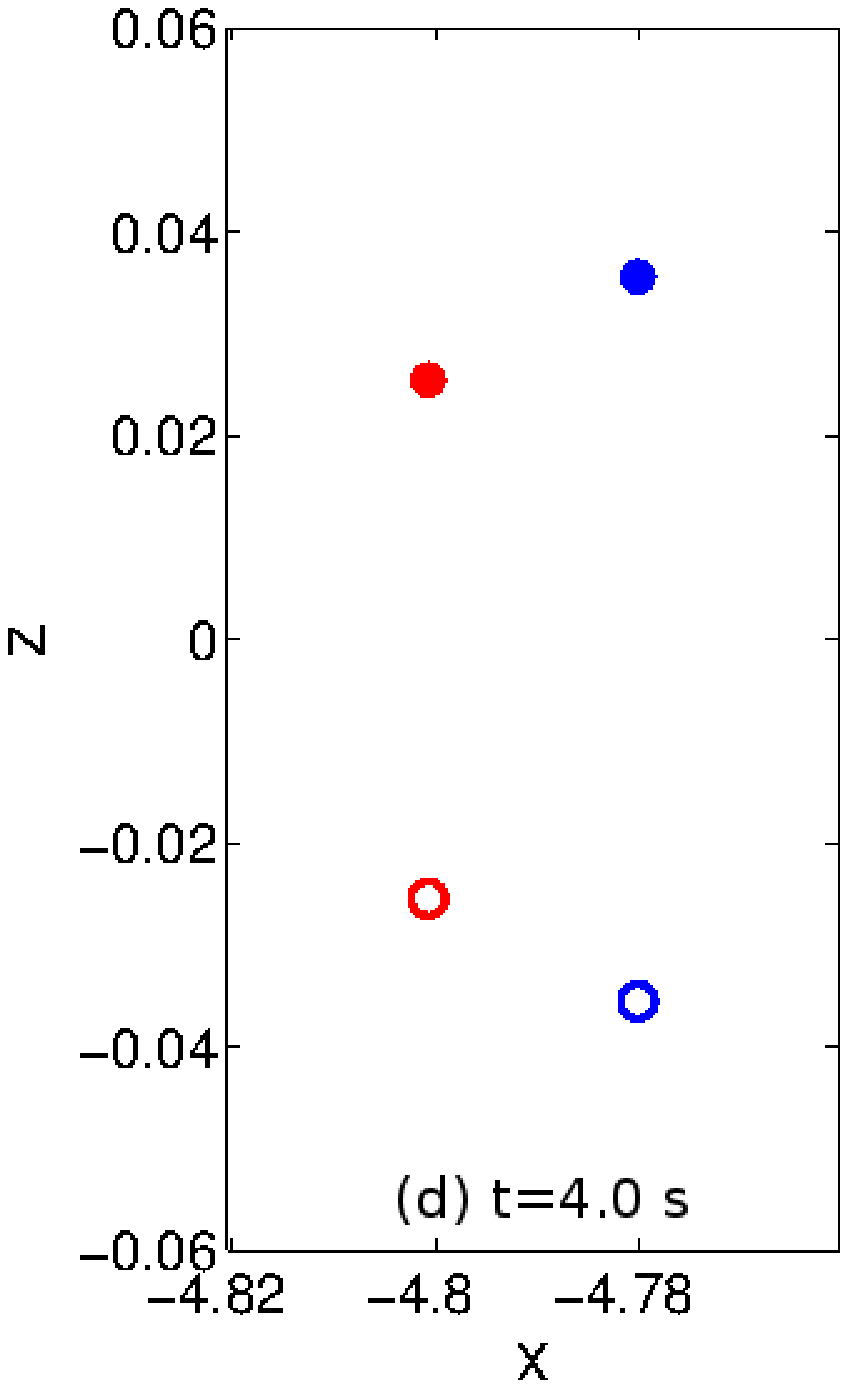}  
  \includegraphics[scale=0.35]{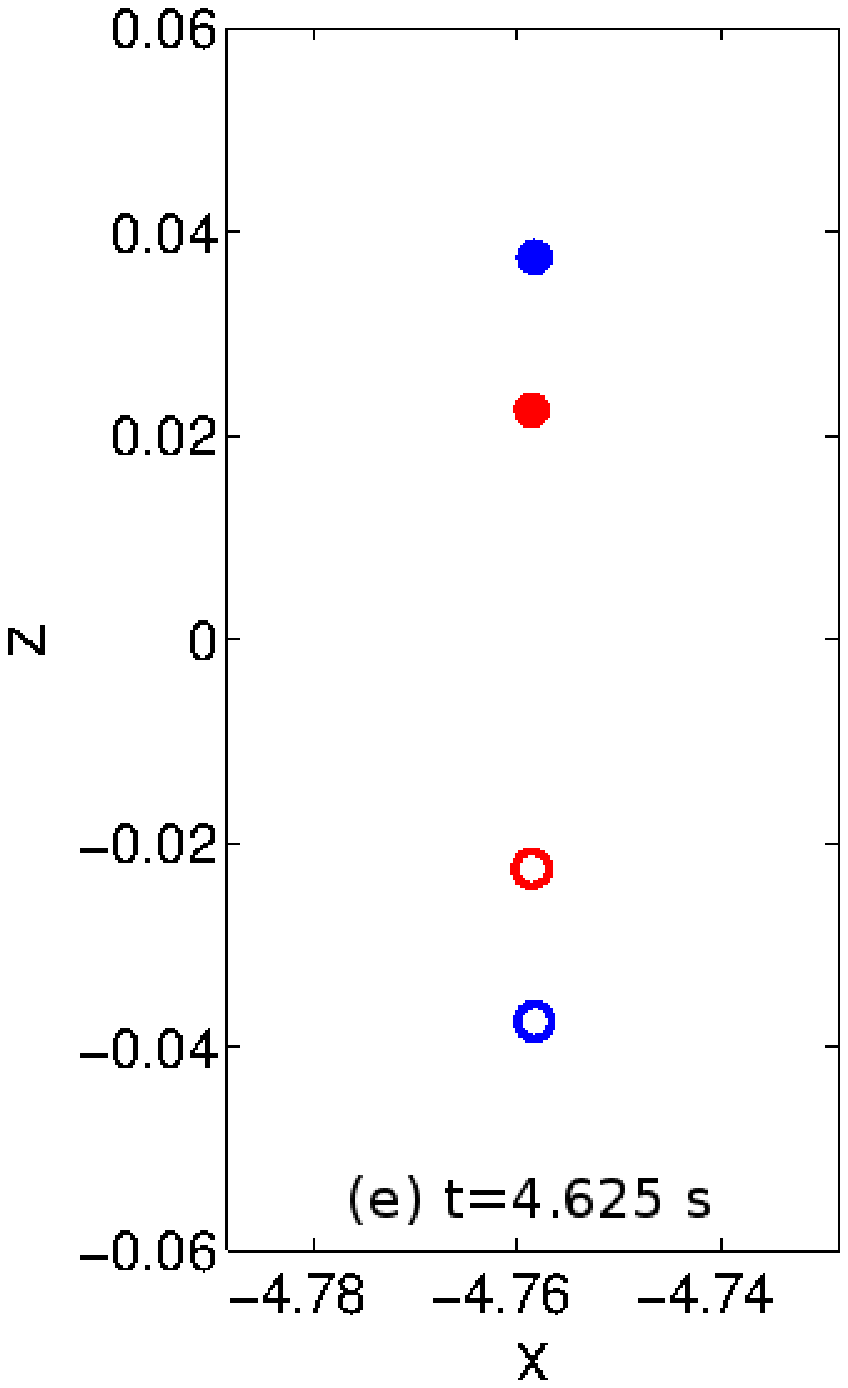}  
  \includegraphics[scale=0.35]{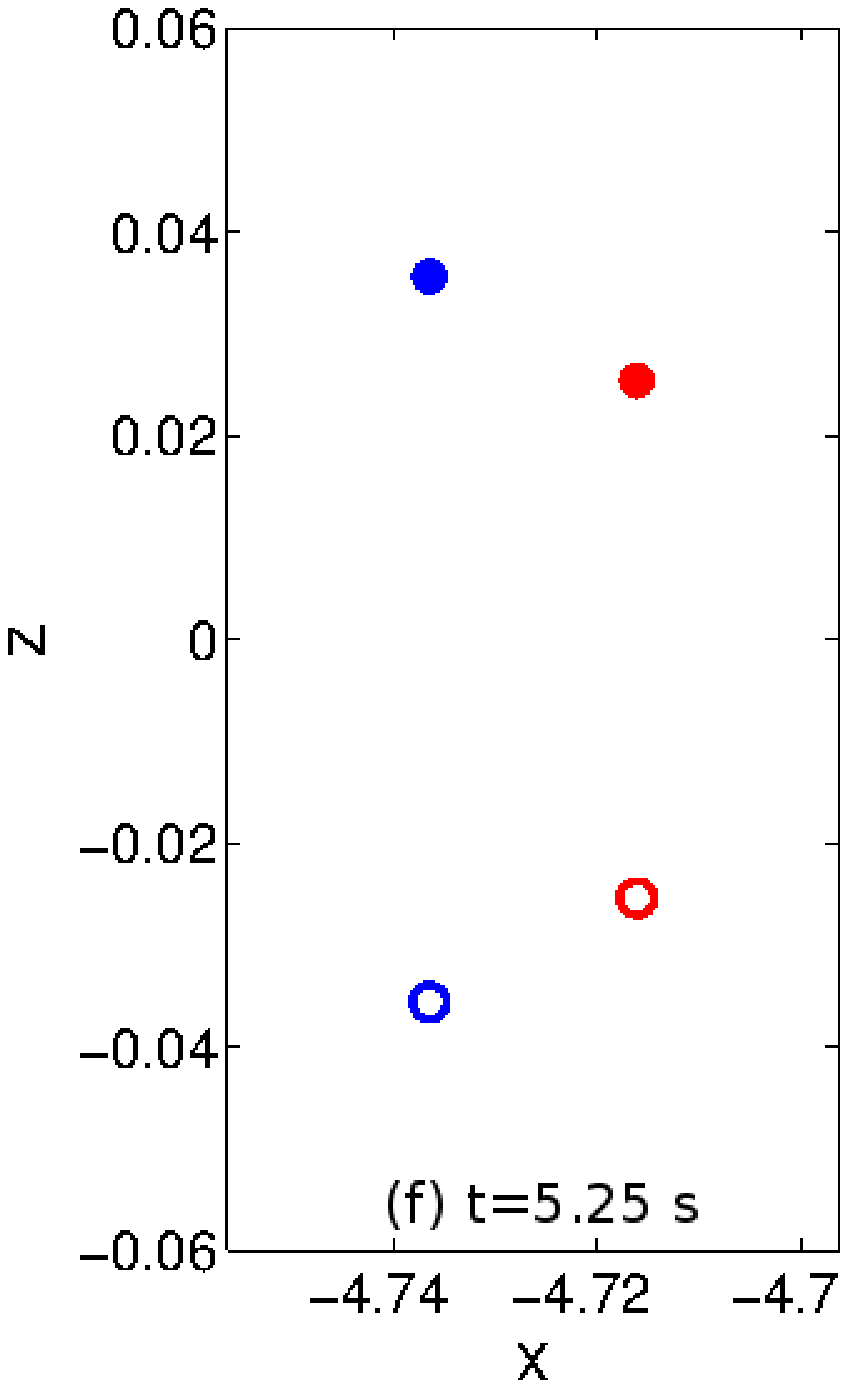}\\
\vskip 1cm
%
  \includegraphics[scale=0.35]{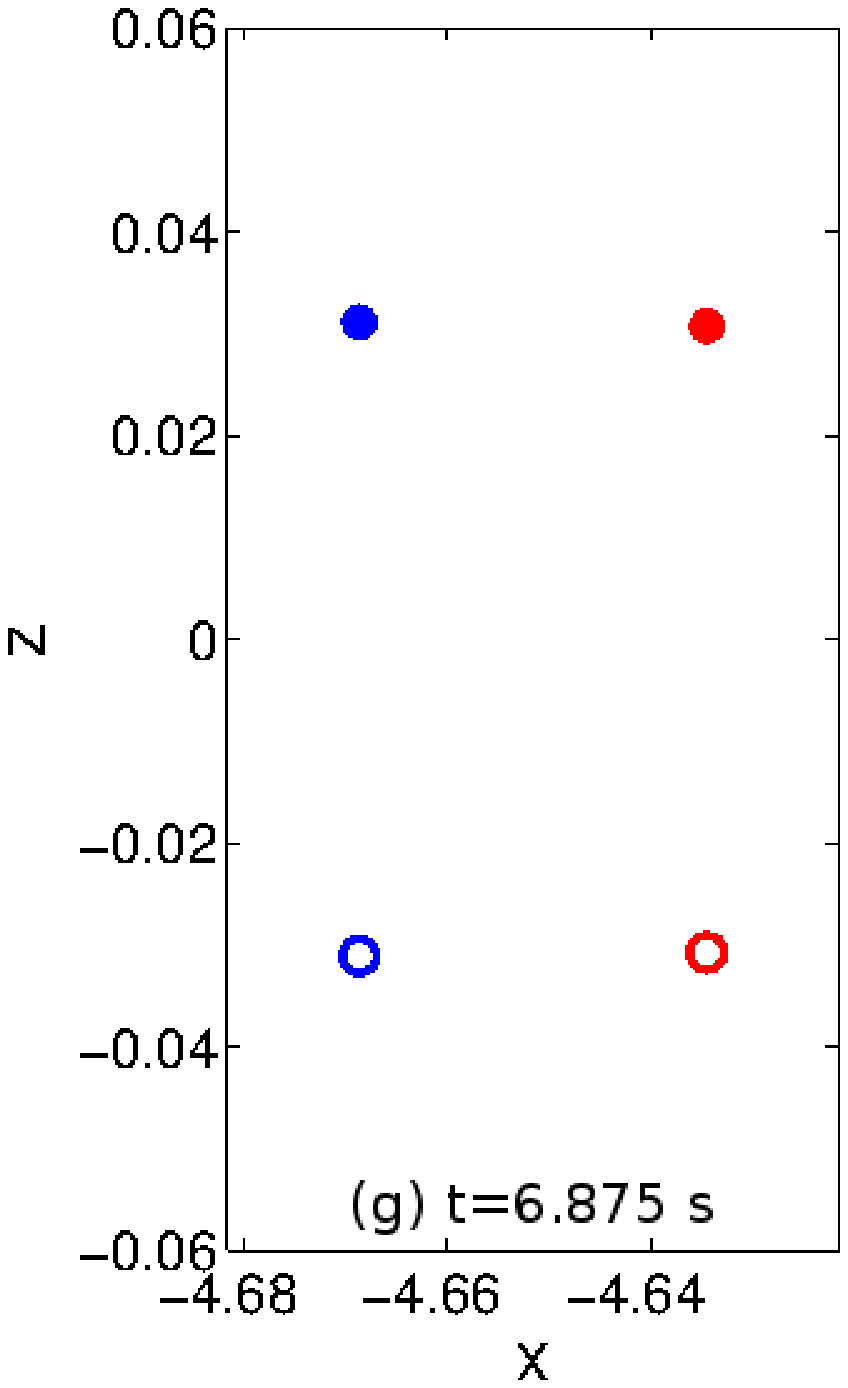}  
  \includegraphics[scale=0.35]{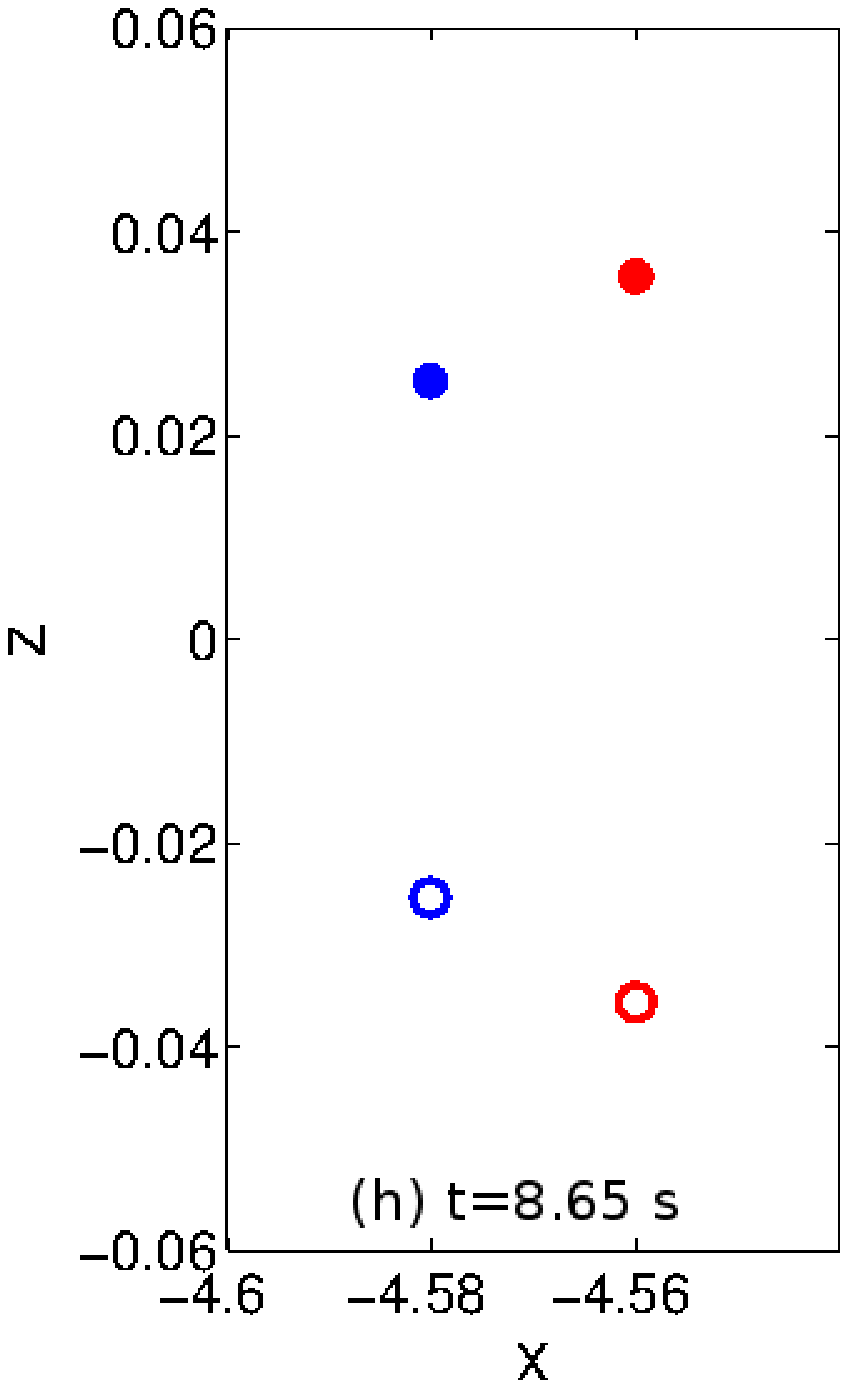}  
  \includegraphics[scale=0.35]{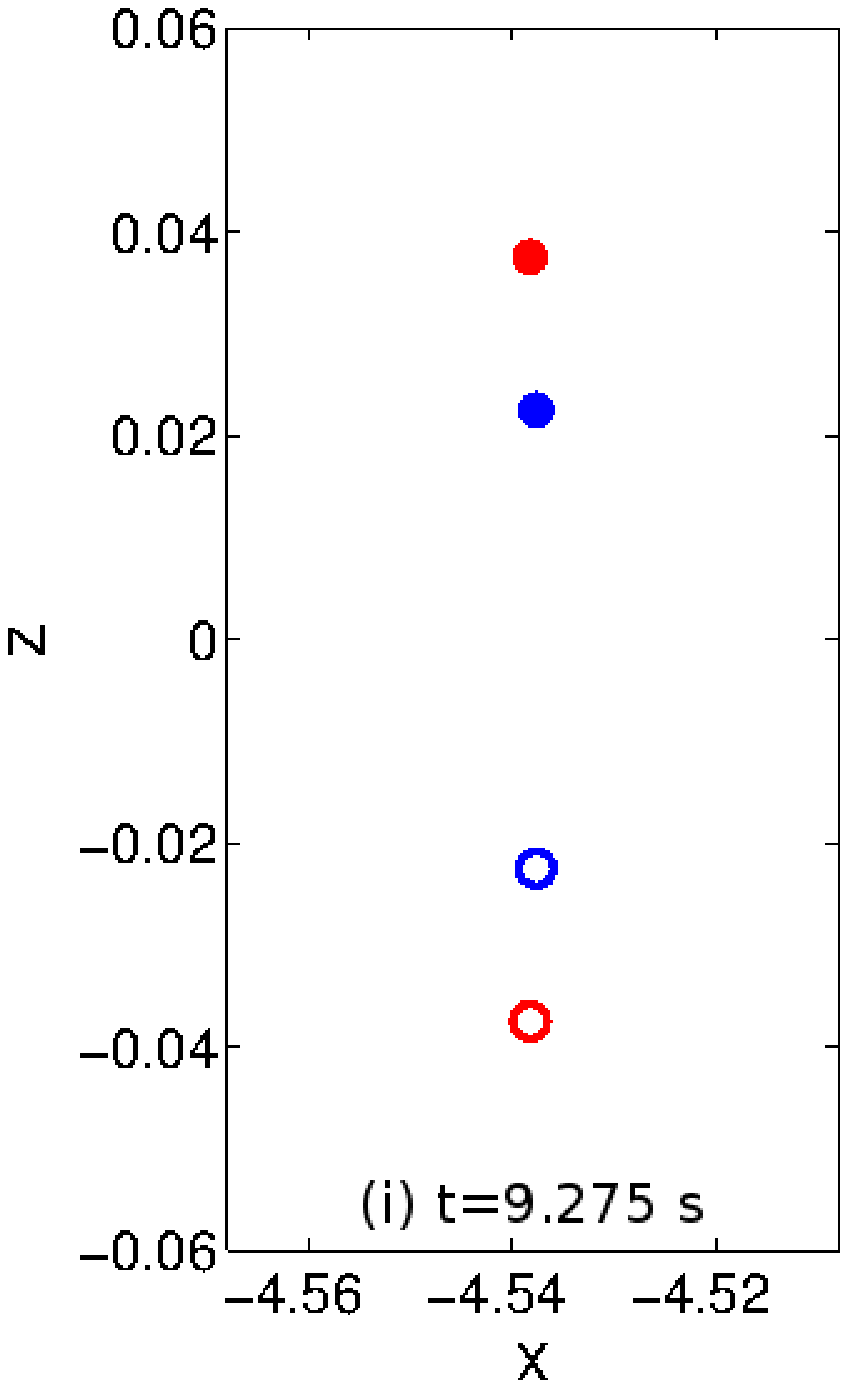}

\vskip 3cm

\caption{Ordinary leapfrogging.
Cross-sectional slice of vortex bundle for $N=2$ at different times. 
The symbols denote the positions where the
vortex rings cut the xz-plane. The symbols' size is arbitrary. Solid
symbols correspond to anticlockwise (positive) circulation, hollow symbols to
clockwise (negative) circulation.
It is apparent that the motion of the vortex positions on this plane
is elliptical. Parameters: $R=0.003~\rm cm$, $a=0.0075~\rm cm$, $\ell=0.015~\rm cm$
and $R/a=4$.}
\label{fig:5}
\end{figure}

The coherent motion of the vortex bundles which we obtain (before
instabilities and reconnections occur) results from a generalised form
of the familiar, ordinary leapfrogging of $N=2$ vortex rings.
As the  toroidal vortex structure moves along
its axis, the individual vortex rings at the back of the torus shrink
and slip through the middle, speeding up their translational
velocity; then they grow in size, slow down, and move round the outer part
of the torus. The ordinary leapfrogging of $N=2$ vortex rings
is demonstrated in Fig.~\ref{fig:5}, 
which shows two-dimensional cross-sectional slices
during a full leapfrogging cycle. 

\begin{figure}
\centering
  \includegraphics[scale=0.38]{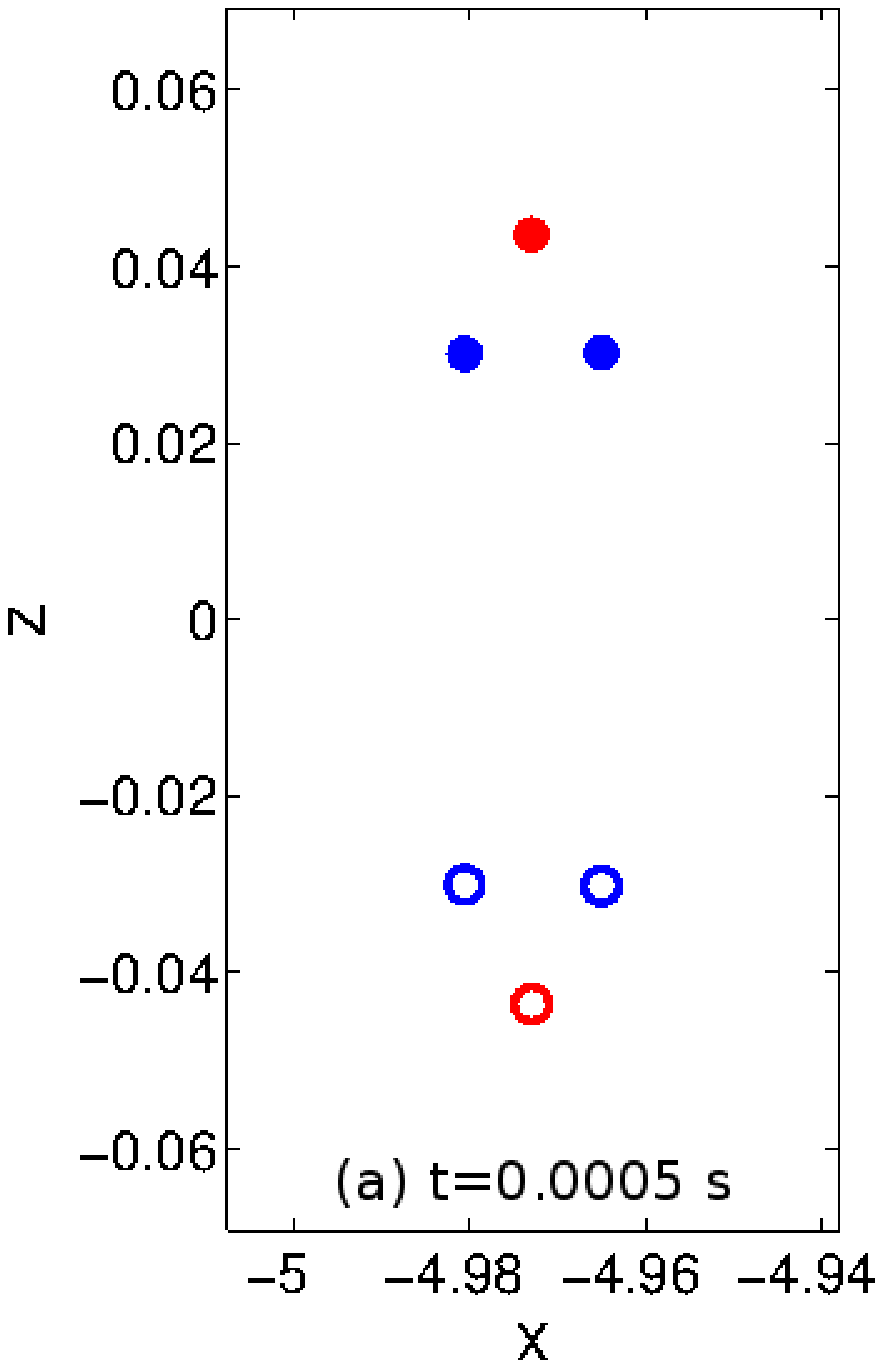}
  \includegraphics[scale=0.38]{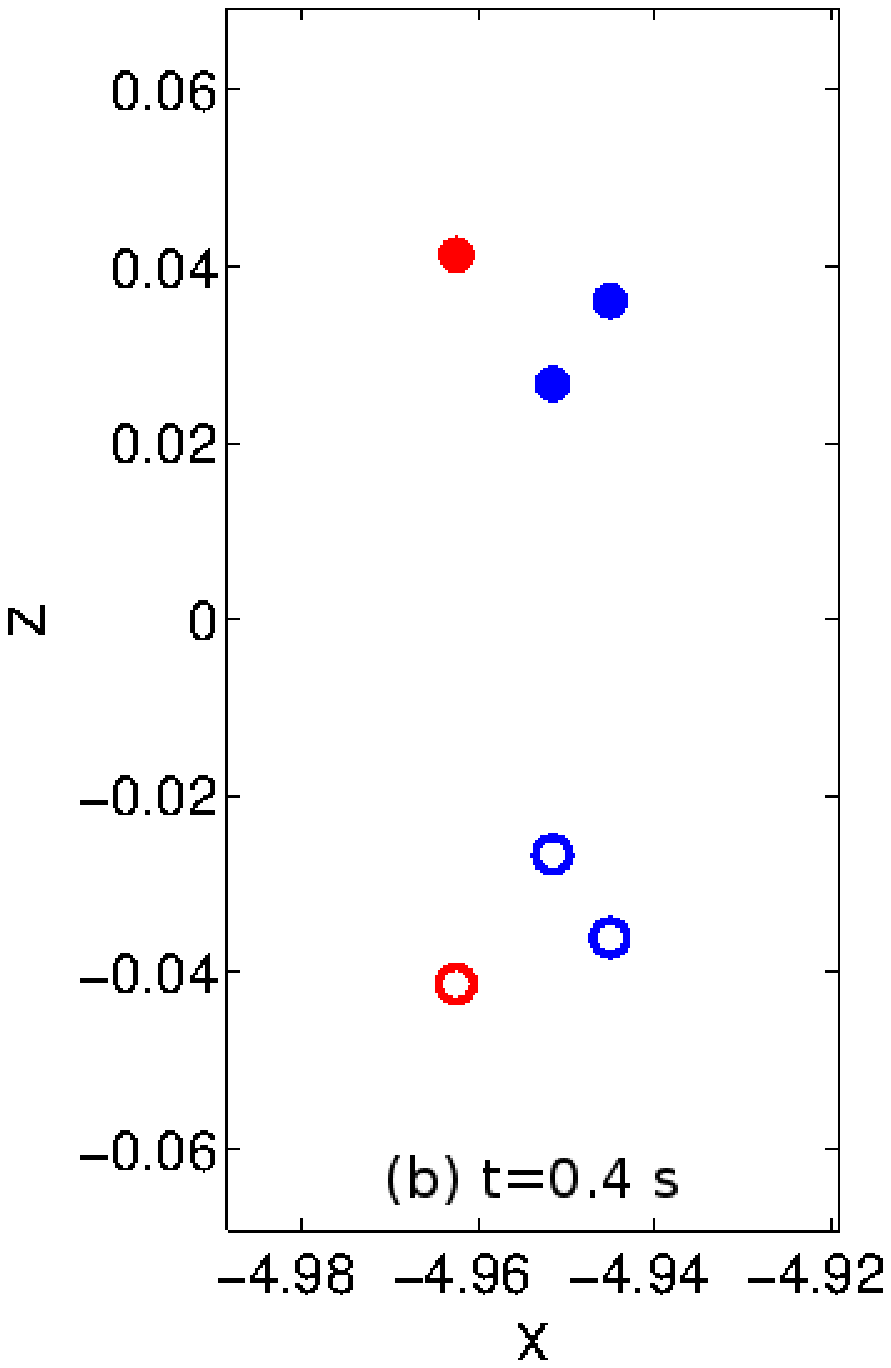}
  \includegraphics[scale=0.38]{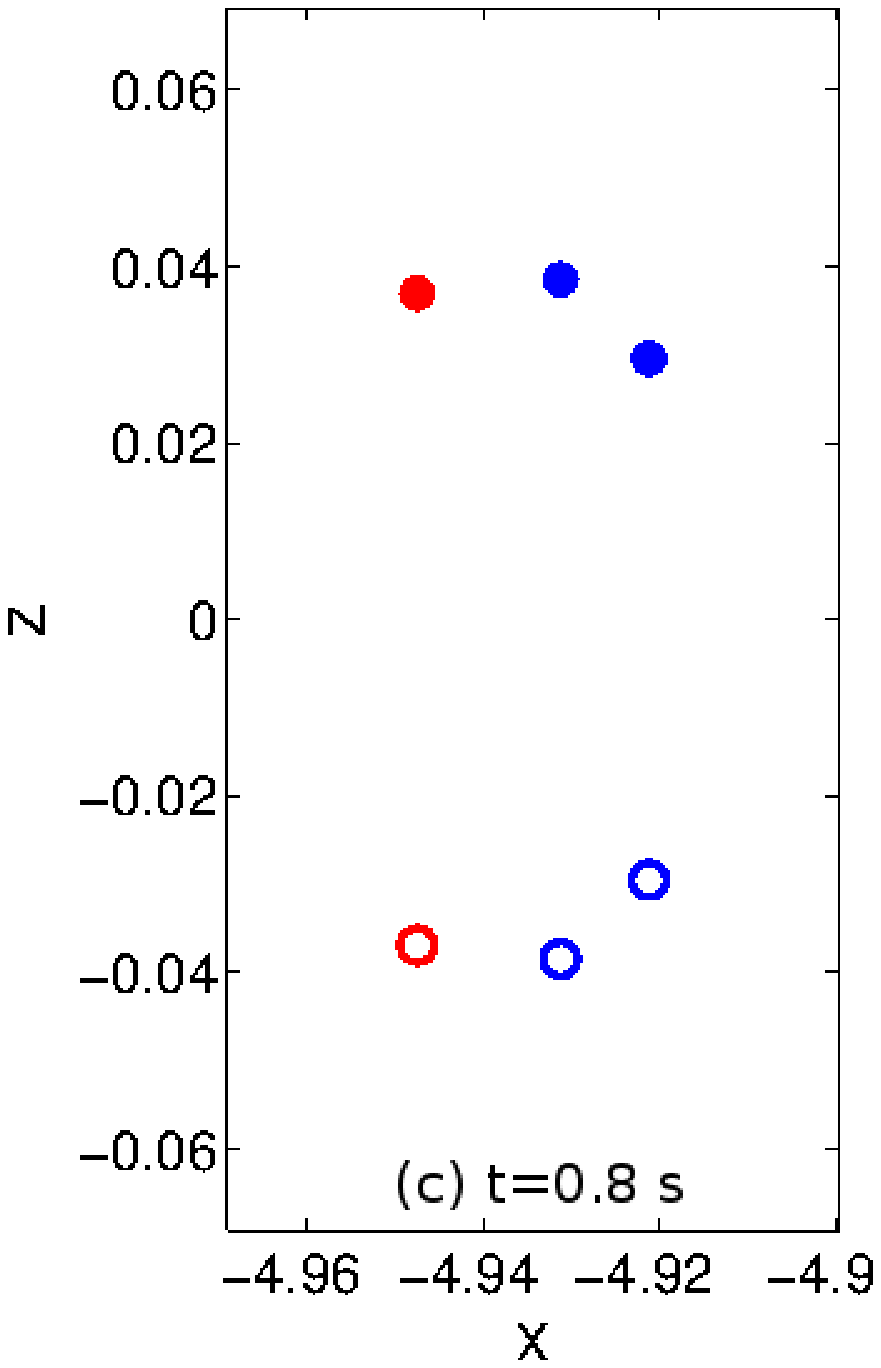}\\
\vskip 1cm
%
  \includegraphics[scale=0.38]{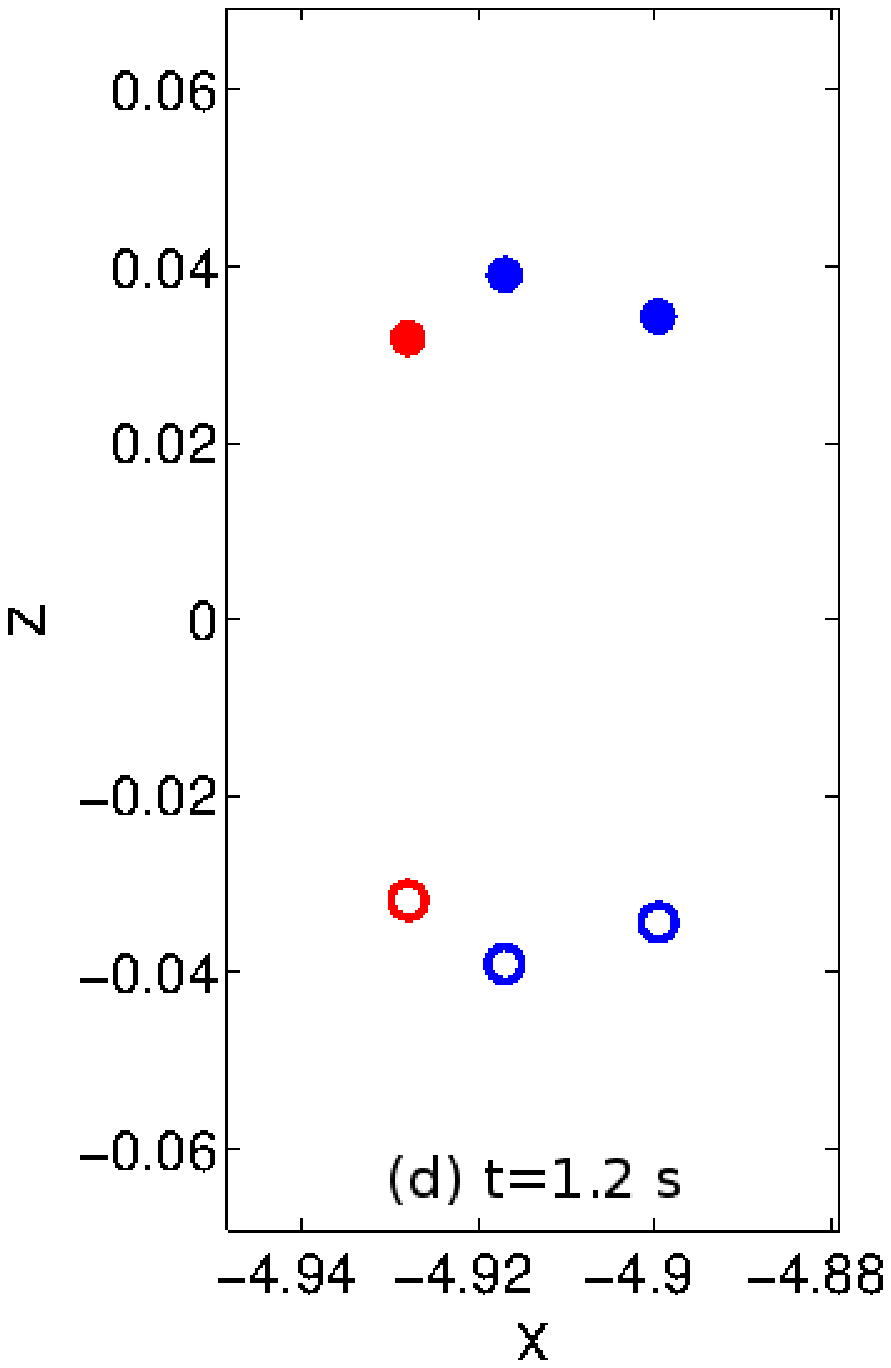}  
  \includegraphics[scale=0.38]{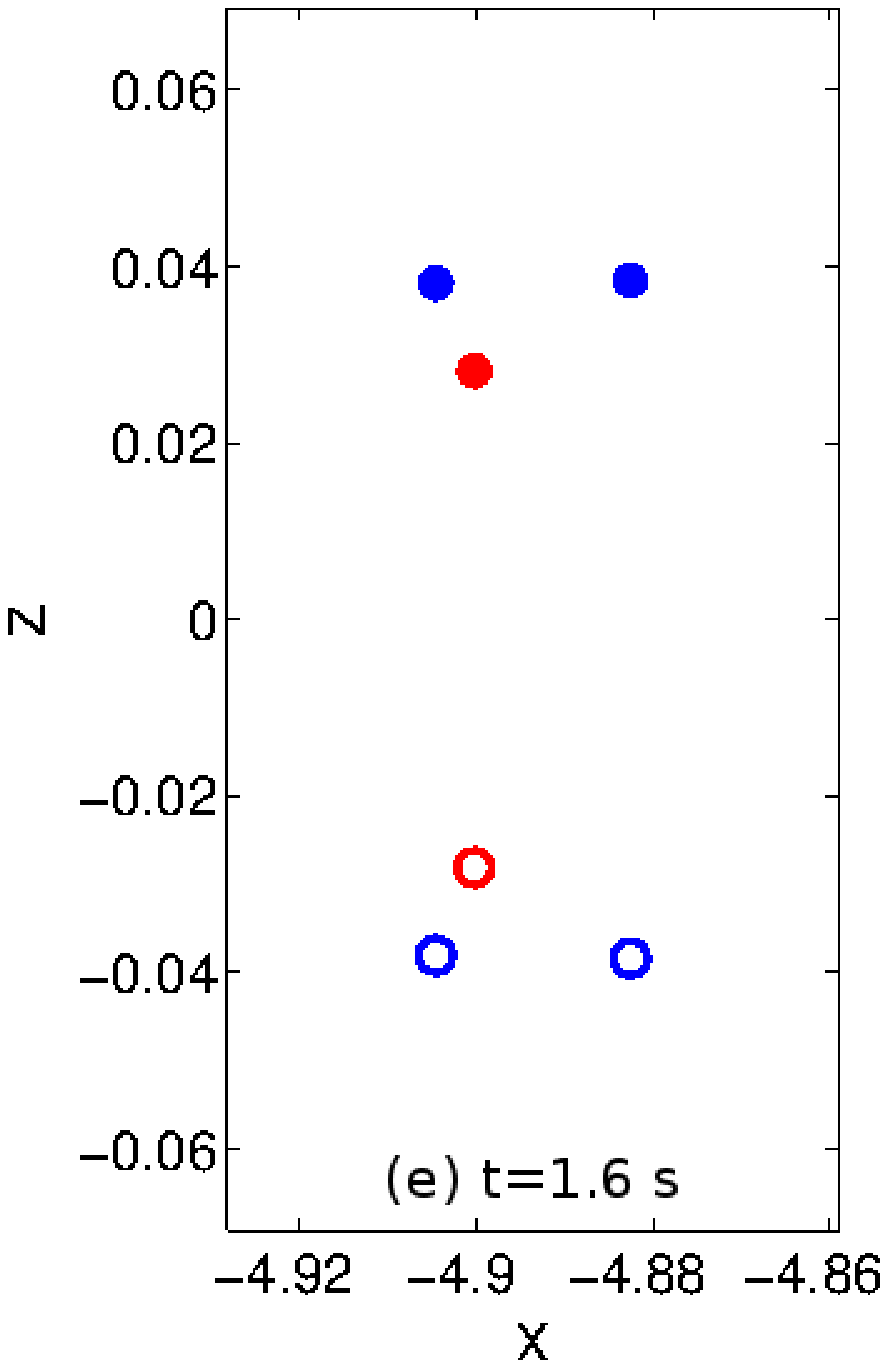}  
  \includegraphics[scale=0.38]{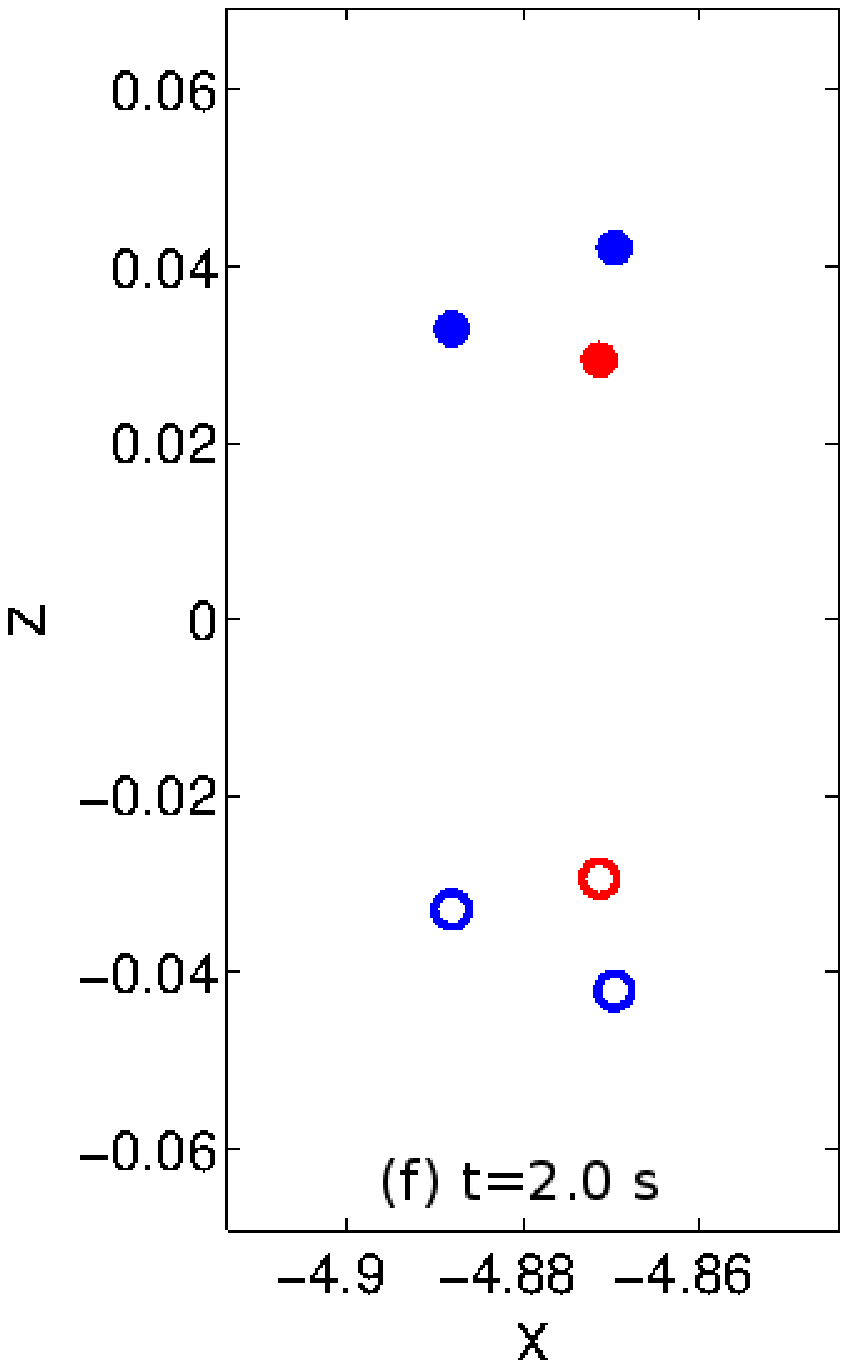} \\
\vskip 1cm
%
  \includegraphics[scale=0.38]{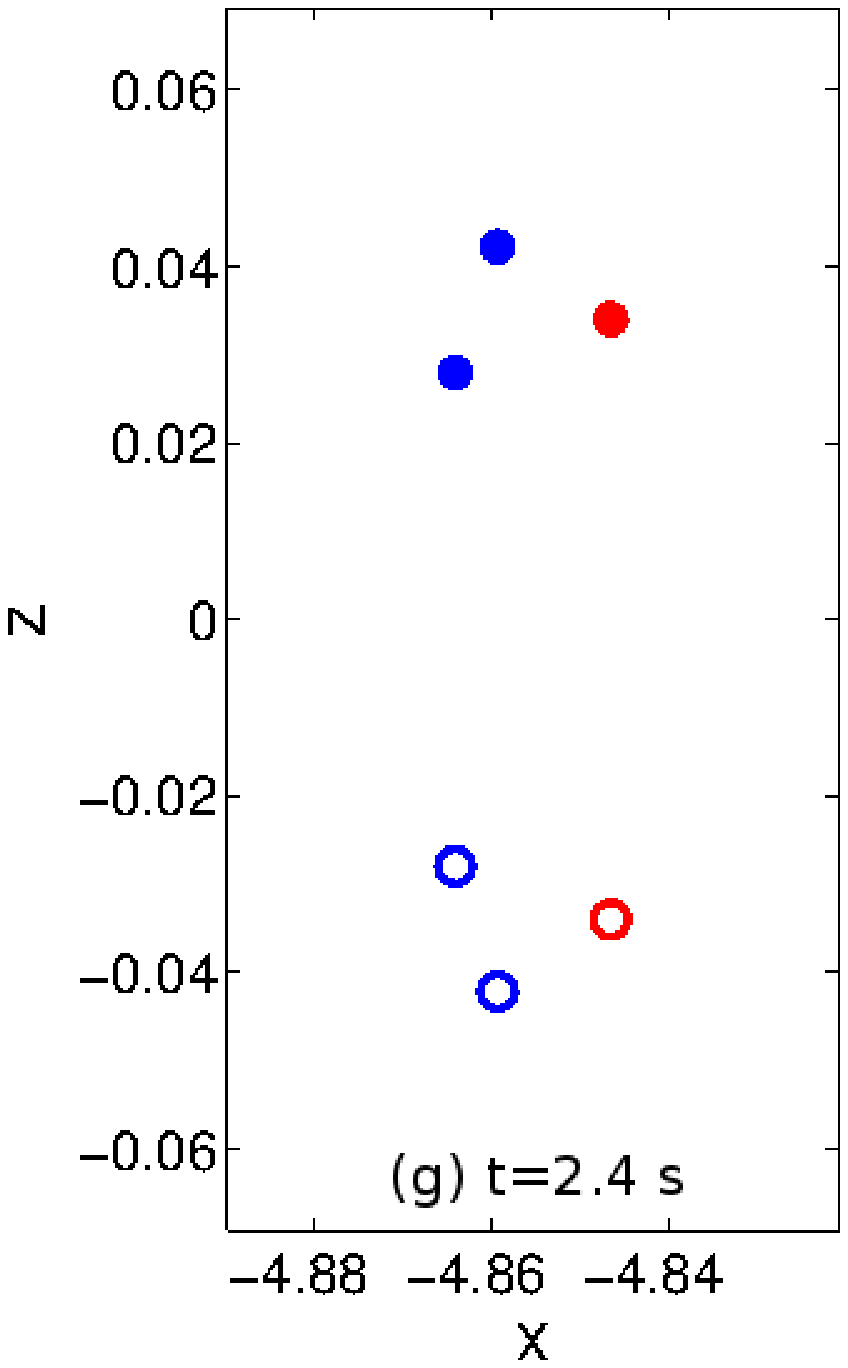}  
  \includegraphics[scale=0.38]{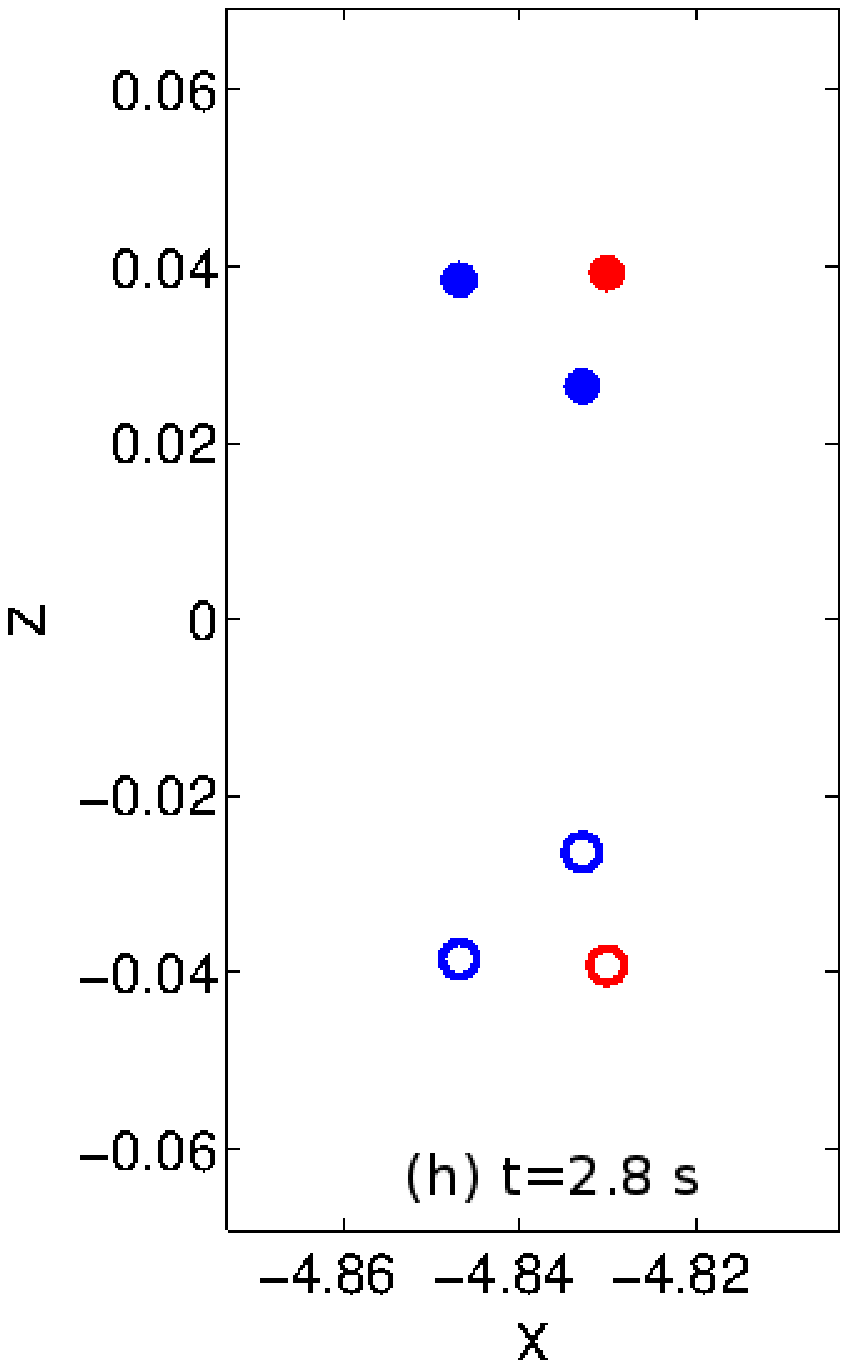}  
  \includegraphics[scale=0.38]{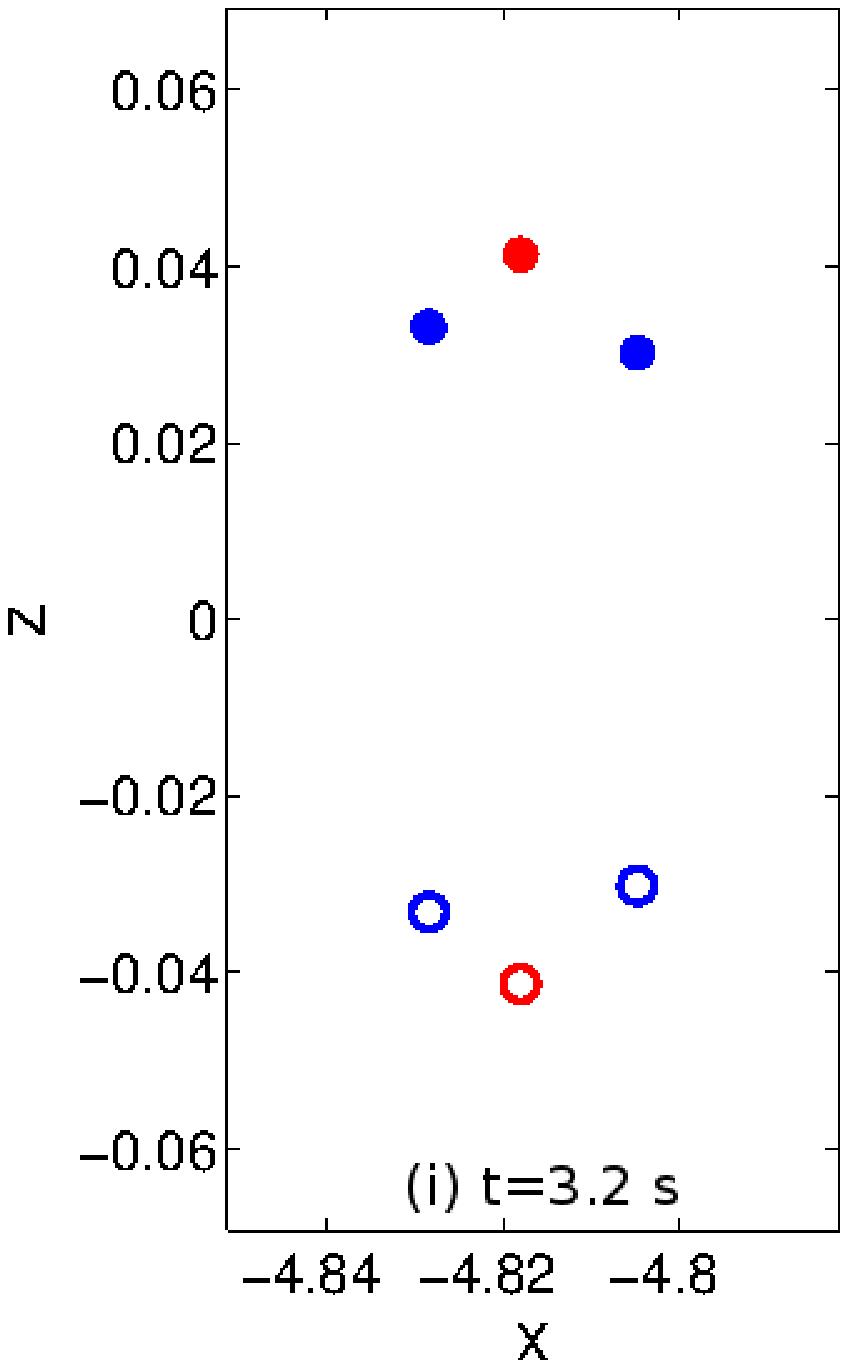}

\vskip 3cm

\caption{Generalised leapfrogging.
As in Fig.~\ref{fig:5}, but $N=3$. Note again the ellipticity of the
trajectories. One vortex has been marked in red colour to follow its motion.
Parameters: $R=0.0346~\rm cm$,
$a=0.00866~\rm cm$, $\ell=0.015~\rm cm$ and $R/a=4$.
}
\label{fig:6}
\end{figure}

\begin{figure}
\centering
\subfloat[$t=0.0005~\rm s$] {
  \includegraphics[scale=0.2]{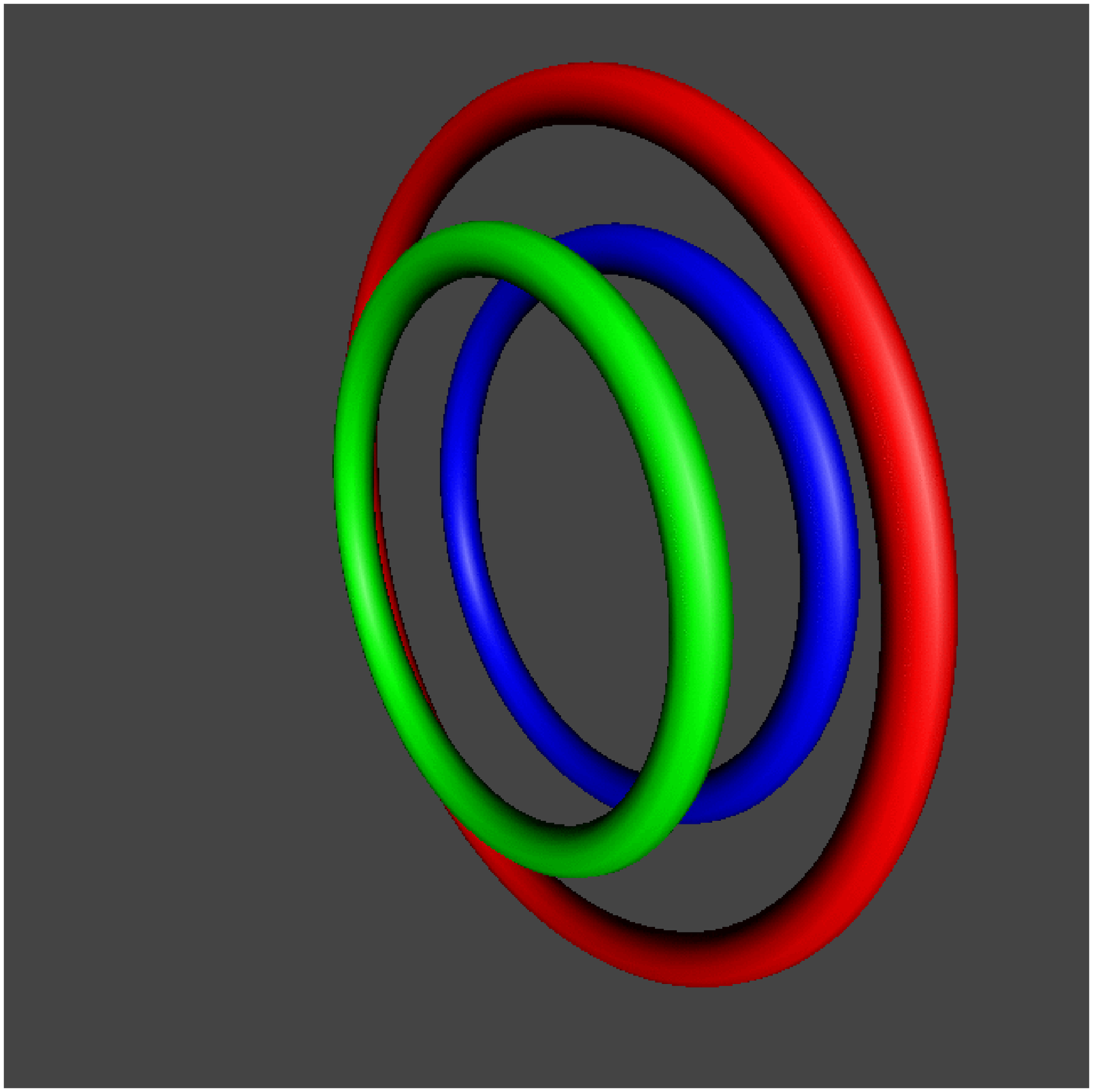}
}
\hspace{1em}
\subfloat[$t=0.4~\rm s$] {
  \includegraphics[scale=0.2]{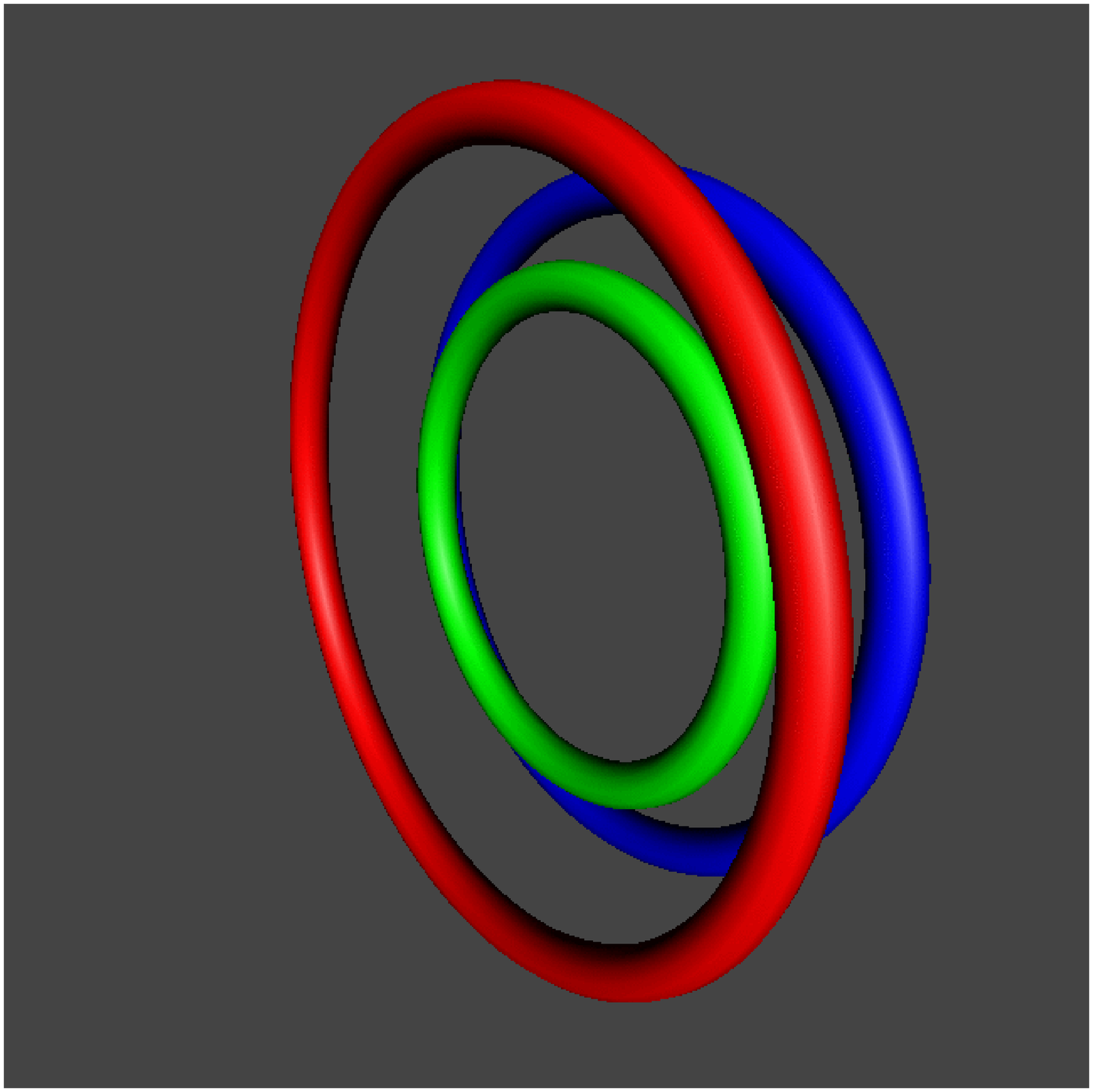}
}
\hspace{1em}
\subfloat[$t=0.8~\rm s$] {
  \includegraphics[scale=0.2]{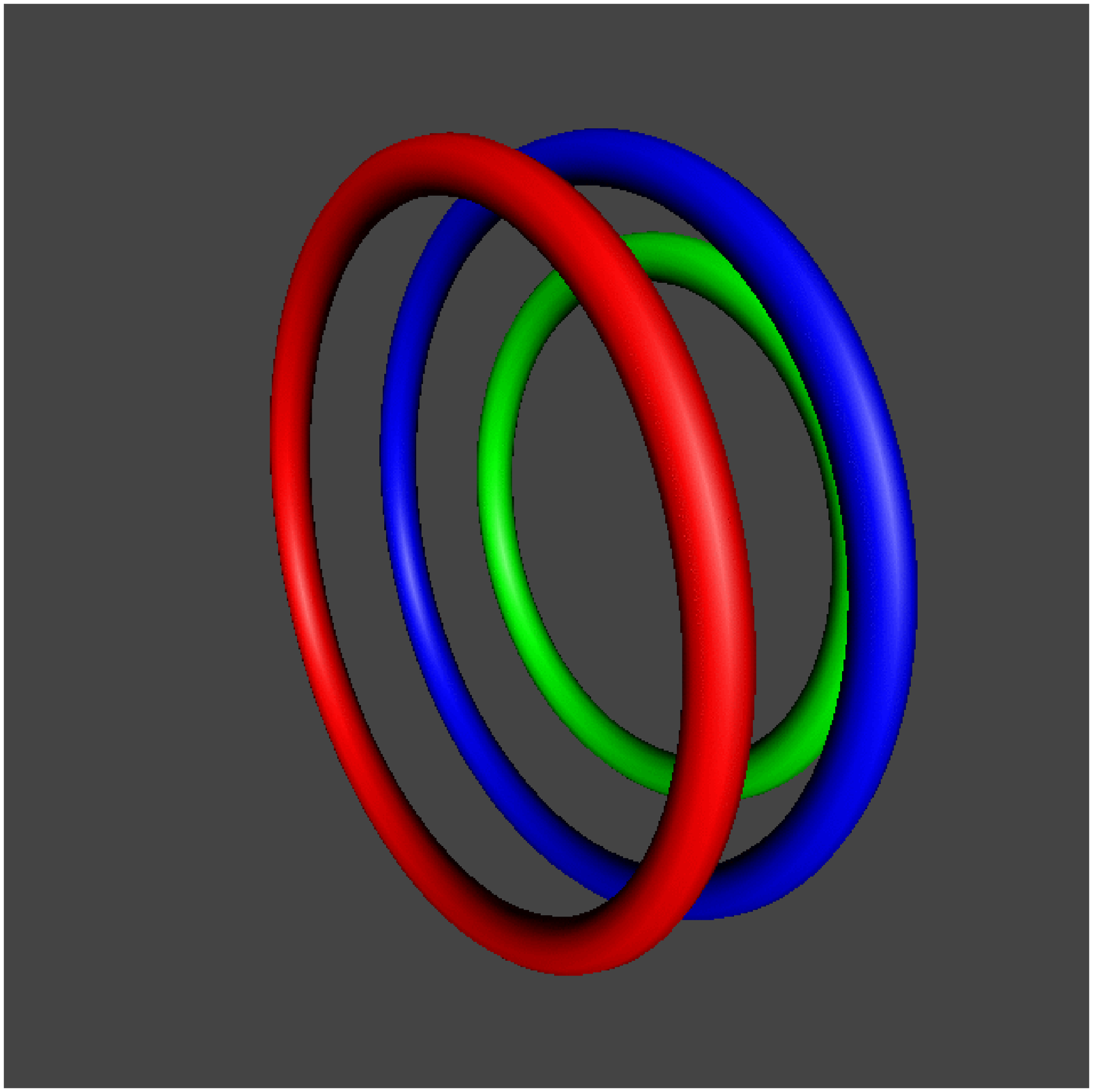}
}\\
\subfloat[$t=1.2~\rm s$] {
  \includegraphics[scale=0.2]{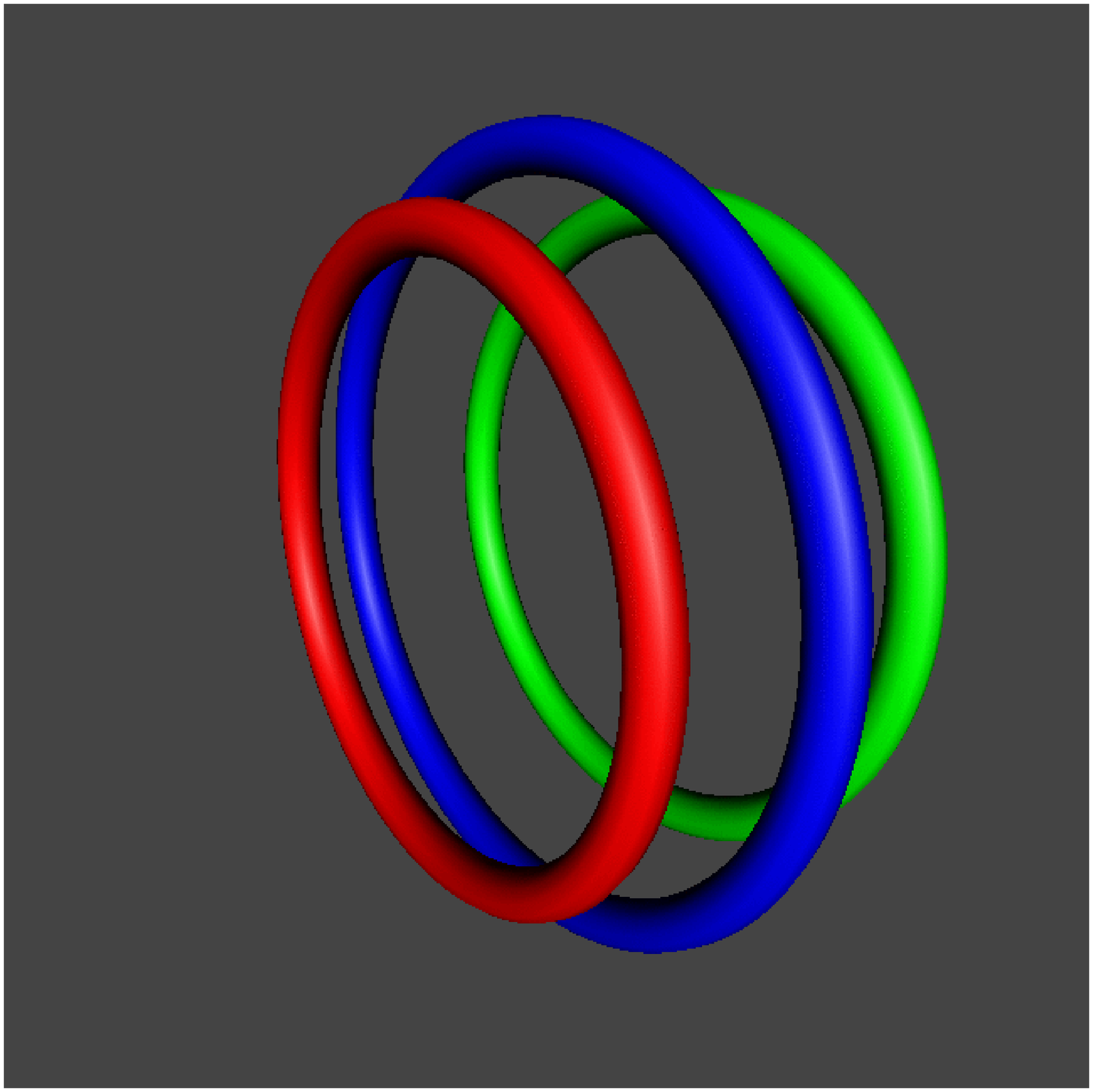}  
}
\hspace{1em}
\subfloat[$t=1.6~\rm s$] {
  \includegraphics[scale=0.2]{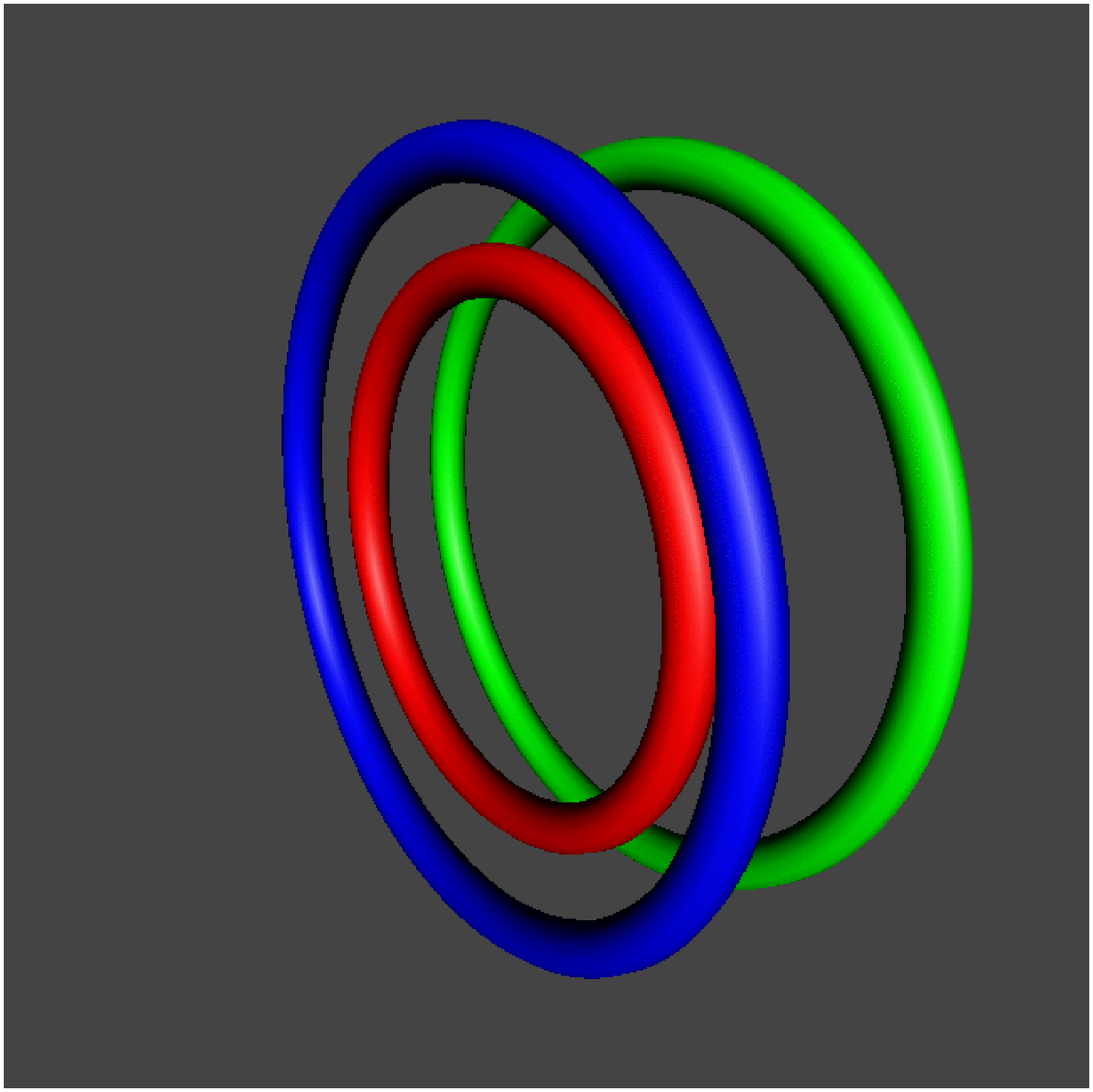}  
}
\hspace{1em}
\subfloat[$t=2.0~\rm s$] {
  \includegraphics[scale=0.2]{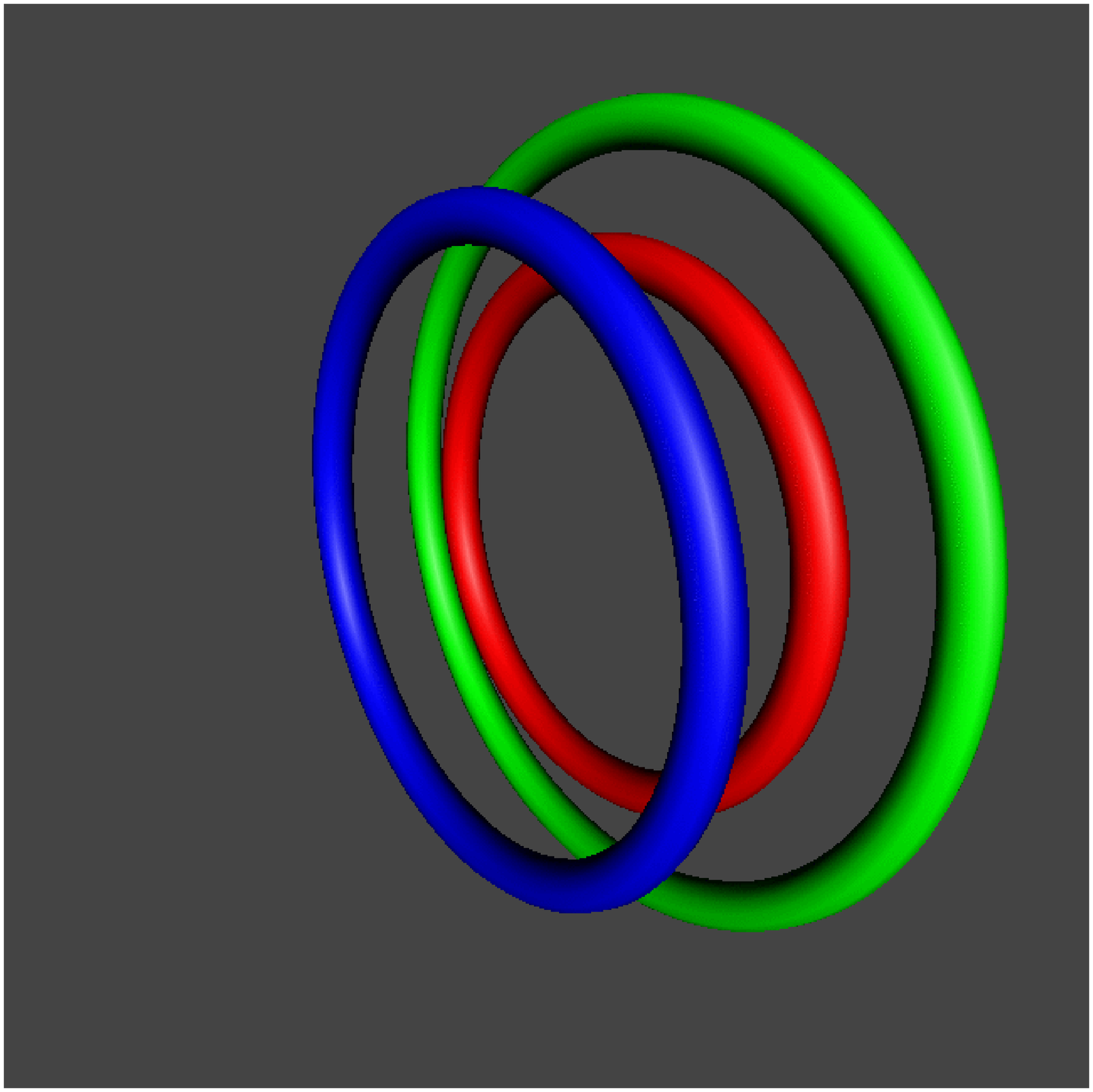}
}\\
\subfloat[$t=2.4~\rm s$] {
  \includegraphics[scale=0.2]{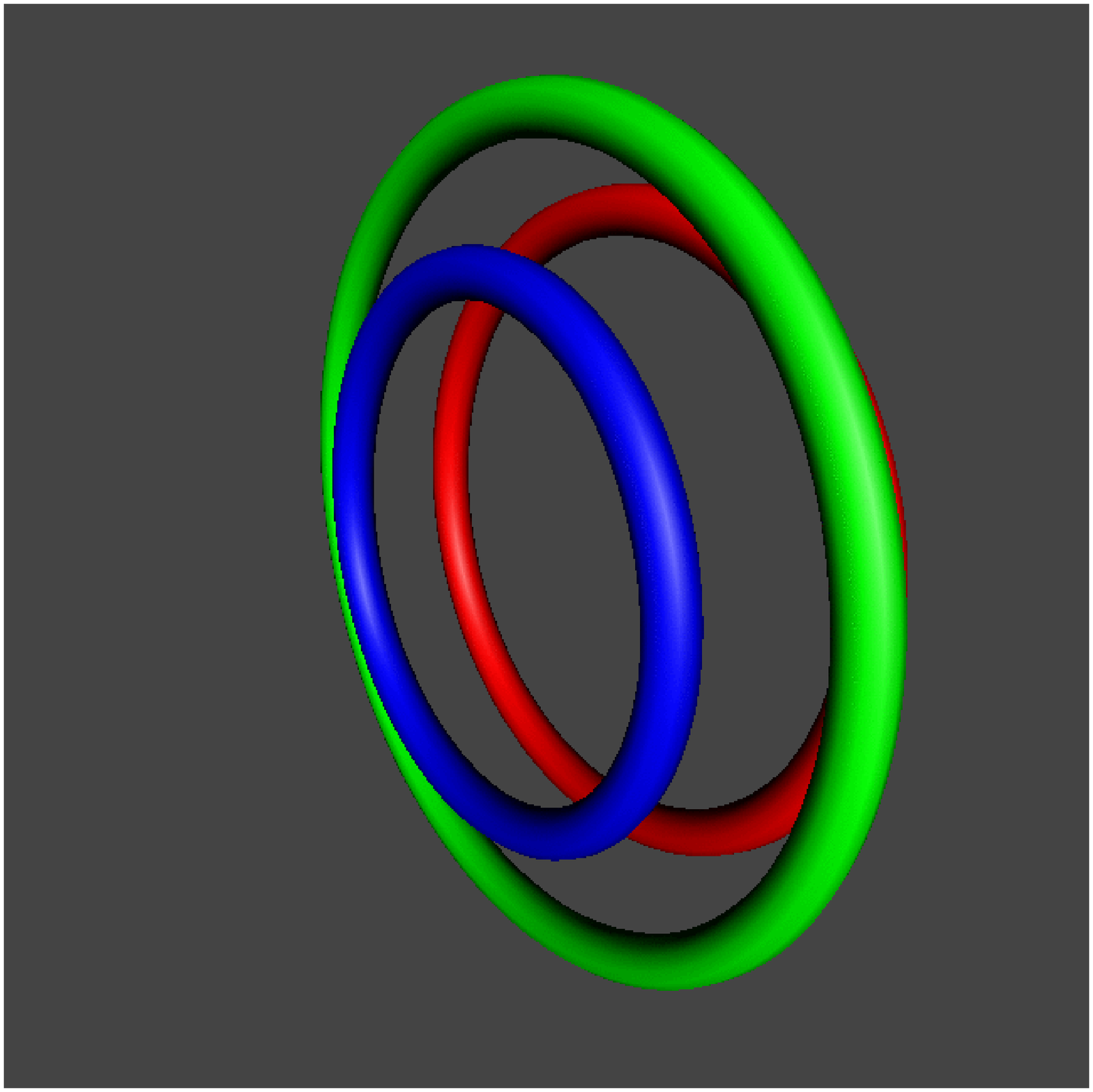}  
}
\hspace{1em}
\subfloat[$t=2.8~\rm s$] {
  \includegraphics[scale=0.2]{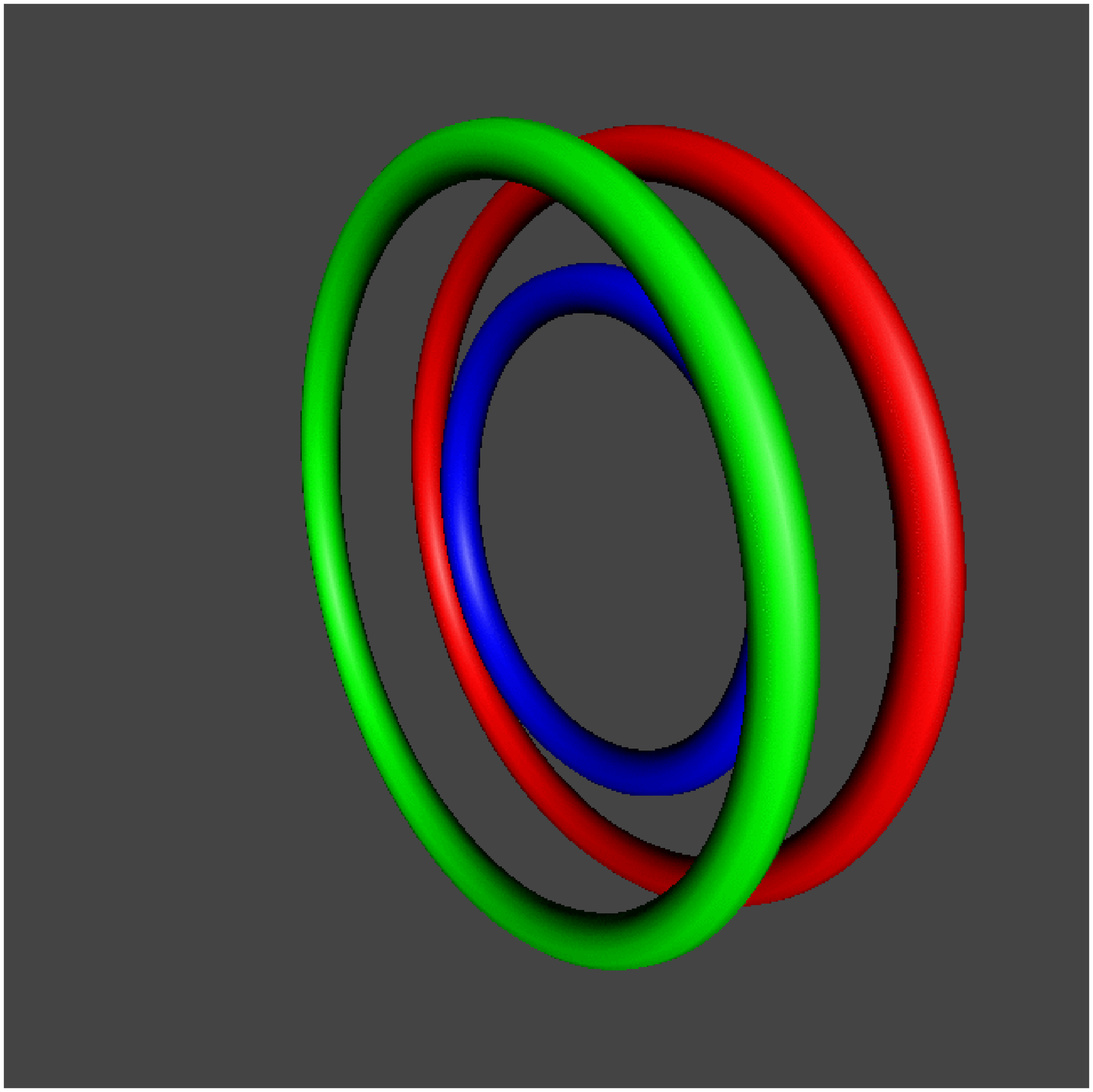}  
}
\hspace{1em}
\subfloat[$t=3.2~\rm s$] {
  \includegraphics[scale=0.2]{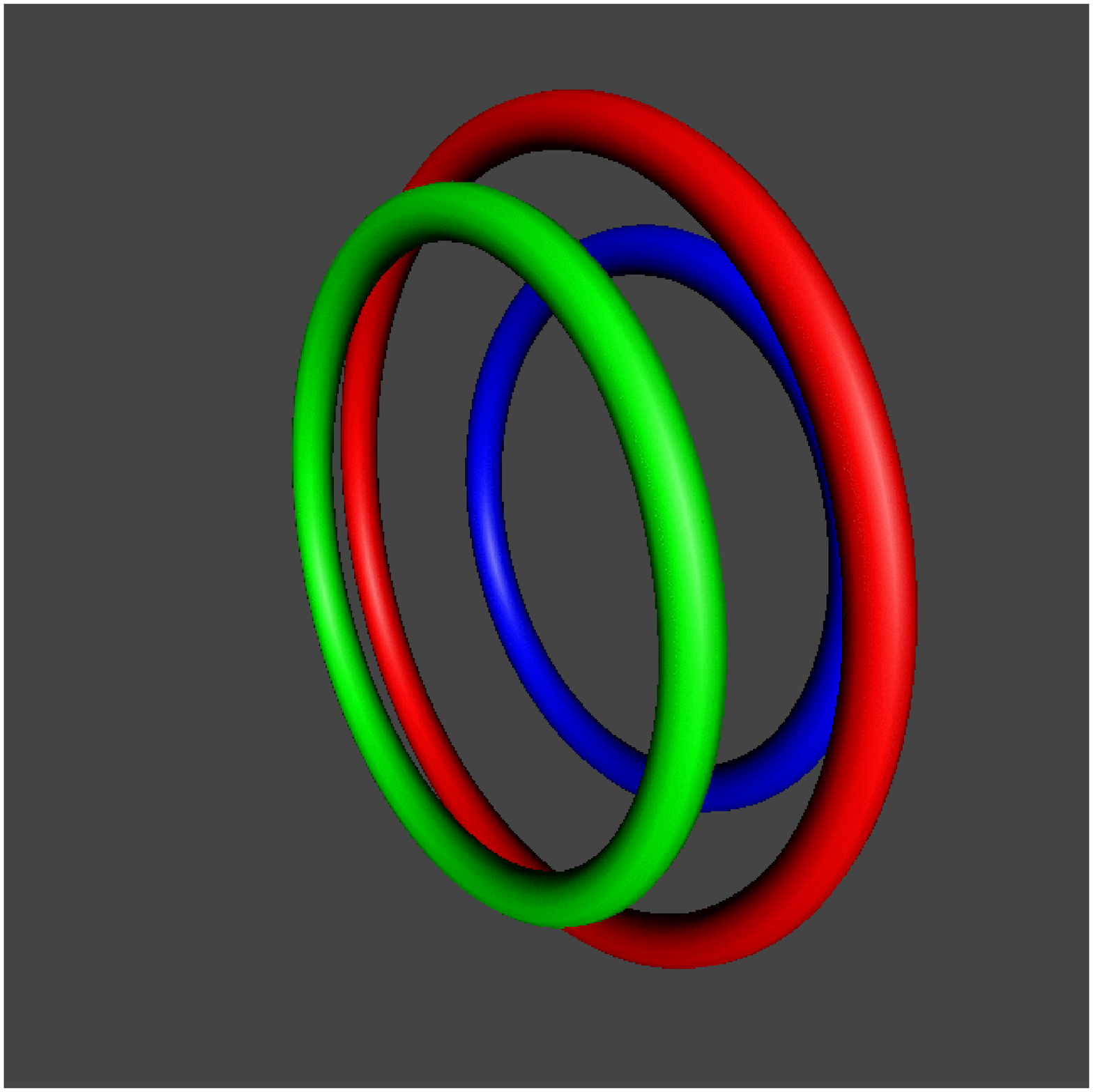}
}
\caption{Generalised leapfrogging.
Three-dimensional images of leapfrogging for $N=3$
corresponding to the two-dimensional cross-sections of Fig.~\ref{fig:6}.
The colours on different images represent the same vortex ring. For 
example, notice the red ring on the outside of the torus
in (a),
which moves around the back of the torus (b), then shrinks (e,f), 
moves ahead of the other rings (g) and grows in size again (h,i).
The arbitrary thickness of the tubes 
which represent the vortex
lines and the size of the box have been adjusted for visualization purpose.
}
\label{fig:7}
\end{figure}

Generalised leapfrogging for $N=3$ is shown in Fig.~\ref{fig:6}.
It is apparent that the trajectories of the
vortices in the cross-sectional plane are not circular but elliptical.
Fig.~\ref{fig:7} shows three-dimensional images of the $N=3$ vortex bundle
during its evolution. 
This and other three-dimensional images have been prepared using 
Bob Scharein's {\it KnotPlot} software \cite{knotplot}. 

\begin{figure}
\centering
\subfloat[$t=0.0075~\rm s$] {
  \includegraphics[scale=0.2]{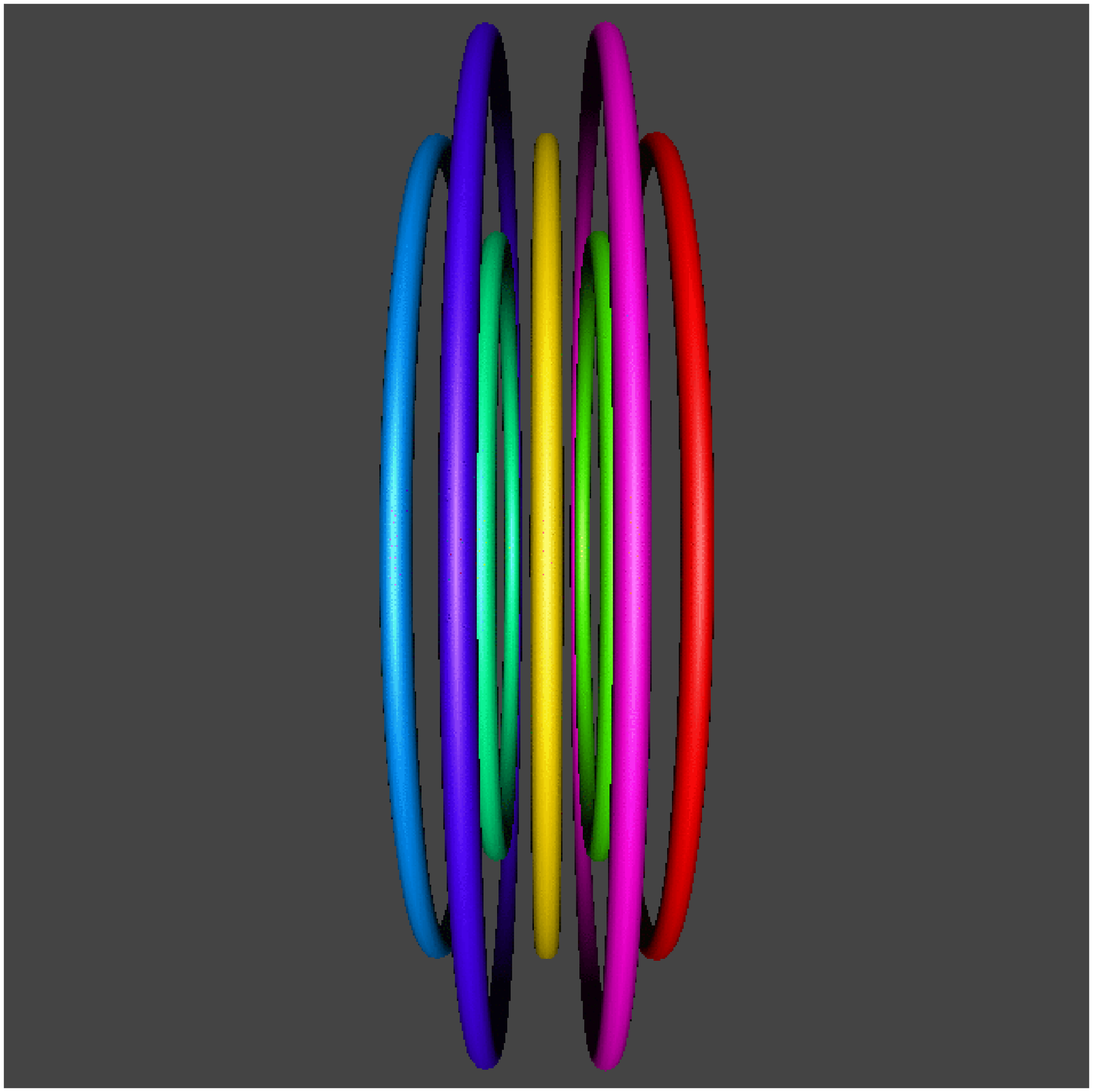}
}
\hspace{1em}
\subfloat[$t=0.75~\rm s$] {
  \includegraphics[scale=0.2]{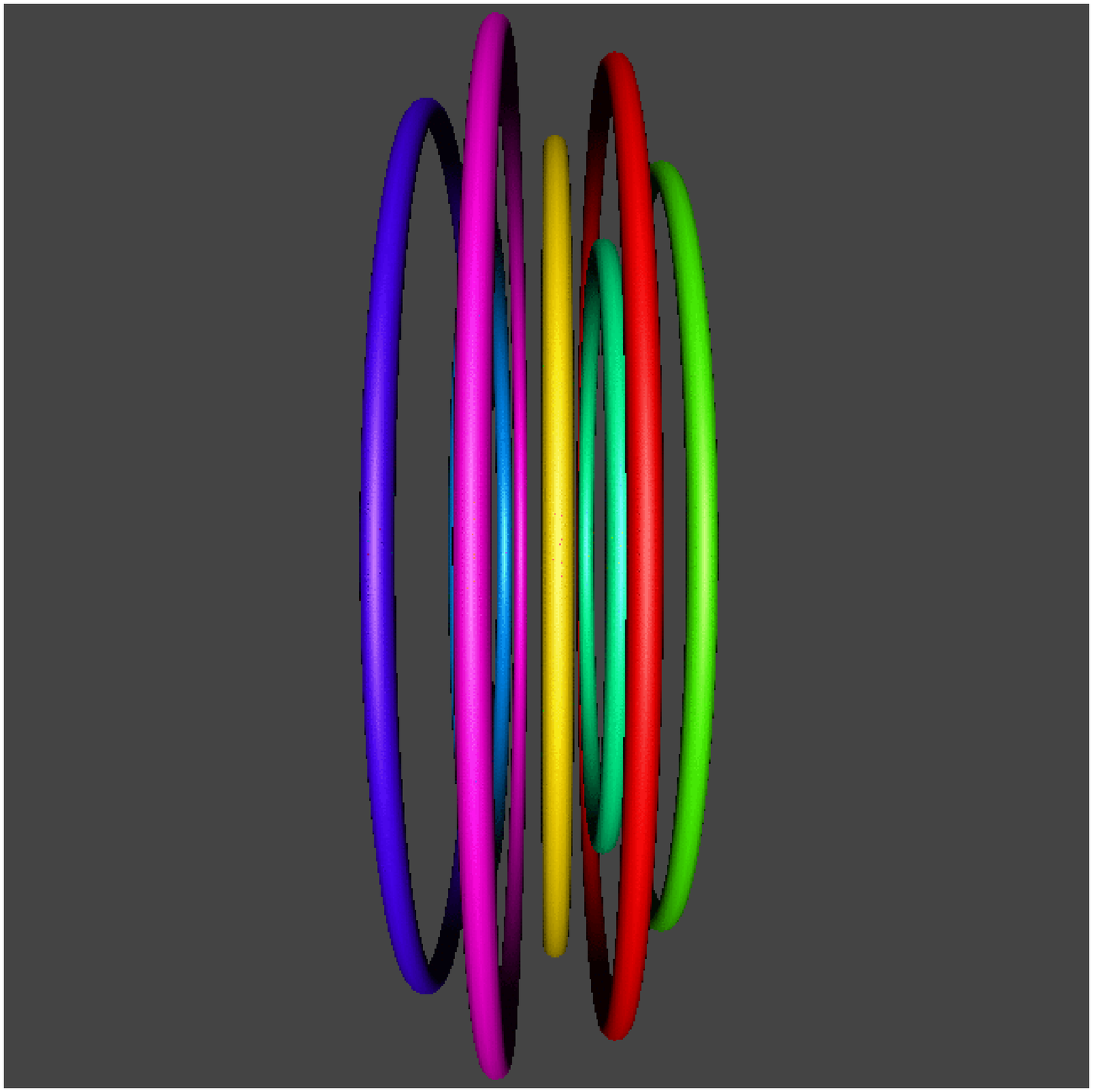}
}
\hspace{1em}
\subfloat[$t=1.5~\rm s$] {
  \includegraphics[scale=0.2]{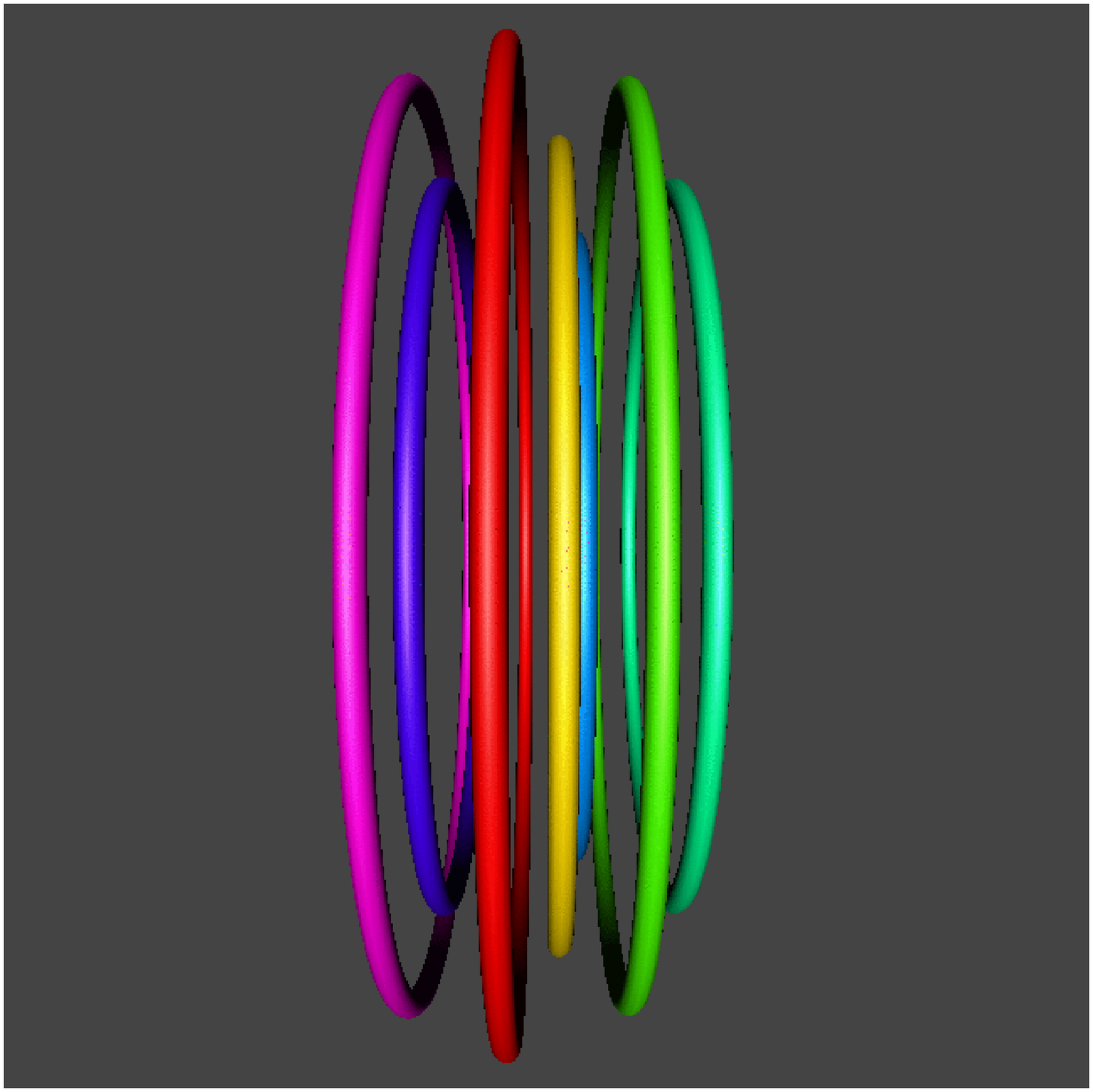}
}\\
\subfloat[$t=2.25~\rm s$] {
  \includegraphics[scale=0.2]{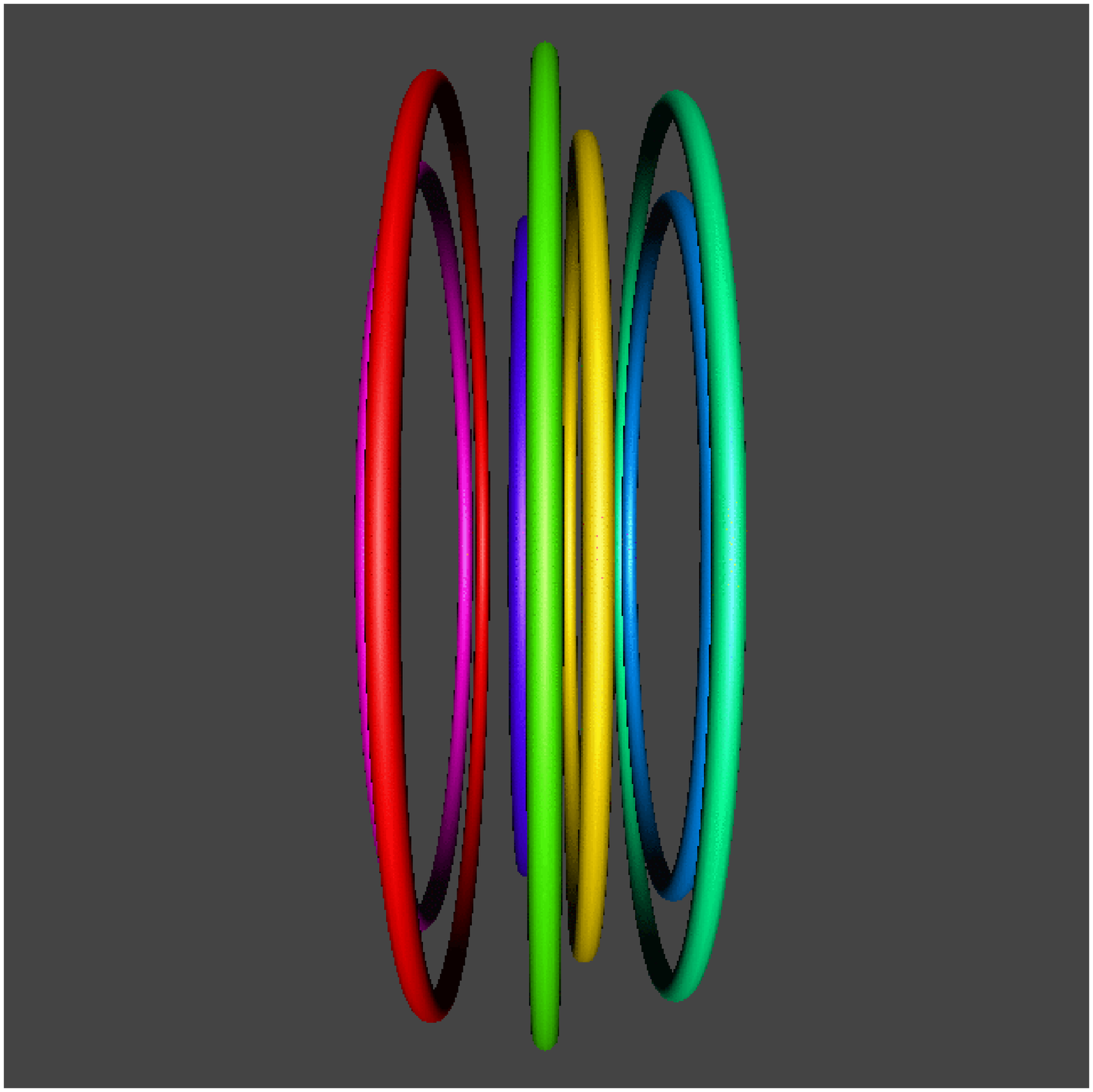}  
}
\hspace{1em}
\subfloat[$t=3.0~\rm s$] {
  \includegraphics[scale=0.2]{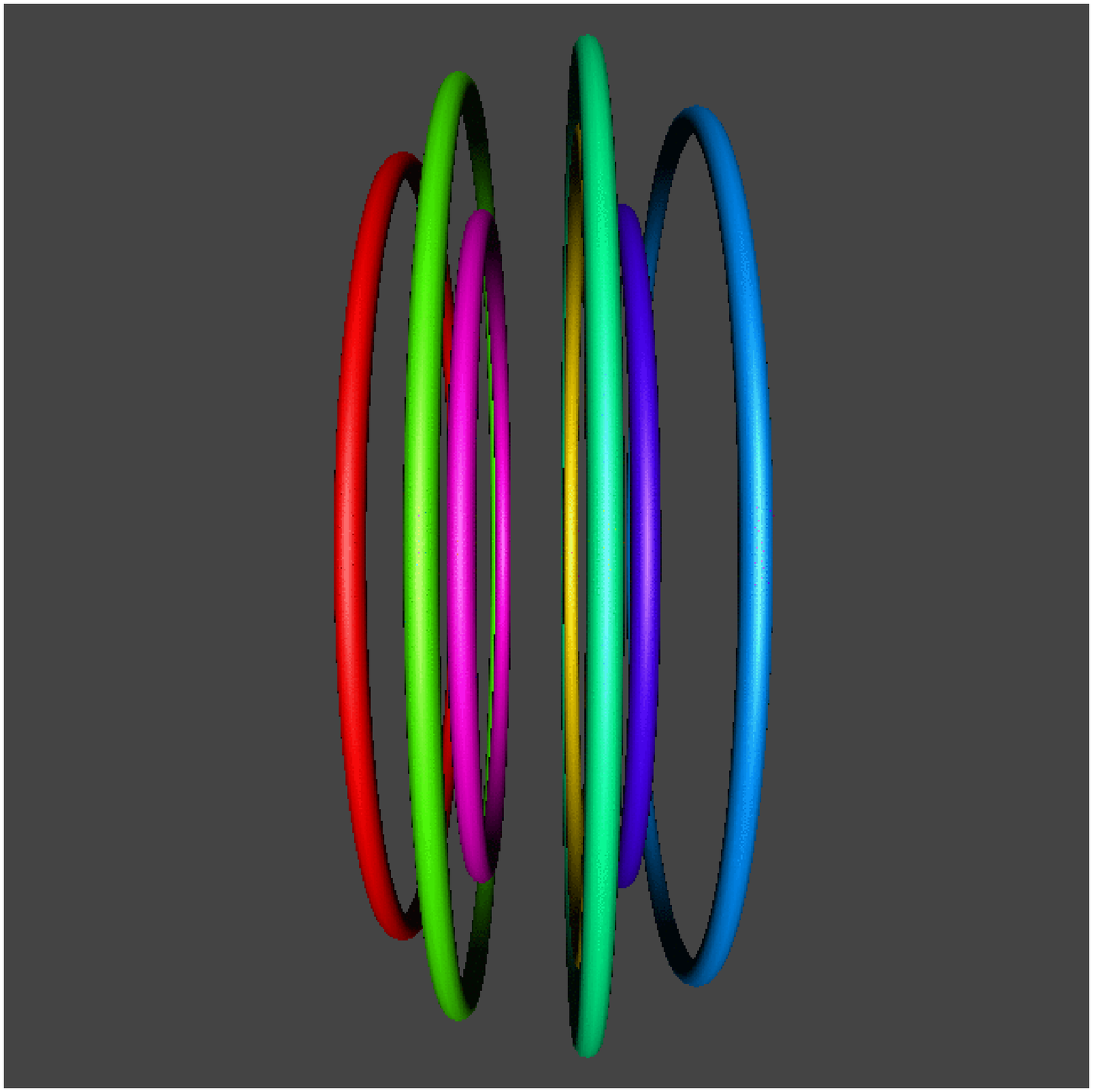}  
}
\hspace{1em}
\subfloat[$t=3.75~\rm s$] {
  \includegraphics[scale=0.2]{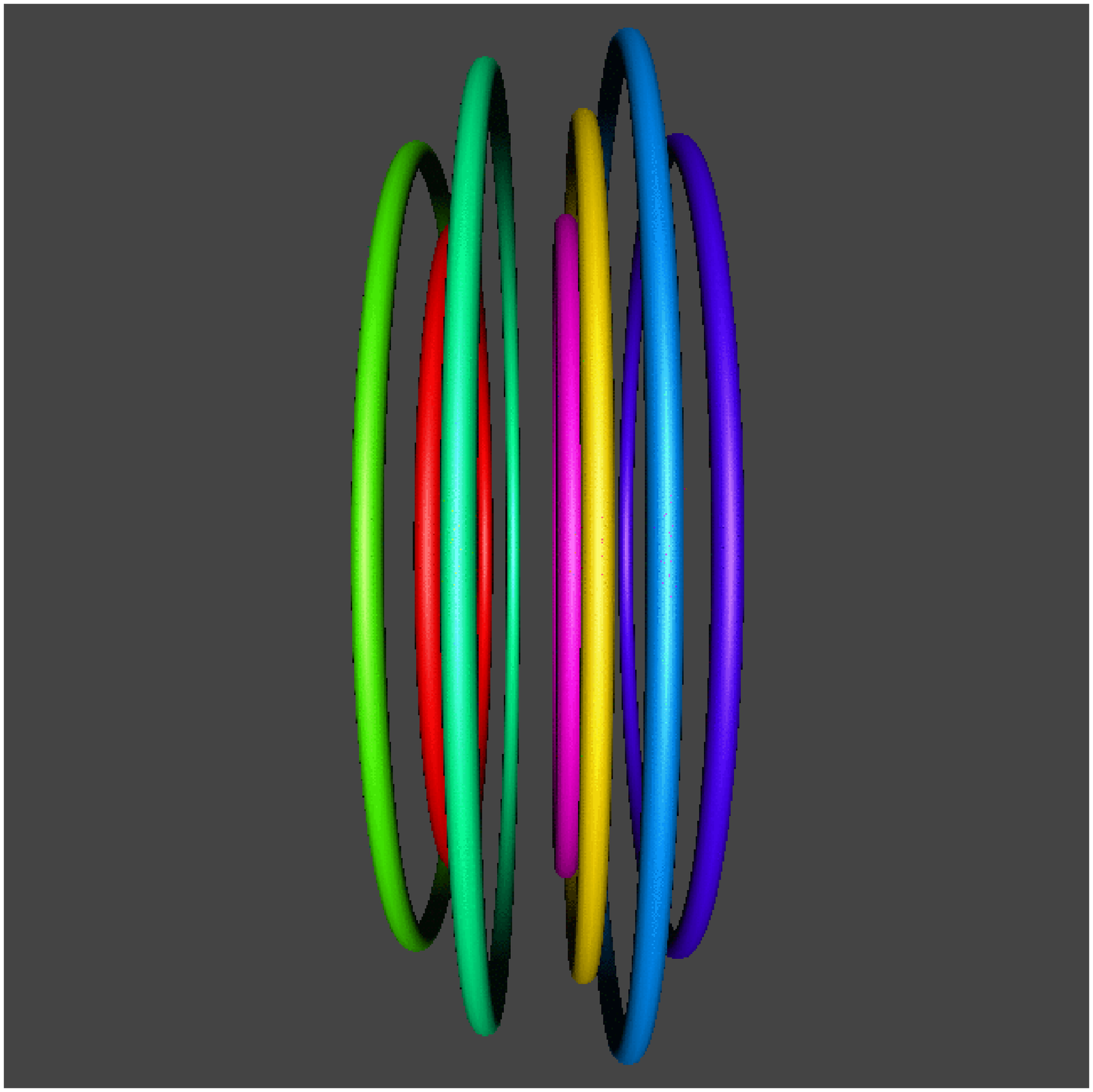}
}\\
\subfloat[$t=4.5~\rm s$] {
  \includegraphics[scale=0.2]{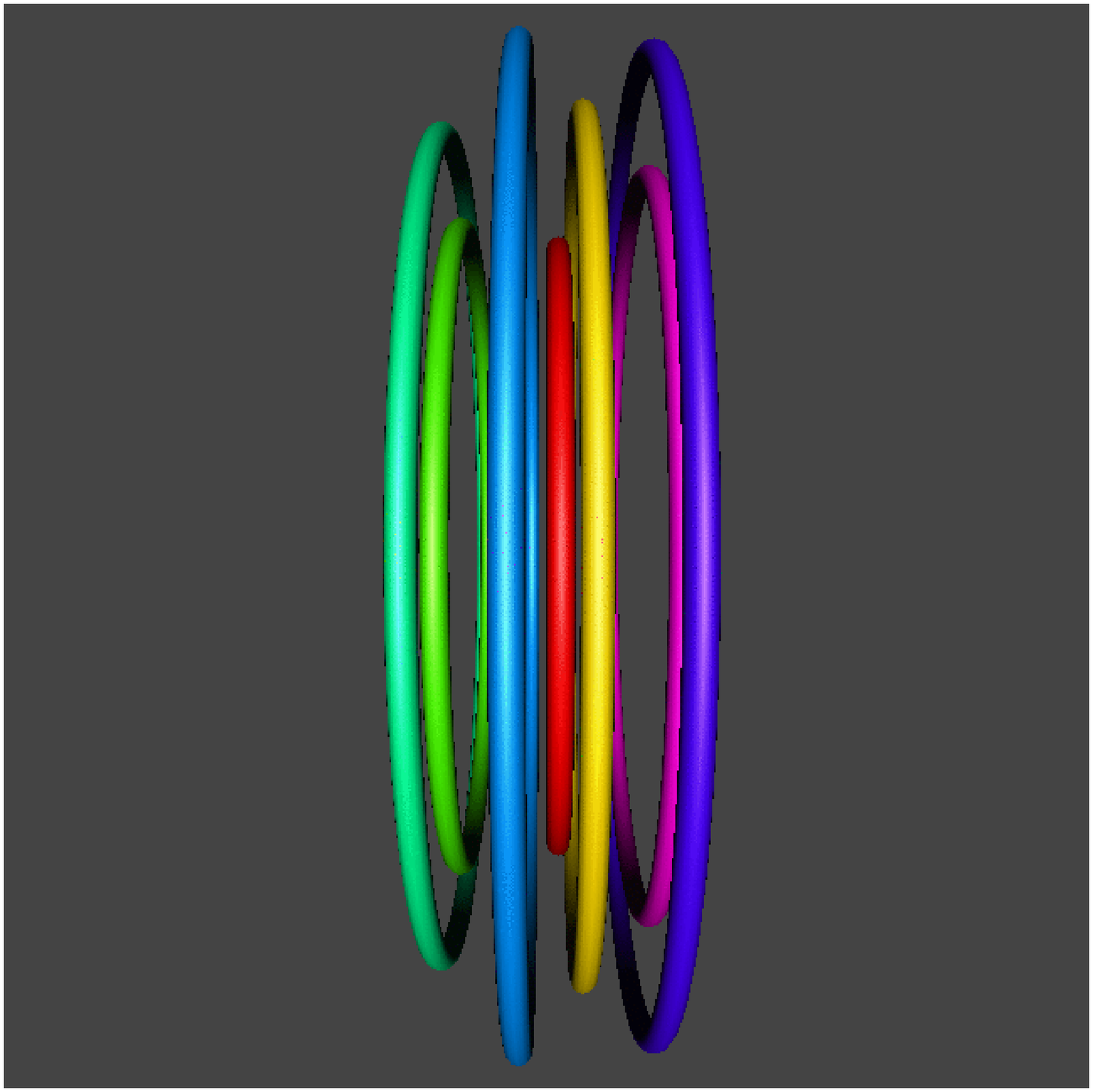}  
}
\hspace{1em}
\subfloat[$t=5.25~\rm s$] {
  \includegraphics[scale=0.2]{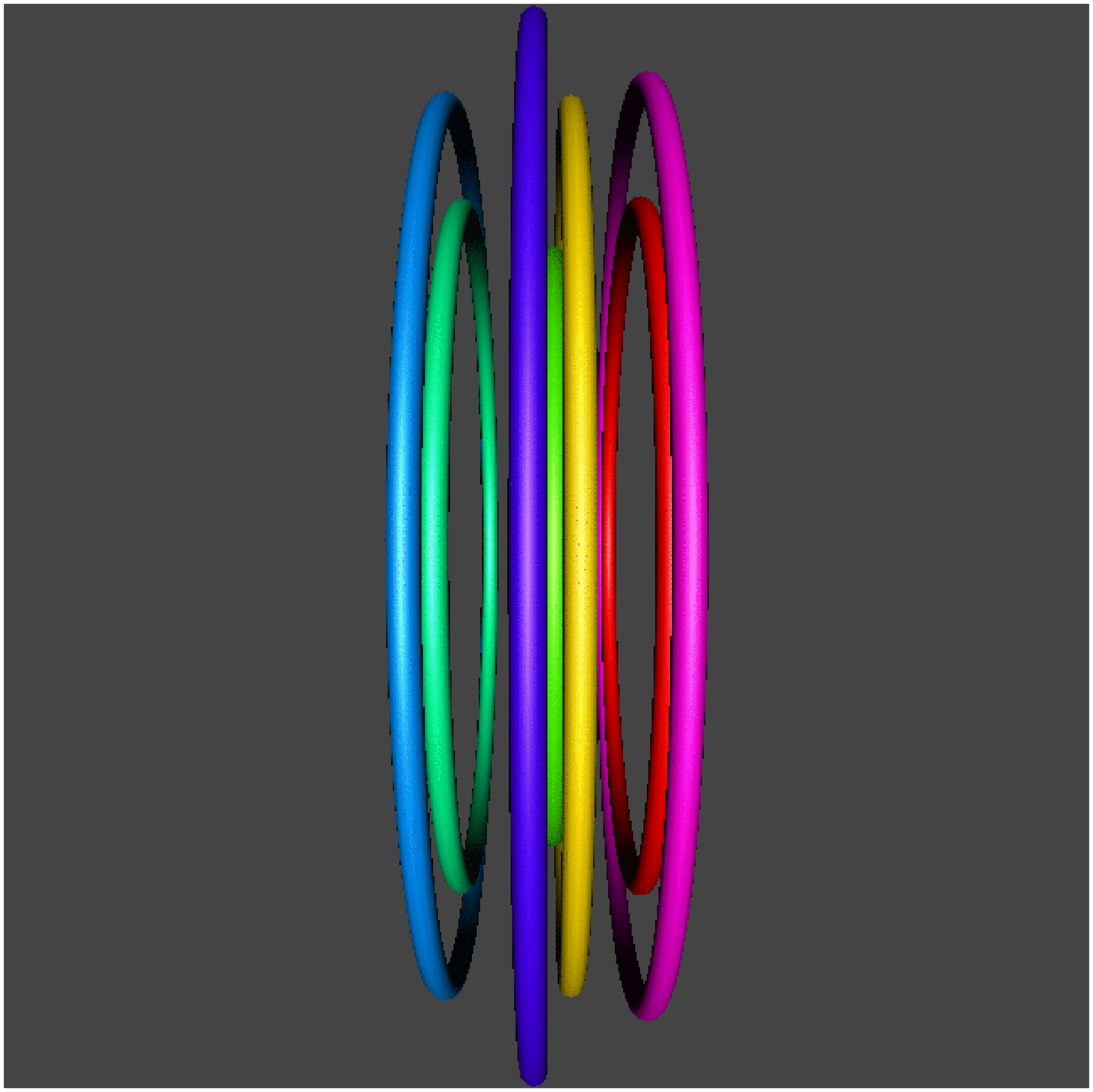}  
}
\hspace{1em}
\subfloat[$t=6.0~\rm s$] {
  \includegraphics[scale=0.2]{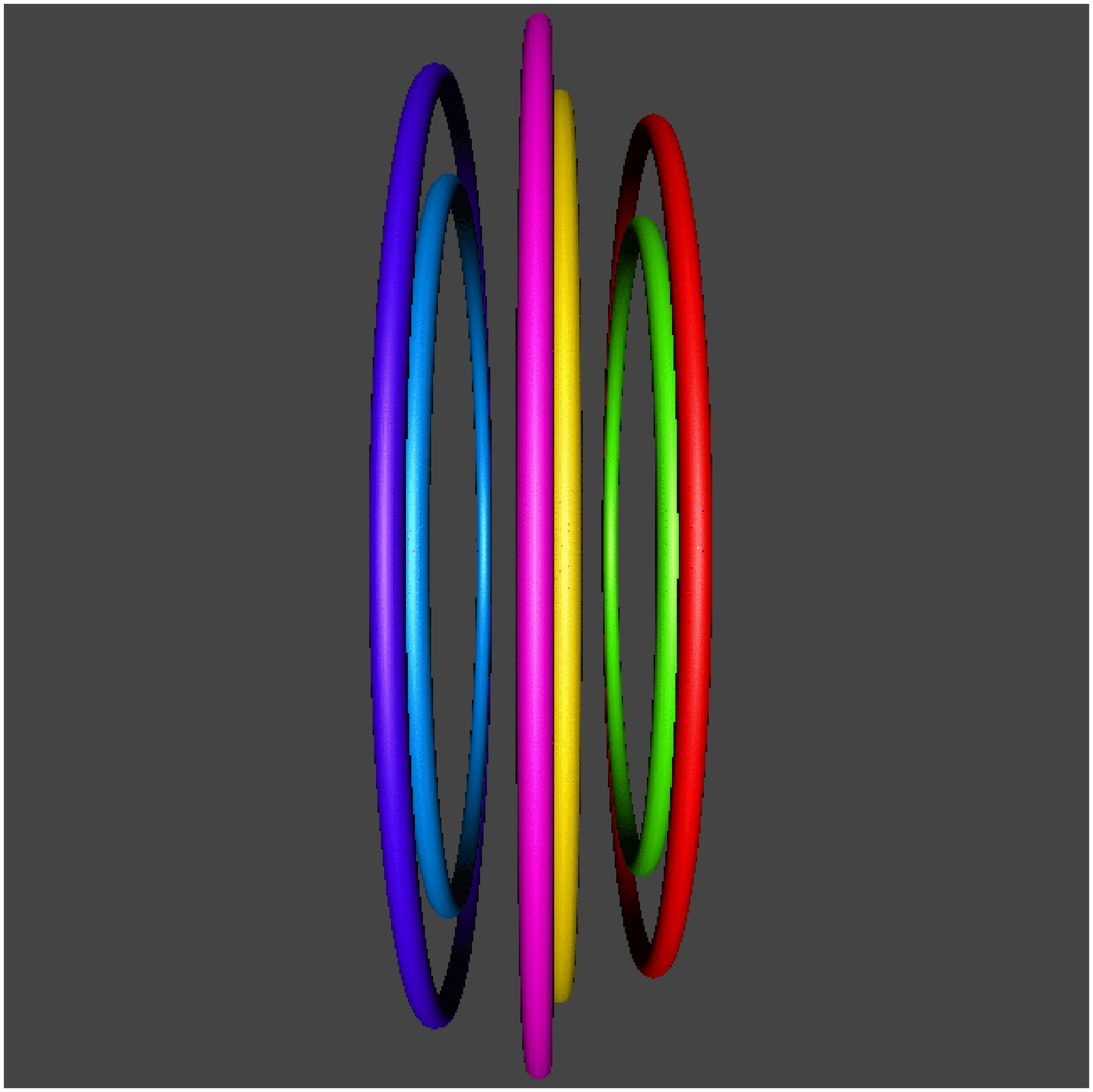}
}
\caption{Generalised leapfrogging.
Three-dimensional images of leapfrogging for $N=7$.
Again, note how the red ring, initially at the front of the torus (a),
moves round to the back (d), then shrinks and overtakes other rings
passing inside them (g), and returns to the front (i).
Parameters: $R=0.0896~\rm cm$, $a=0.0223~\rm cm$, $\ell=0.0223~\rm cm$
and $R/a=4$.
}
\label{fig:8}
\end{figure}

\begin{figure}
\centering
  \includegraphics[scale=0.38]{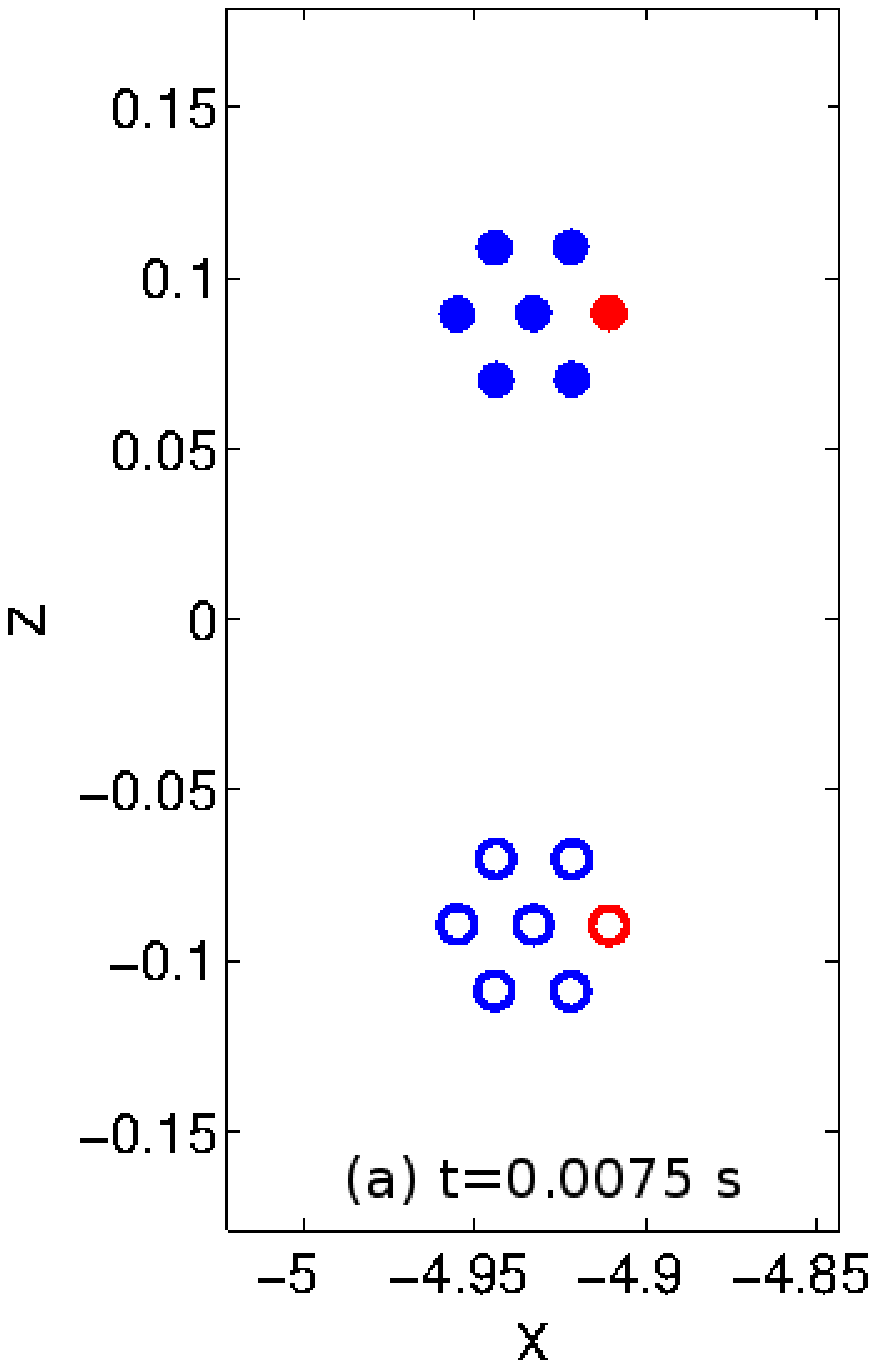}
  \includegraphics[scale=0.38]{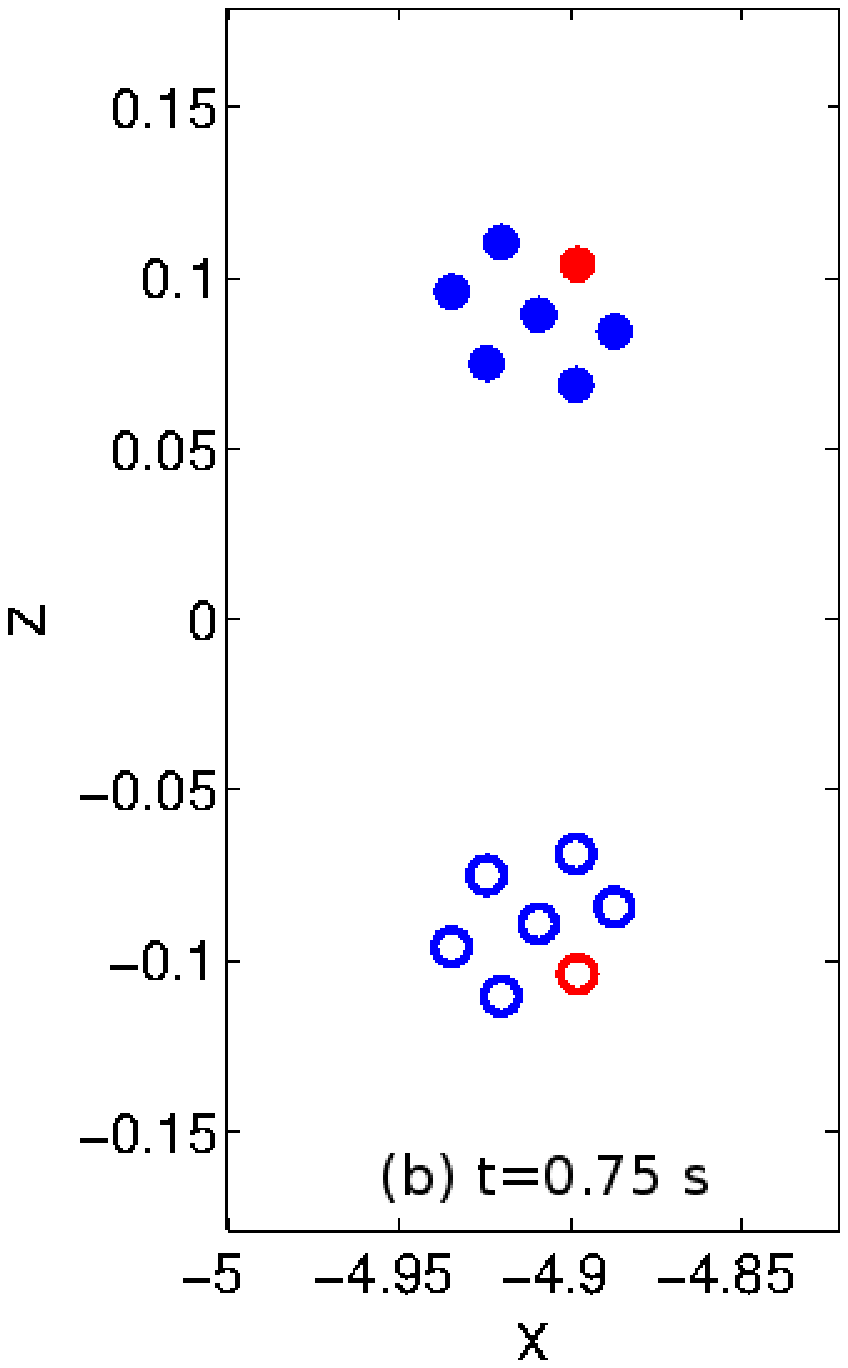}
  \includegraphics[scale=0.38]{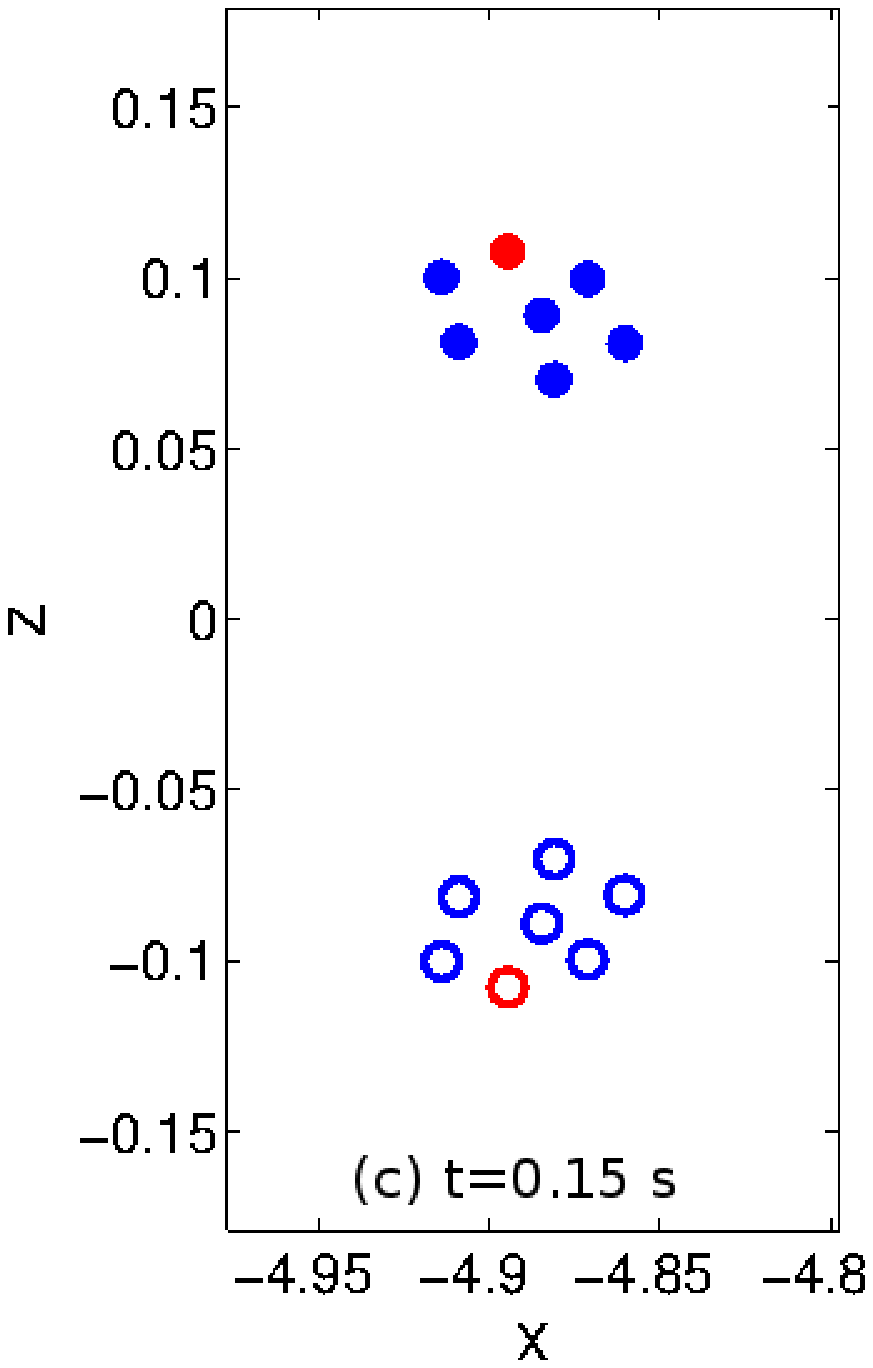} \\
\vskip 1cm
%
  \includegraphics[scale=0.38]{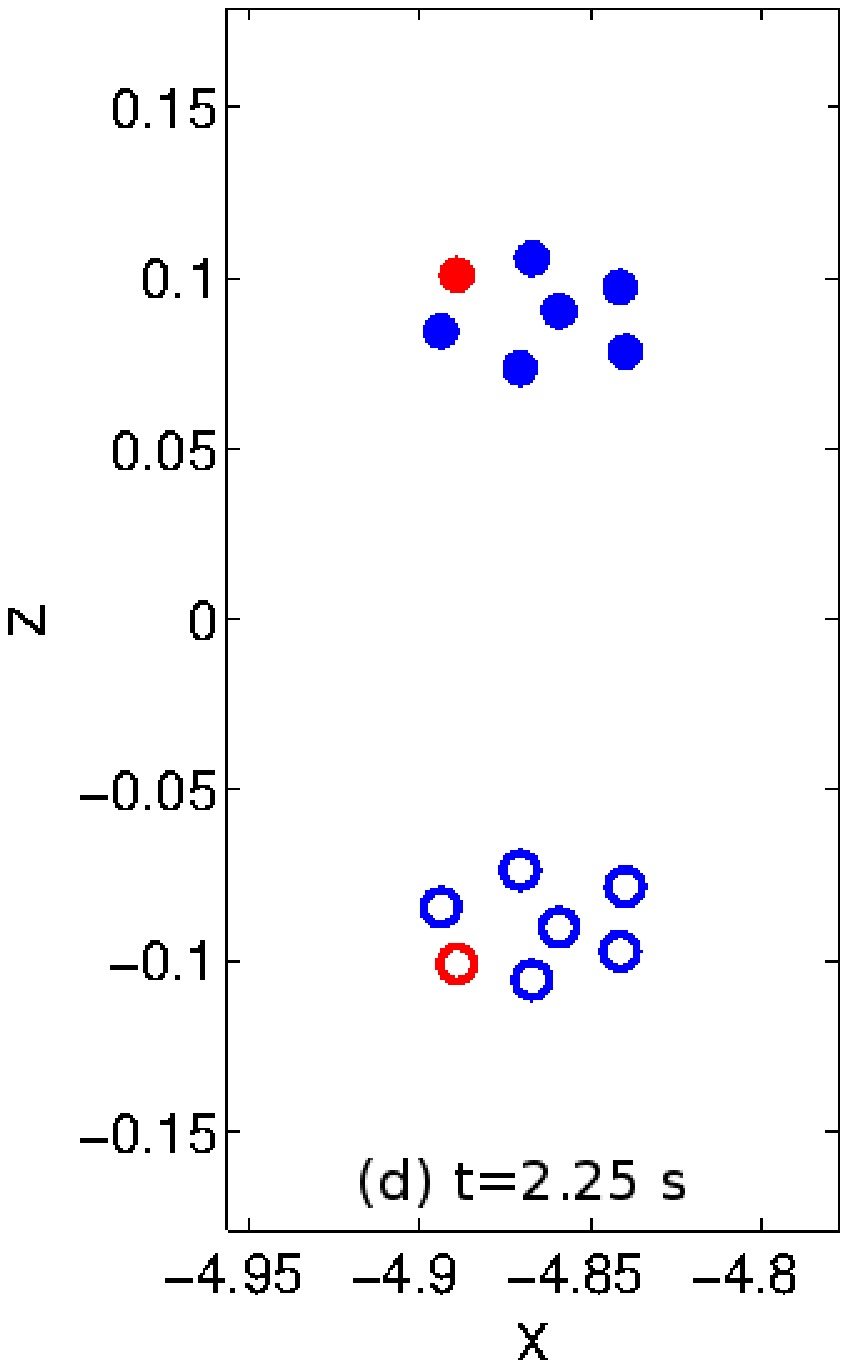}  
  \includegraphics[scale=0.38]{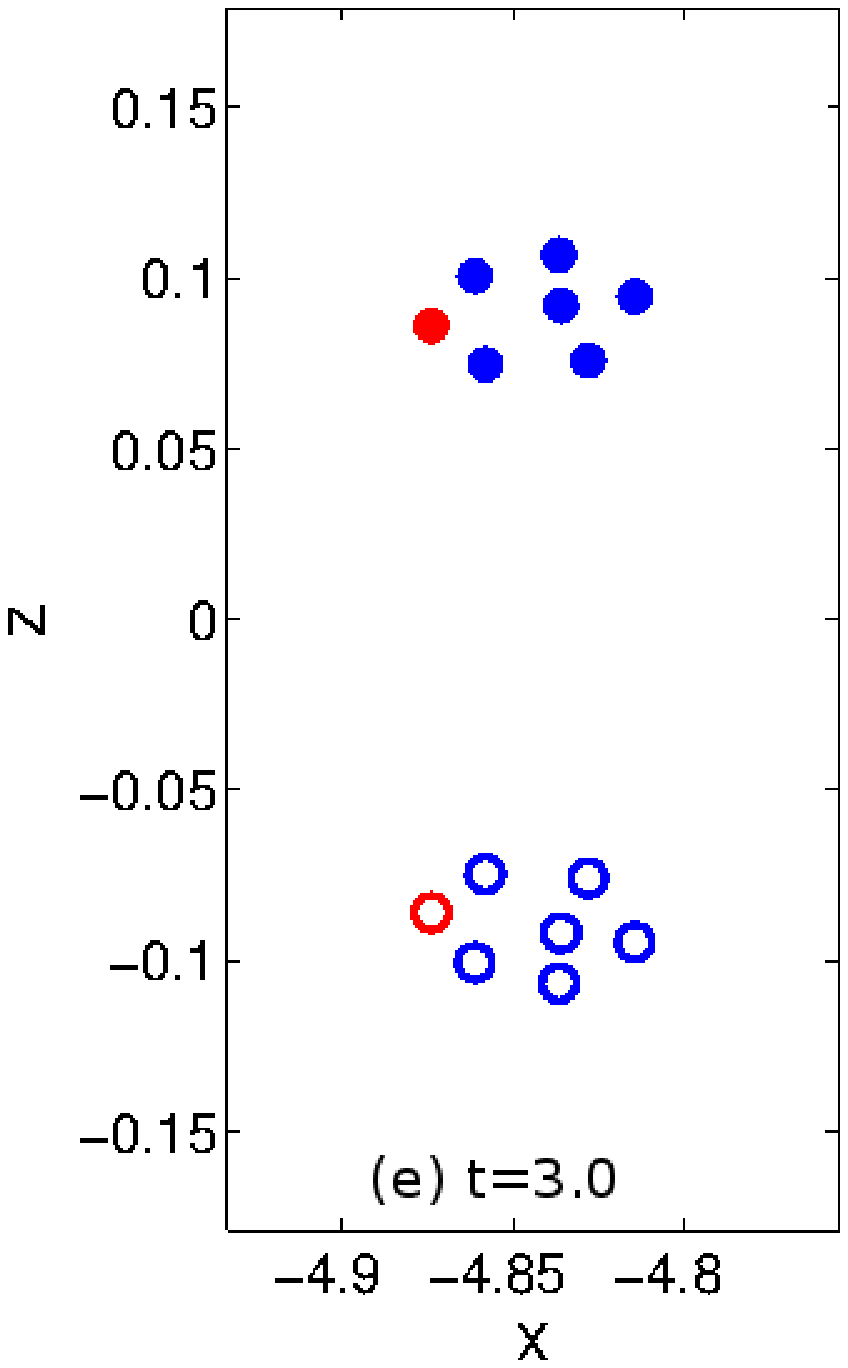}  
  \includegraphics[scale=0.38]{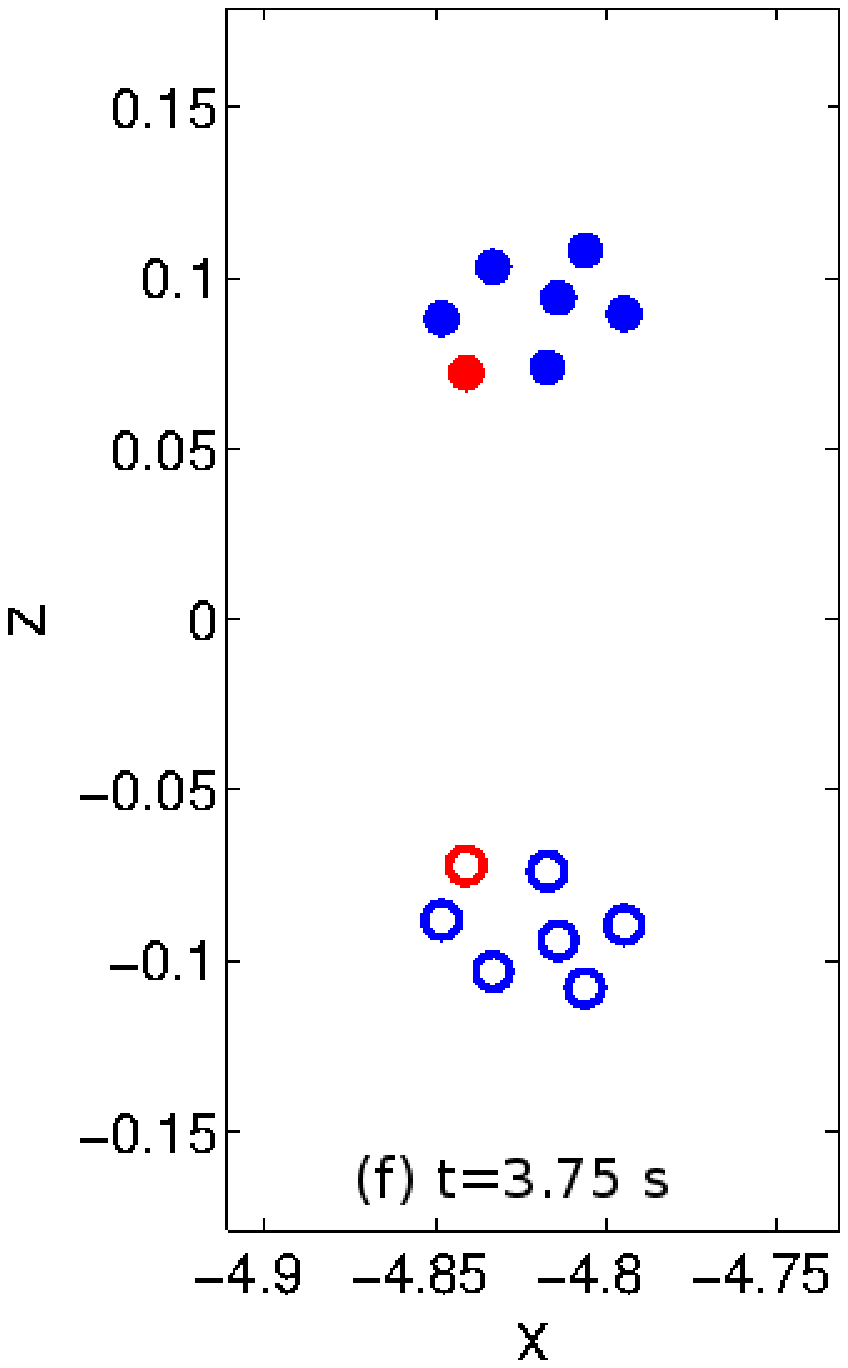} \\
\vskip 1cm
%
  \includegraphics[scale=0.38]{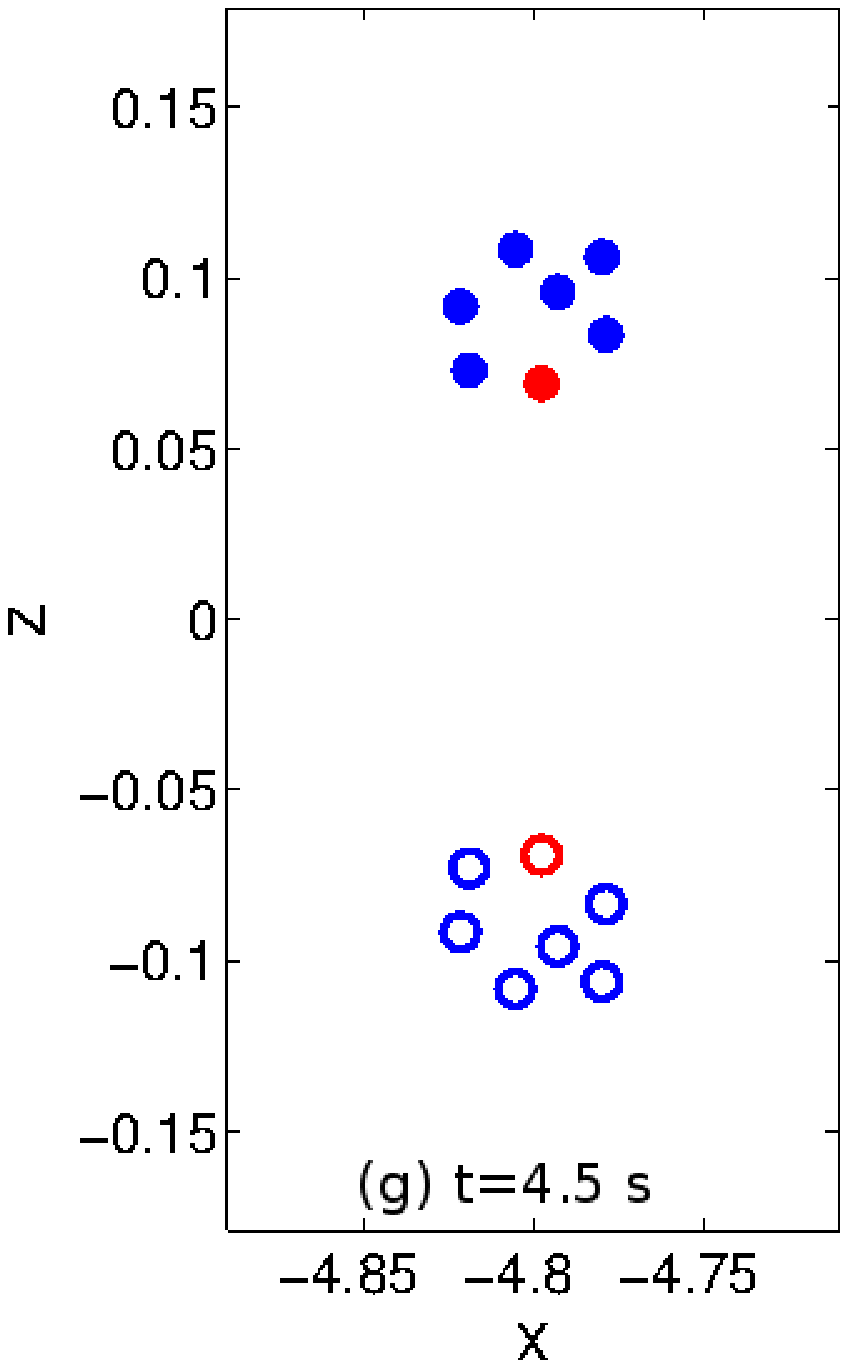}  
  \includegraphics[scale=0.38]{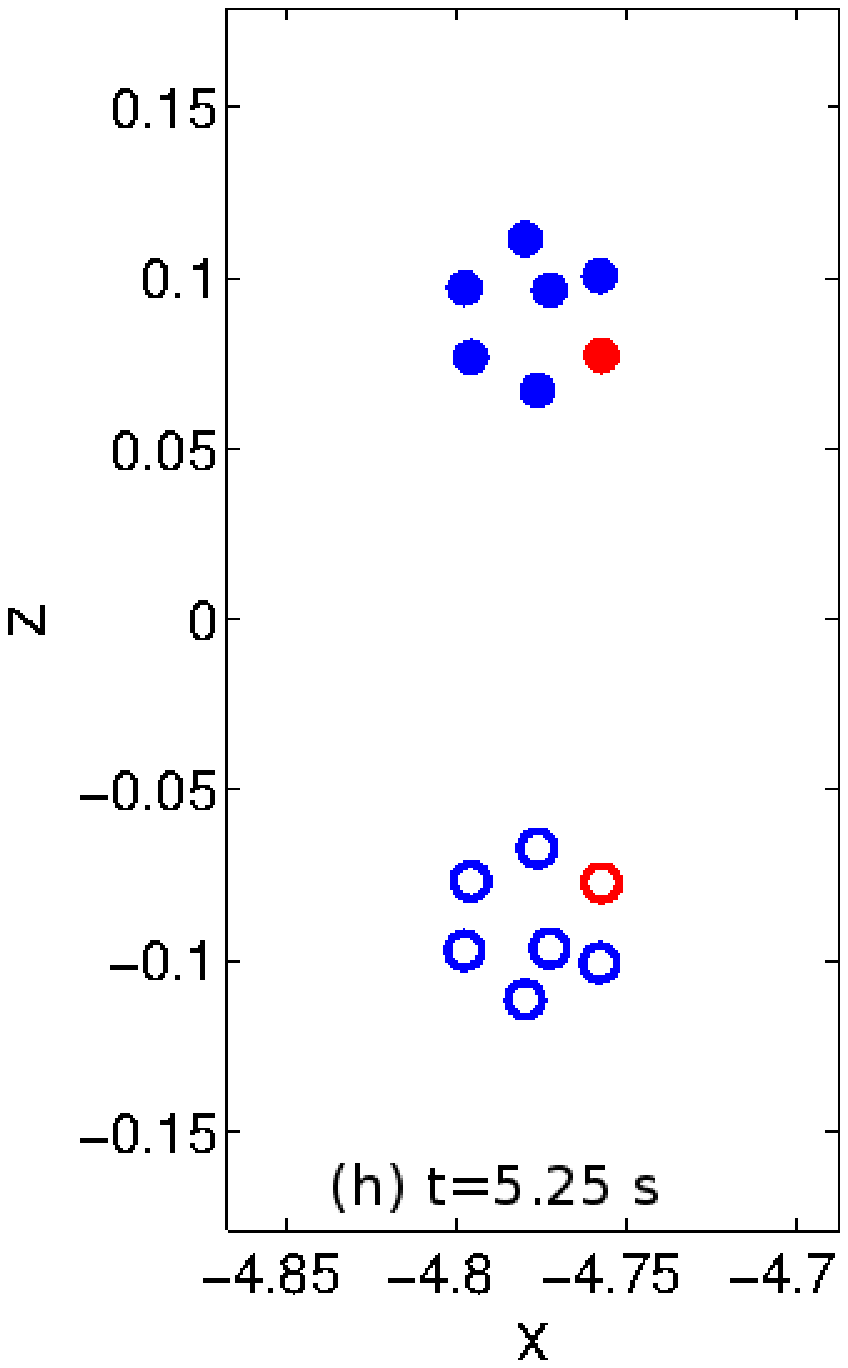}  
  \includegraphics[scale=0.38]{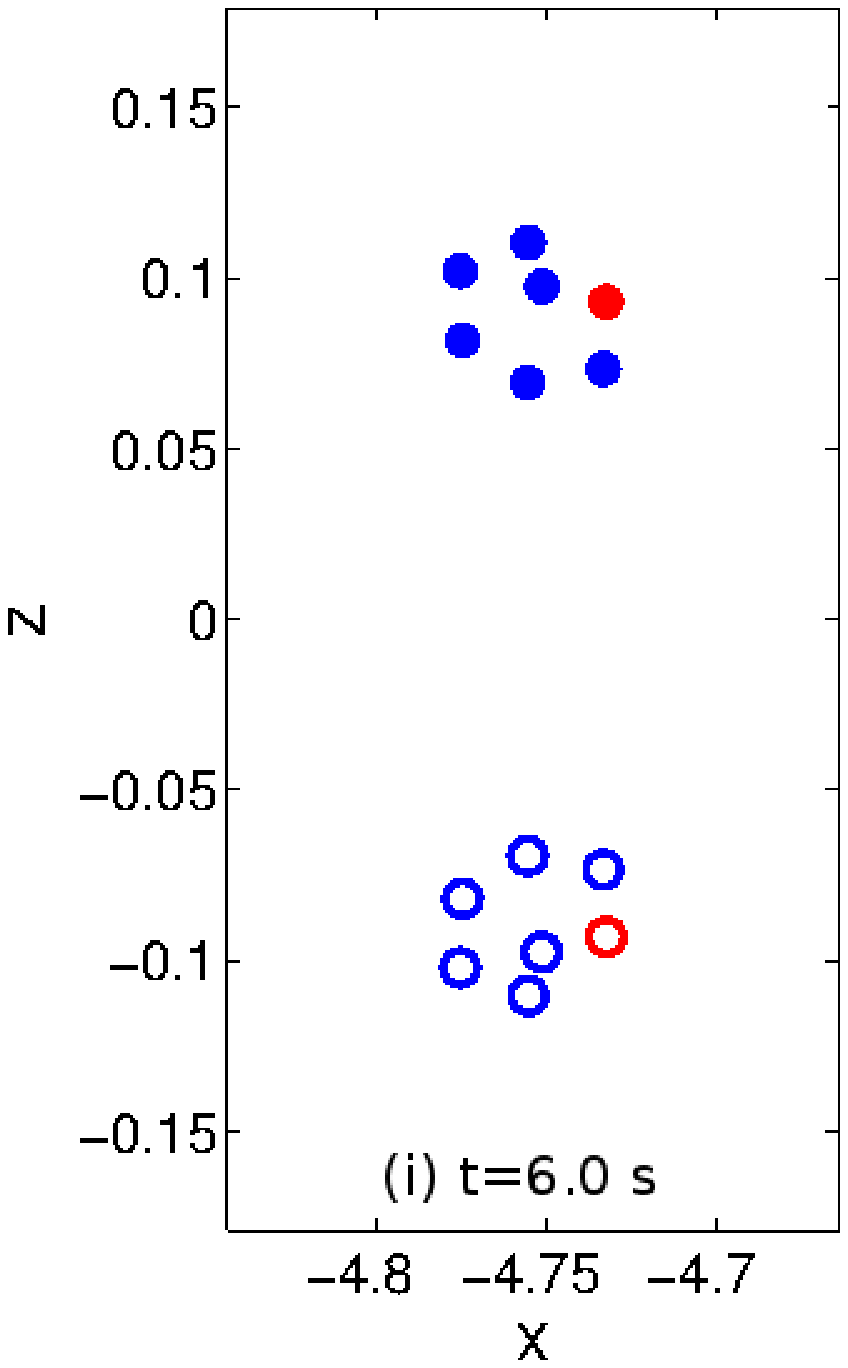}

\vskip 3cm

\caption{Generalised leapfrogging.
As in Fig.~\ref{fig:6}, but $N=7$. Note again the ellipticity of the
trajectories. One vortex has been marked in red colour to follow its motion.
}
\label{fig:9}
\end{figure}

\begin{figure}
\centering
  \includegraphics[scale=0.33]{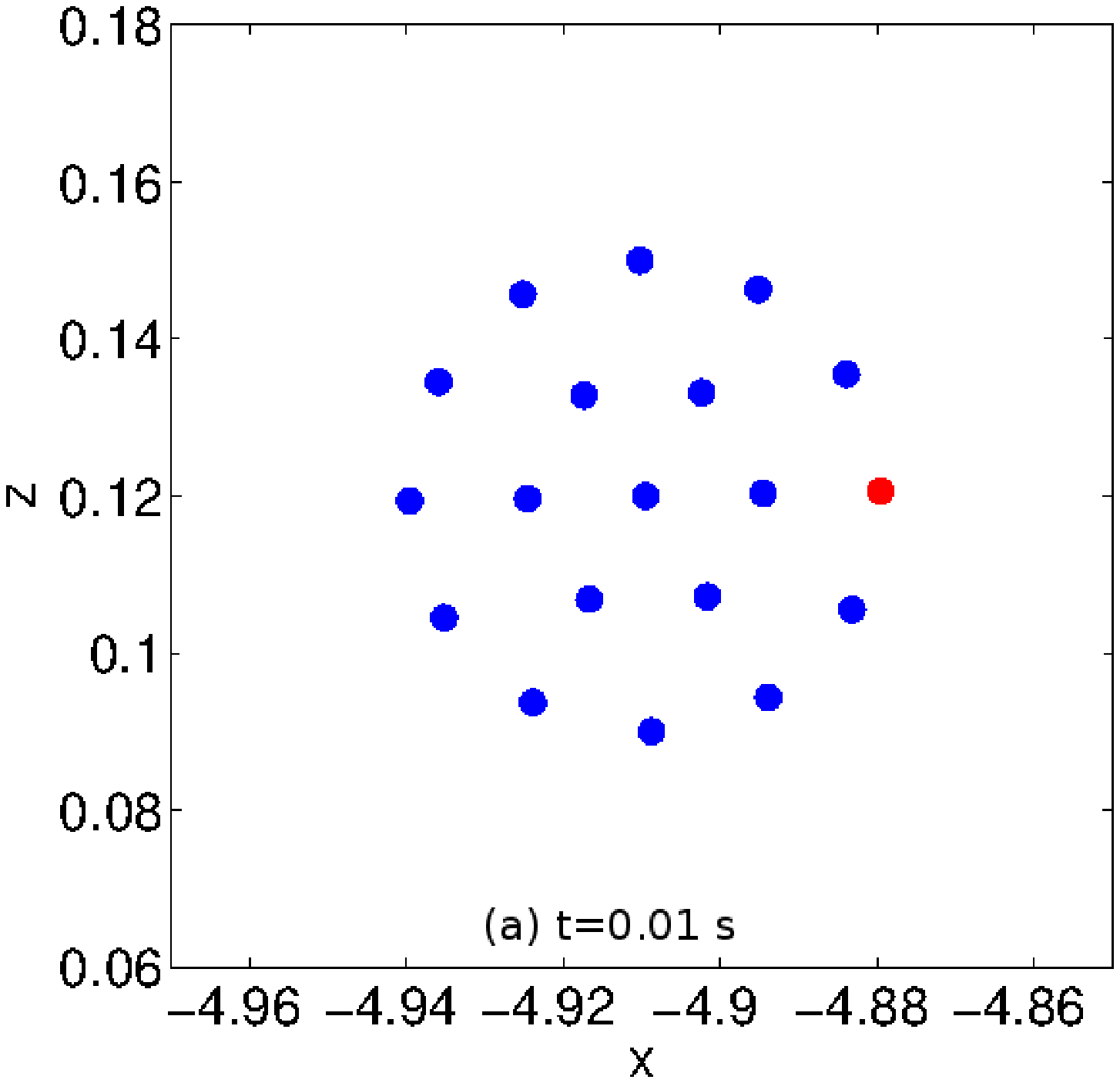}
  \includegraphics[scale=0.33]{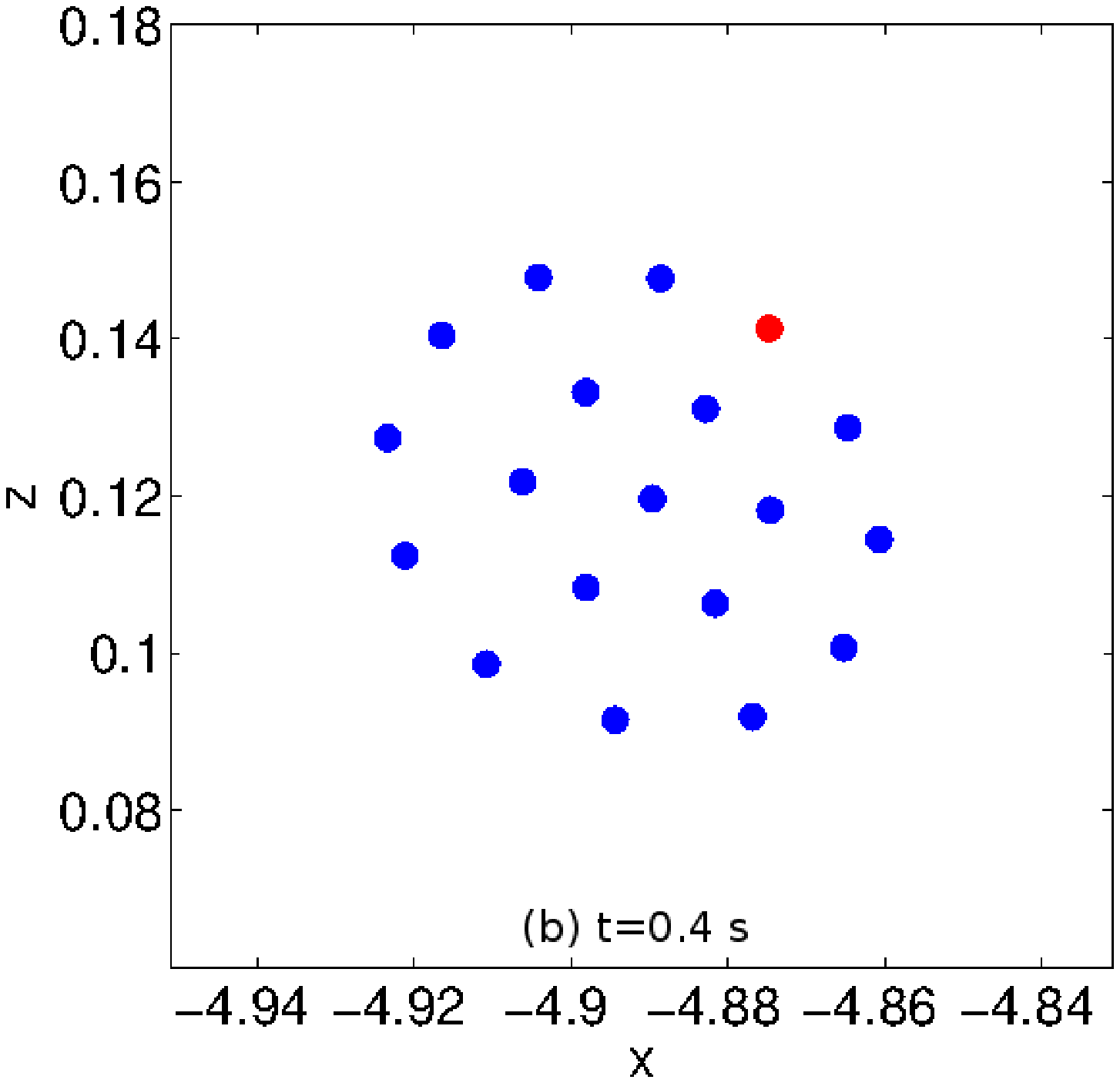}
  \includegraphics[scale=0.33]{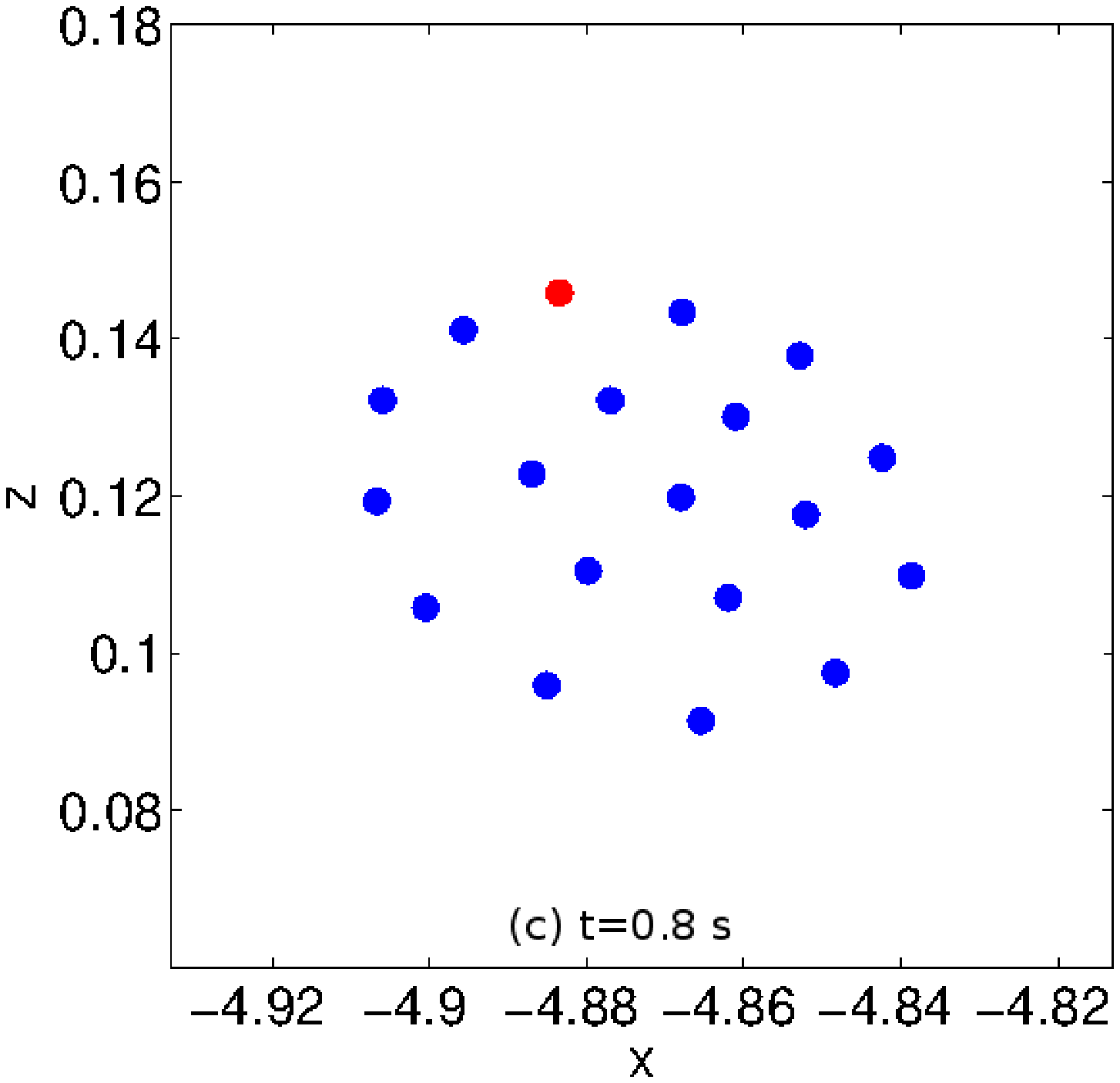} \\
\vskip 1cm
%
  \includegraphics[scale=0.33]{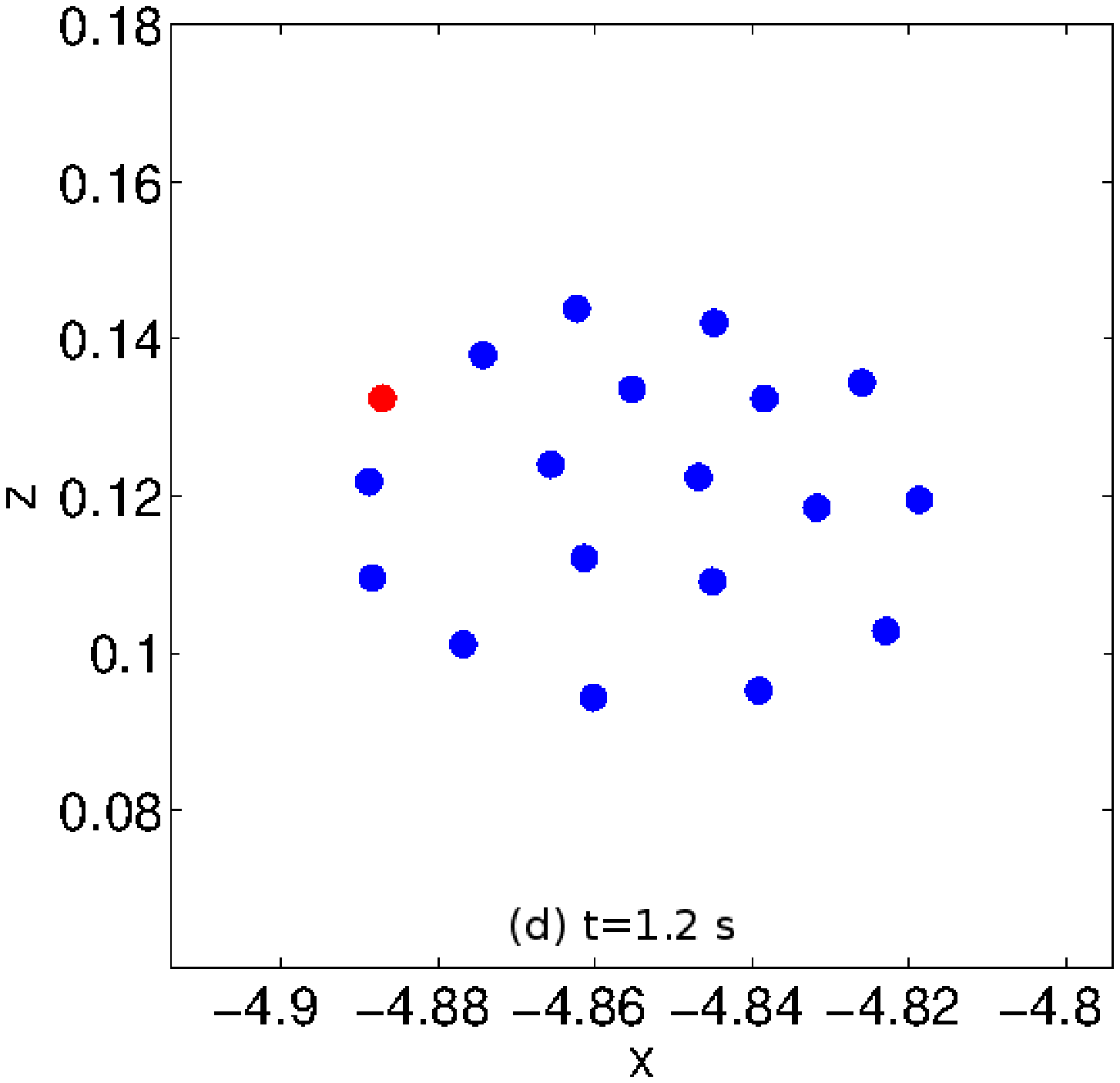}  
  \includegraphics[scale=0.33]{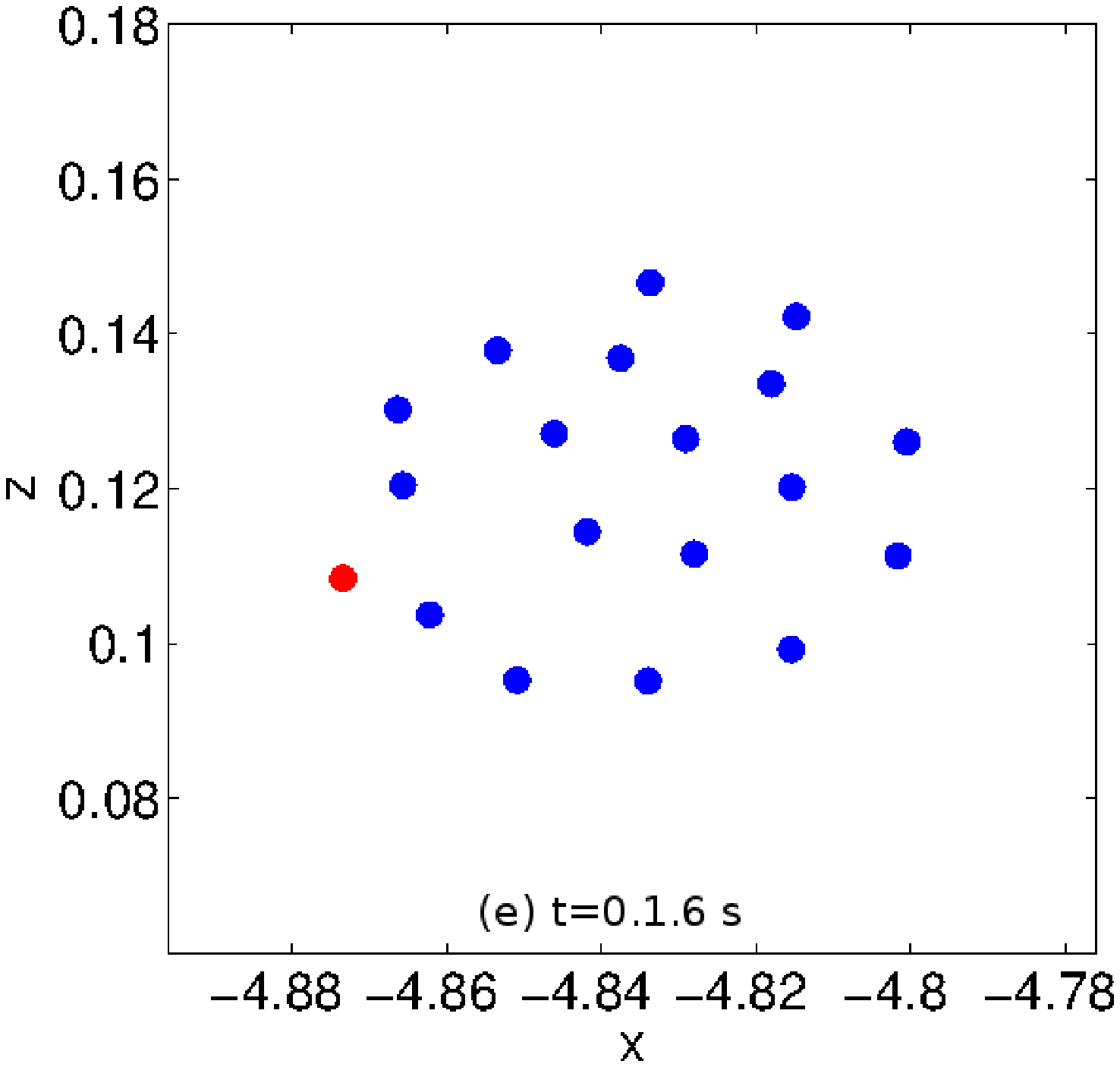}  
  \includegraphics[scale=0.33]{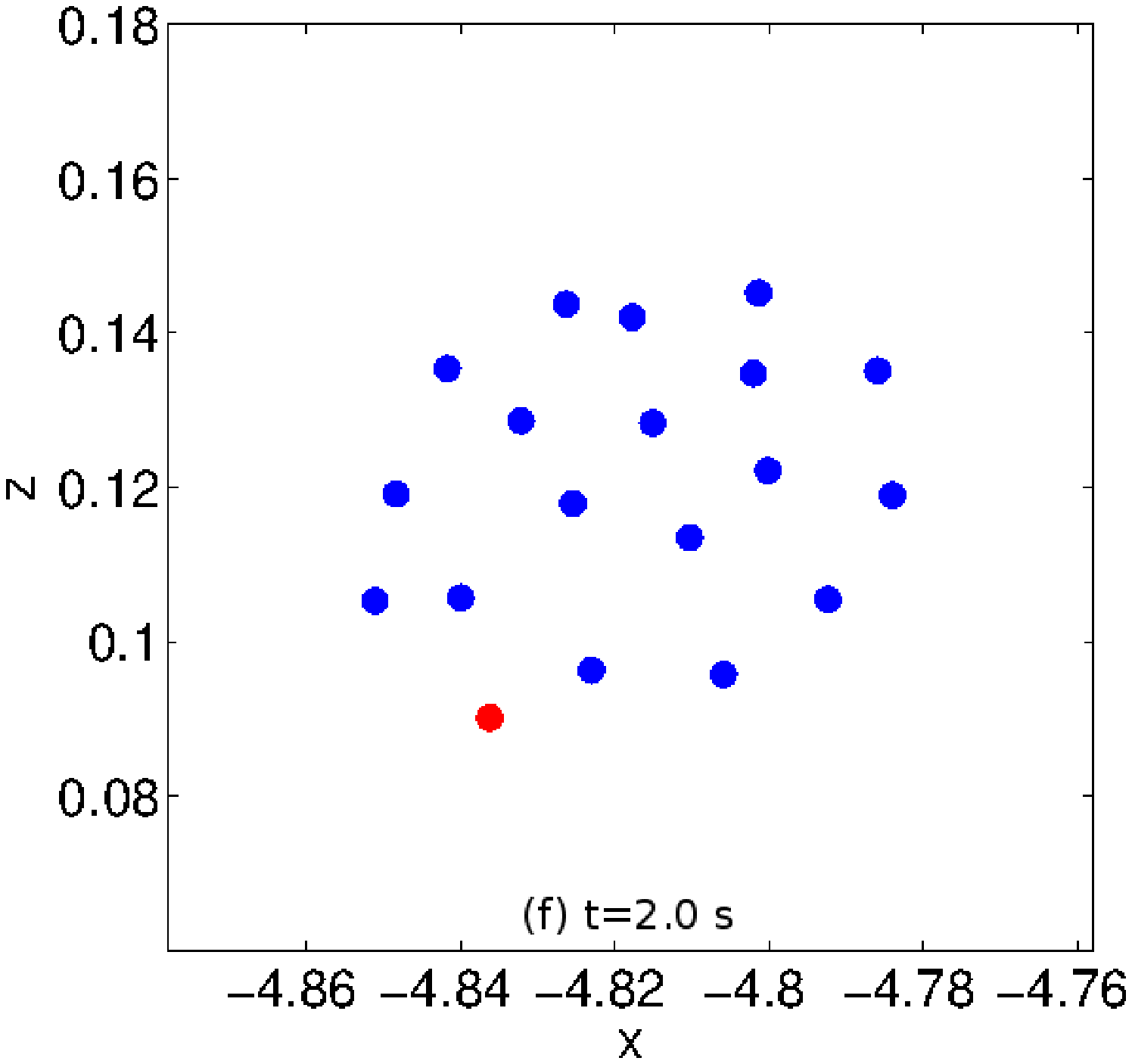} \\
\vskip 1cm
%
  \includegraphics[scale=0.33]{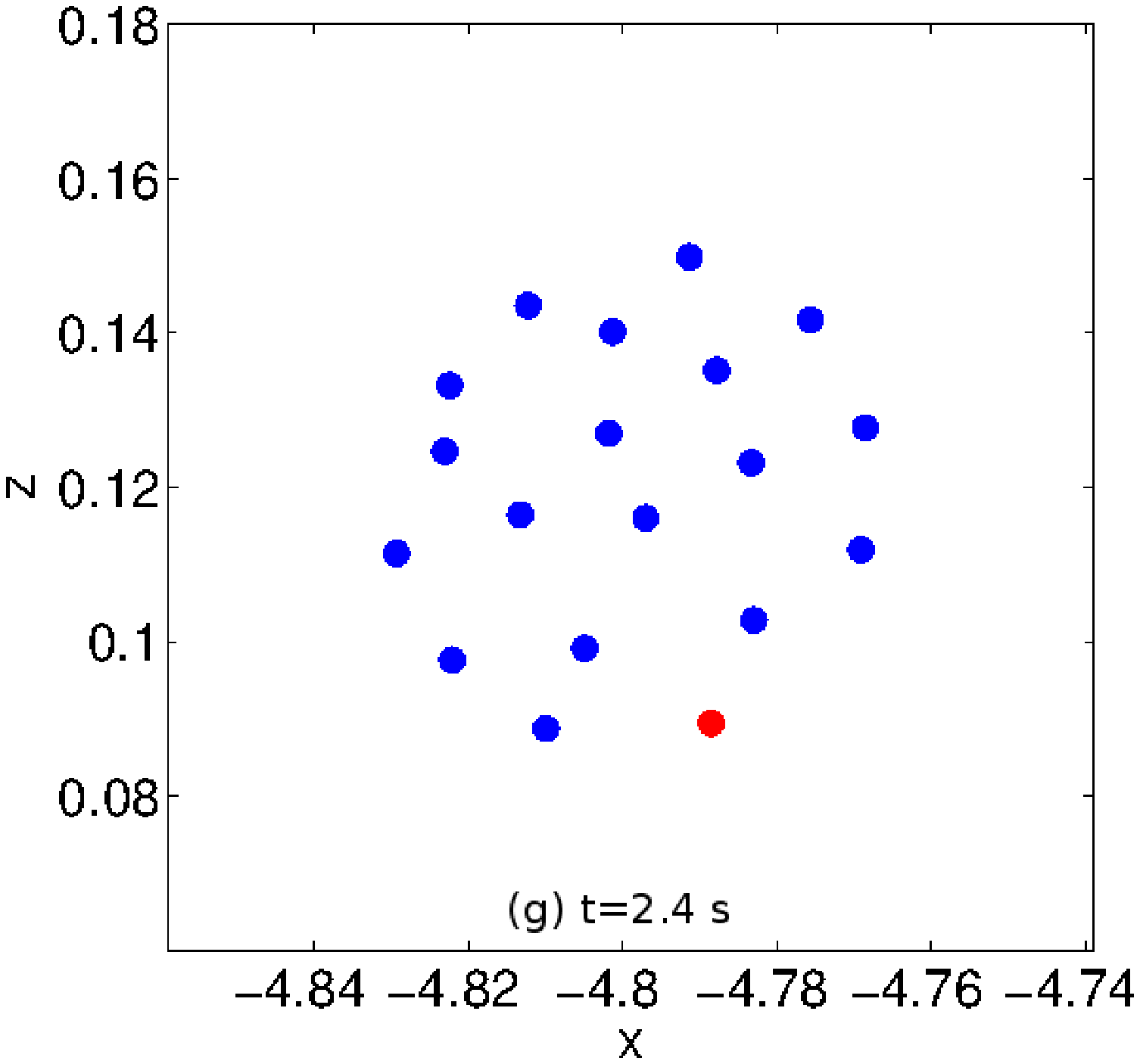}  
  \includegraphics[scale=0.33]{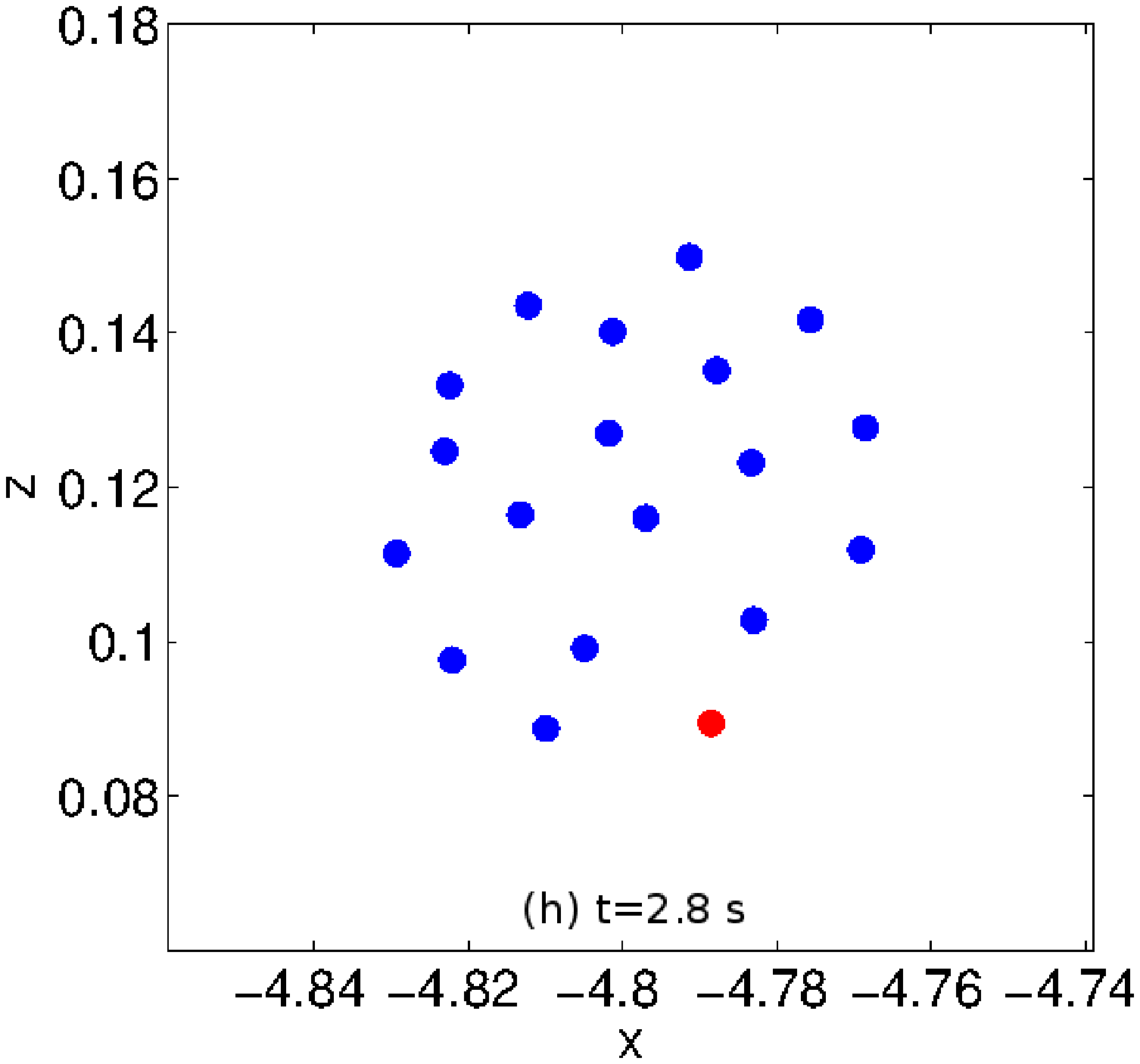}  
  \includegraphics[scale=0.33]{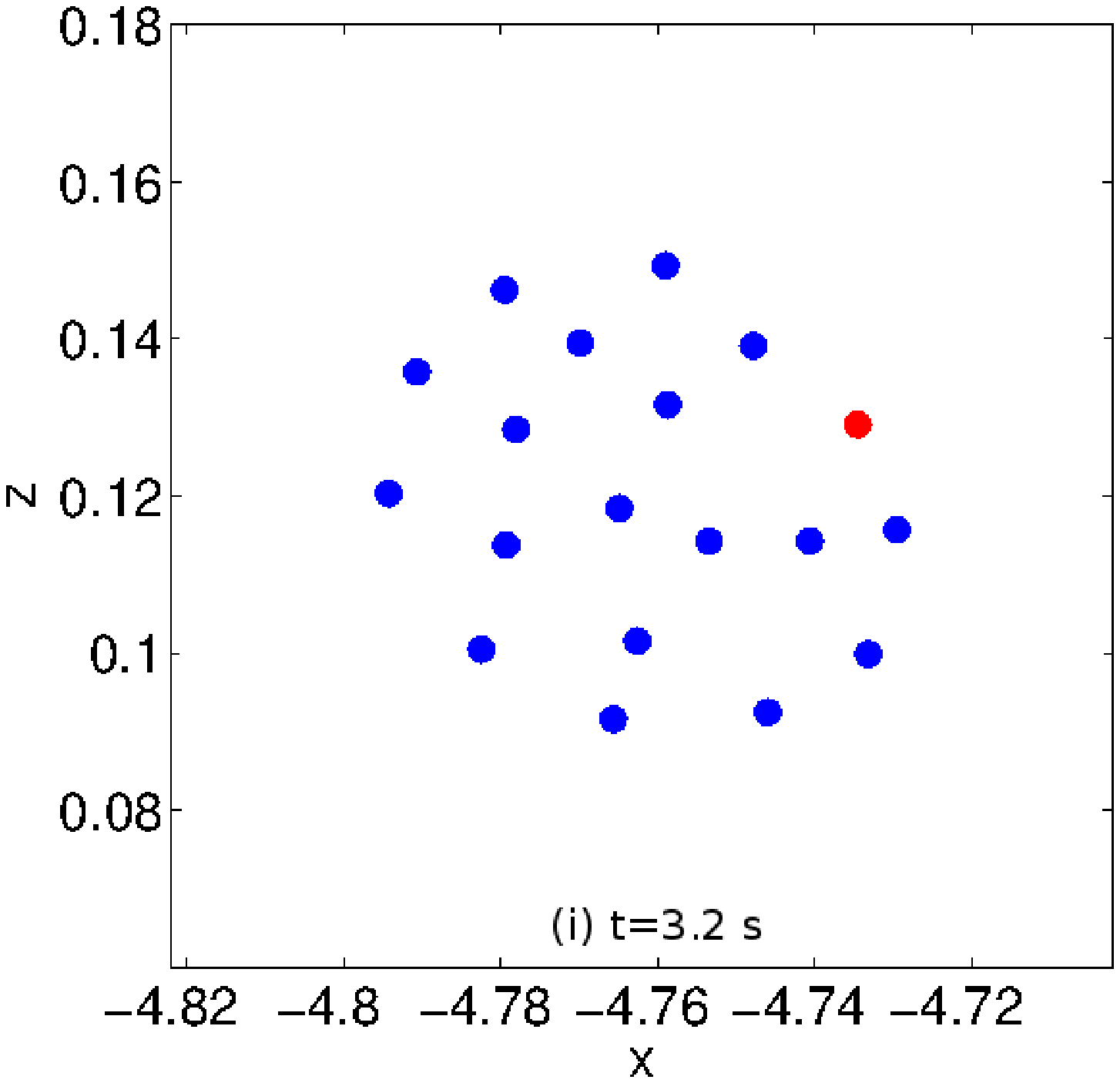}

\caption{Generalised leapfrogging.
As in Fig.~\ref{fig:6}, but $N=19$. Only one cross-section is shown. Again,
one vortex has been marked in red colour to follow its motion.
Parameters: $R=0.12~\rm cm$, $a=0.03~\rm cm$, $\ell=0.015~\rm cm$, $R/a=4$.
}
\label{fig:10}
\end{figure}

Generalised leapfrogging of vortex bundles 
with $N=7$ and $N=19$ is shown in 
Fig.~\ref{fig:8},~\ref{fig:9} and ~\ref{fig:10}. 
The evolution is qualitatively the same for all values of $N$. 
Fig.~\ref{fig:10} shows the cross-section of an $N=19$ bundle at different
times: clearly the motion of individual vortex lines on the plane is 
more disordered.

It must be stressed that, after the first reconnection, the meaning of 
the term leapfrogging
is lost: the number of vortex rings may not remain constant, as the constituent
vortex rings do not have a well-defined identity any longer.

\subsection{Comparison between two-dimensional and three-dimensional 
leapfrogging}

Over $100$ years ago,  Love \cite{Love} showed analytically
that two vortex-antivortex pairs
(the two-dimensional analogue of two coaxial vortex rings)
will perform leapfrogging motion only if
the parameter $\hat{\alpha}$, defined as the ratio of diameters of the
smaller to the larger pair,
satisfies the condition $\hat{\alpha}>3-2\sqrt{2}\approx0.172=\hat{\alpha}_c$.
If $\hat{\alpha}<\hat{\alpha}_c$, leapfrogging does not occur:
the smaller, faster ring 'escapes' the influence of the larger, slower ring 
and disappears off to infinity.
More recently, using numerical methods, Acheson \cite{Acheson} extended 
Love's work and identified three regimes: (i) when $\hat{\alpha}<0.172$ 
leapfrogging does not occur;
(ii) when $0.172<\hat{\alpha}<0.382$ leapfrogging is possible, but is unstable.
(iii) when $\hat{\alpha}>0.382$, leapfrogging occurs without instabilities: 
the system is unaffected by small perturbations.

The comparison of Love's and Acheson's two-dimensional work with
our three-dimensional calculation raises the question 
as to whether
there exists a critical $\hat{\alpha}$ for a system of $N=2$ (initially) 
coplanar, coaxial rings, and, if so, what is its value.

%
We have performed  numerical simulations to identify the range of 
leapfrogging for $N=2$
in our three-dimensional case.  We use the criterion that
leapfrogging occurs only when $dR/dt=\dot{R}$ is non-zero: 
if $\hat{\alpha}$ is such that $\dot{R}=0$
the vortex rings have moved so far apart that one ring has no effect 
on the other.
To make connection to the two-dimensional work of Love and Acheson we relate
their parameter $\hat{\alpha}$ to our $R/a$, obtaining

\begin{equation}
\hat{\alpha}=\frac{R-a}{R+a}=\frac{R/a-1}{R/a+1}.
\end{equation}

To determine the range in which $\left(R/a\right)_c$ 
(hence  $\hat{\alpha}_c$) lies,
first we proceed by fixing $R$ and varying $a$, then by fixing $a$ 
and varying $R$, obtaining the same results. 
We conclude that $3.5<\left(R/a\right)_c<3.75$ or,
equivalently, $0.556<\hat{\alpha}_c<0.579$,
as opposed to $\hat{\alpha}_c=0.172$ in Acheson's and Love's
two-dimensional case. Three-dimensional leapfrogging seems
more robust than two-dimensional one.

\subsection{Turbulent vortex bundles}
\label{sec:turbo}

Table~\ref{tab:2} shows that the $N=7$
and $N=19$ vortex bundles underwent
reconnections before reaching the distance $\Delta x \approx 10 D$
relevant to Borner's experiment, but nevertheless
remained coherent and travelled
at essentially the same constant speed up to this distance,
as shown in Fig.~\ref{fig:3}. 
In this section we examine in more detail the long-term
evolution of perturbations. We concentrate our attention to
the vortex bundle with $N=7$ rings (we use slightly different 
parameters than in Section~\ref{subsec:coherent}).

\begin{figure}
\centering
\subfloat[$t=55.875~\rm s$] {
  \includegraphics[scale=0.16]{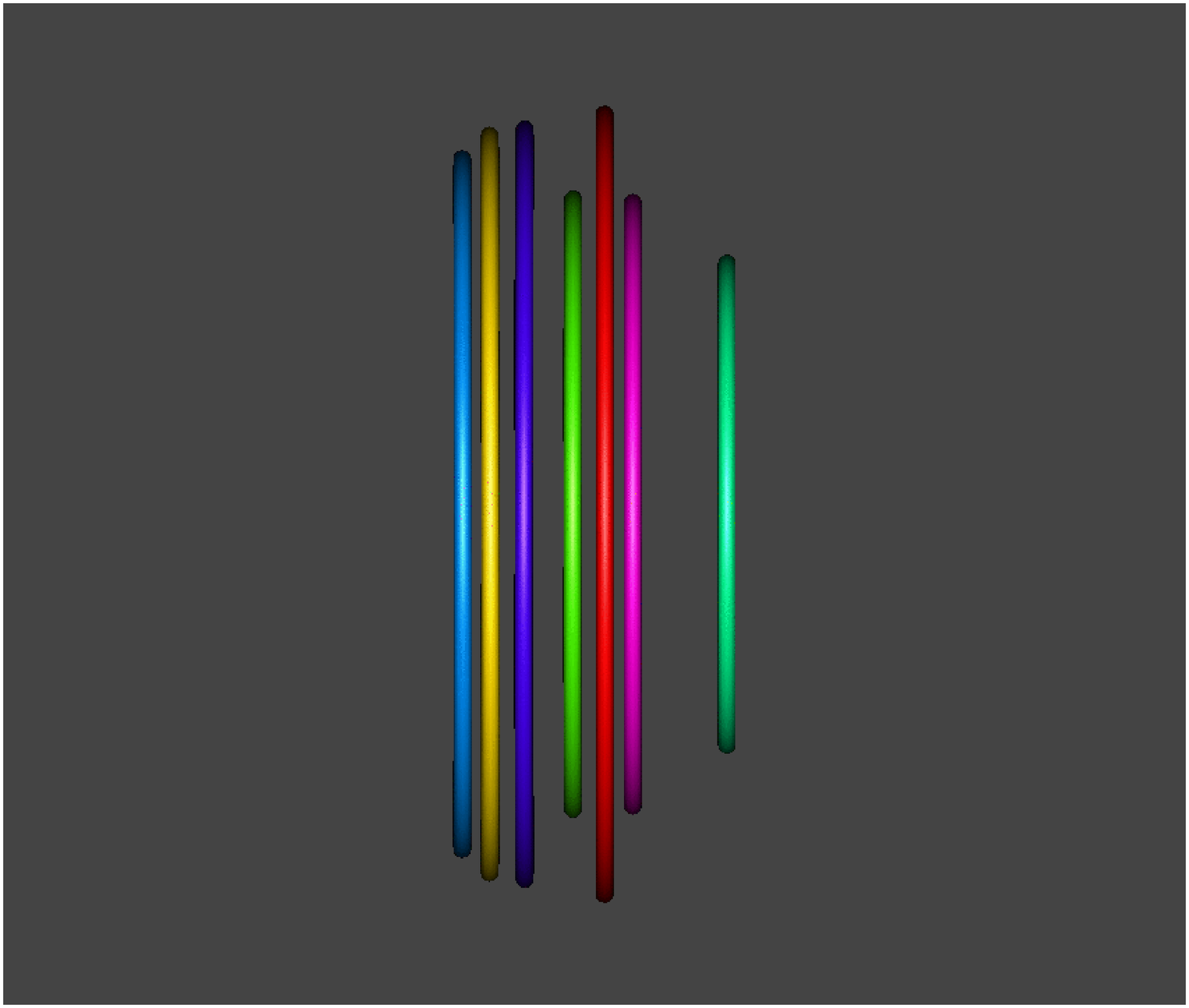}
}
\hspace{1em}
\subfloat[$t=55.75~\rm s$] {
  \includegraphics[scale=0.16]{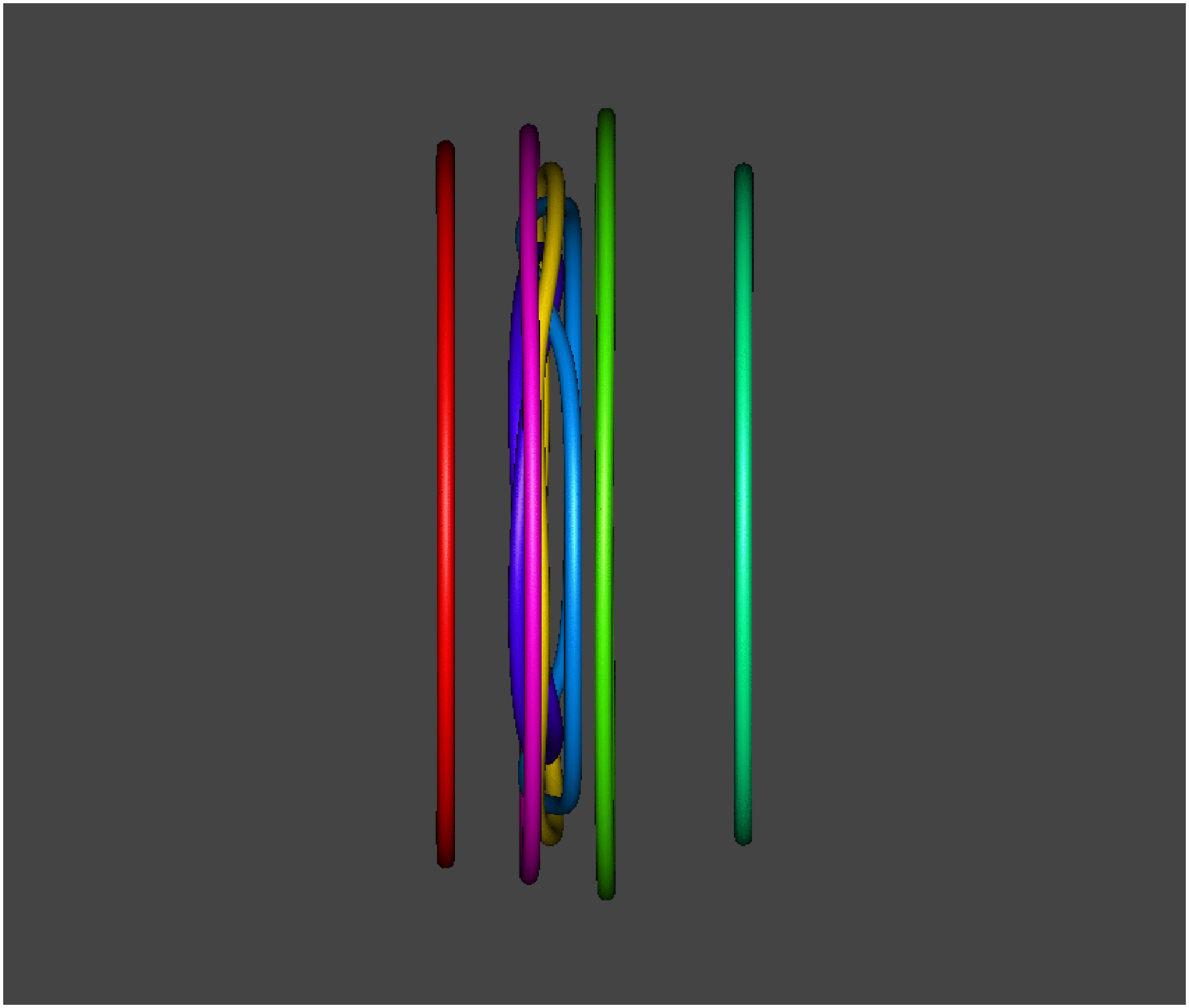}
}
\hspace{1em}
\subfloat[$t=58.50~\rm s$] {
  \includegraphics[scale=0.16]{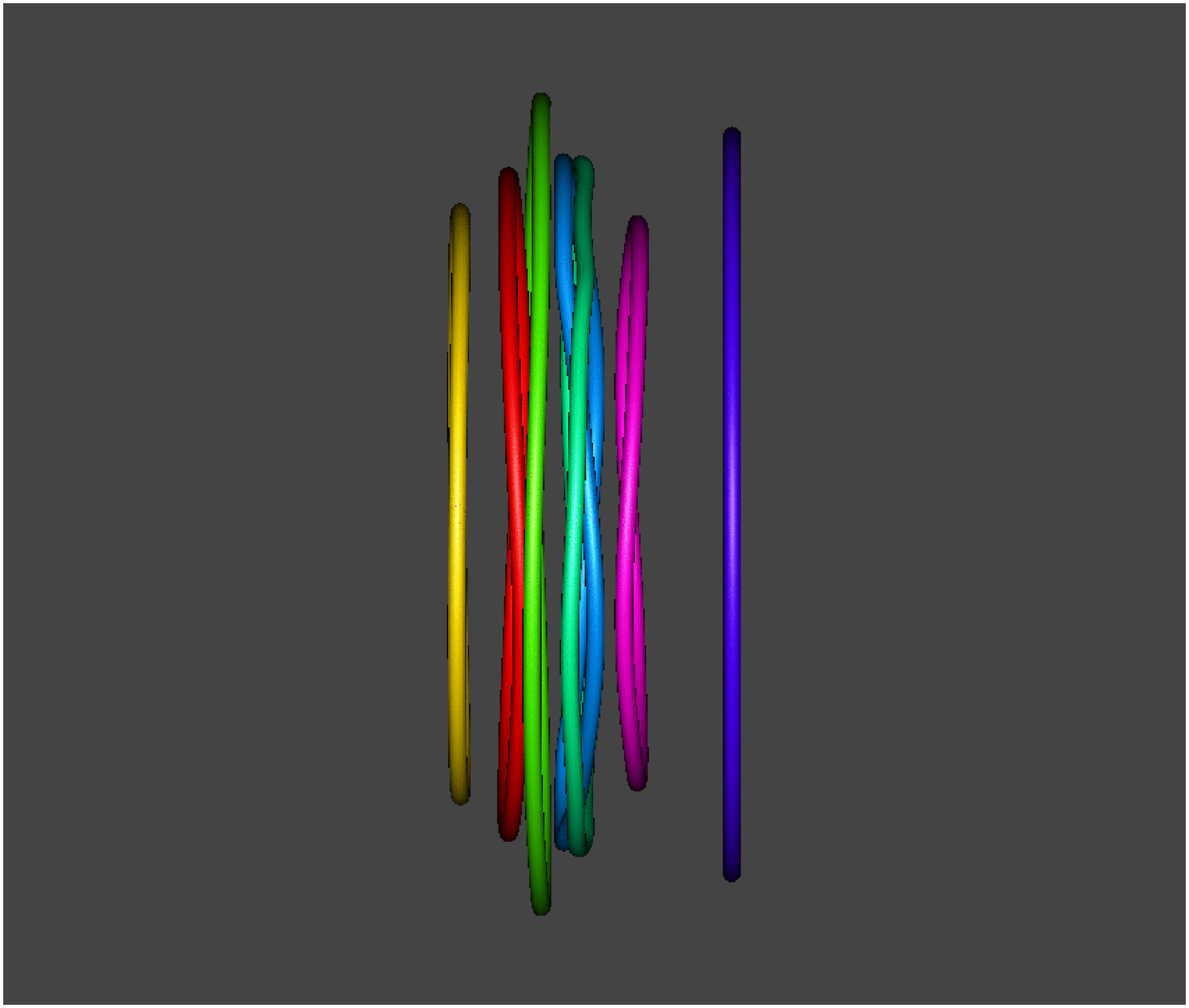}
}\\
\subfloat[$t=60.0~\rm s$] {
  \includegraphics[scale=0.16]{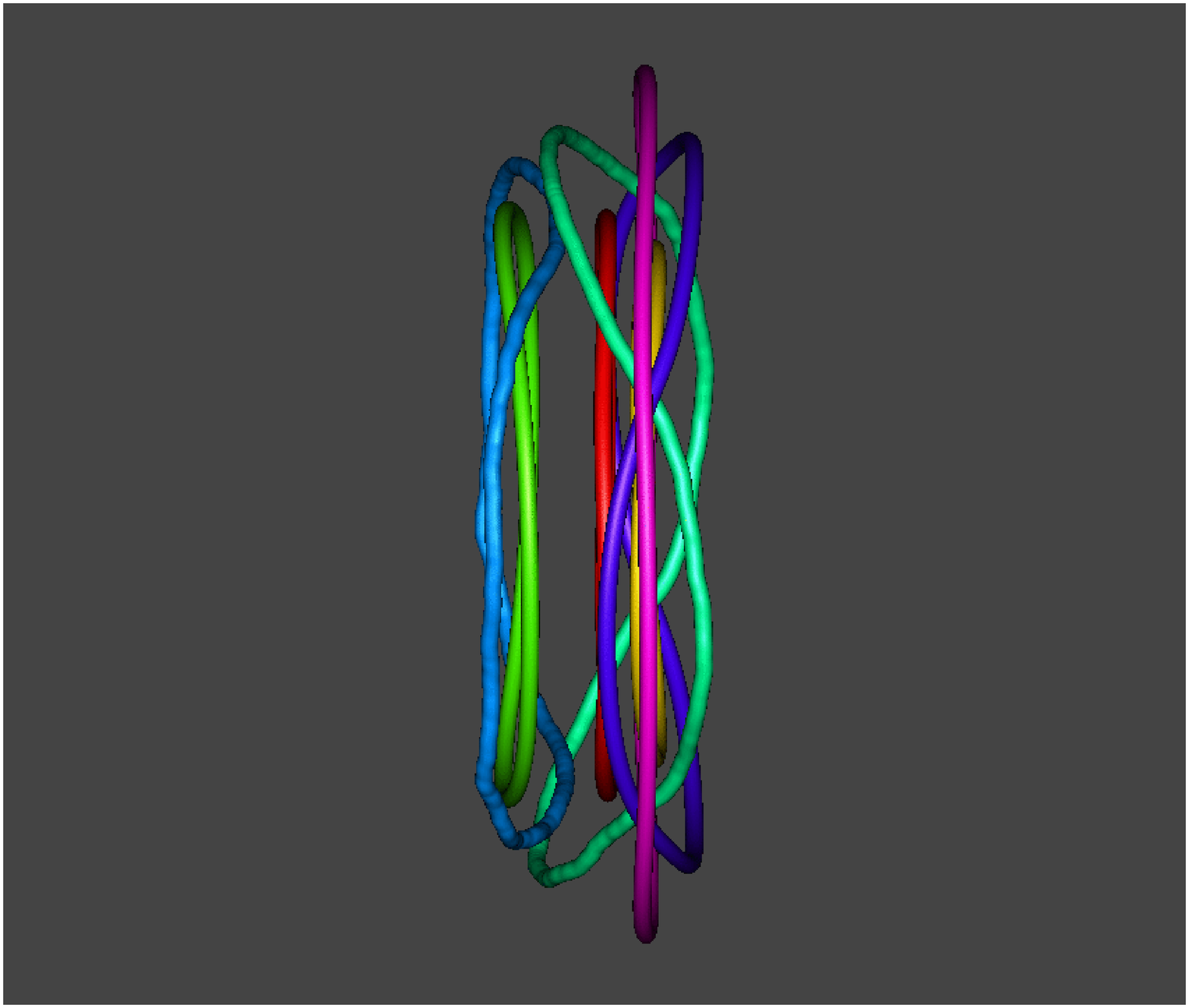}
}
\hspace{1em}
\subfloat[$t=63.75~\rm s$] {
  \includegraphics[scale=0.16]{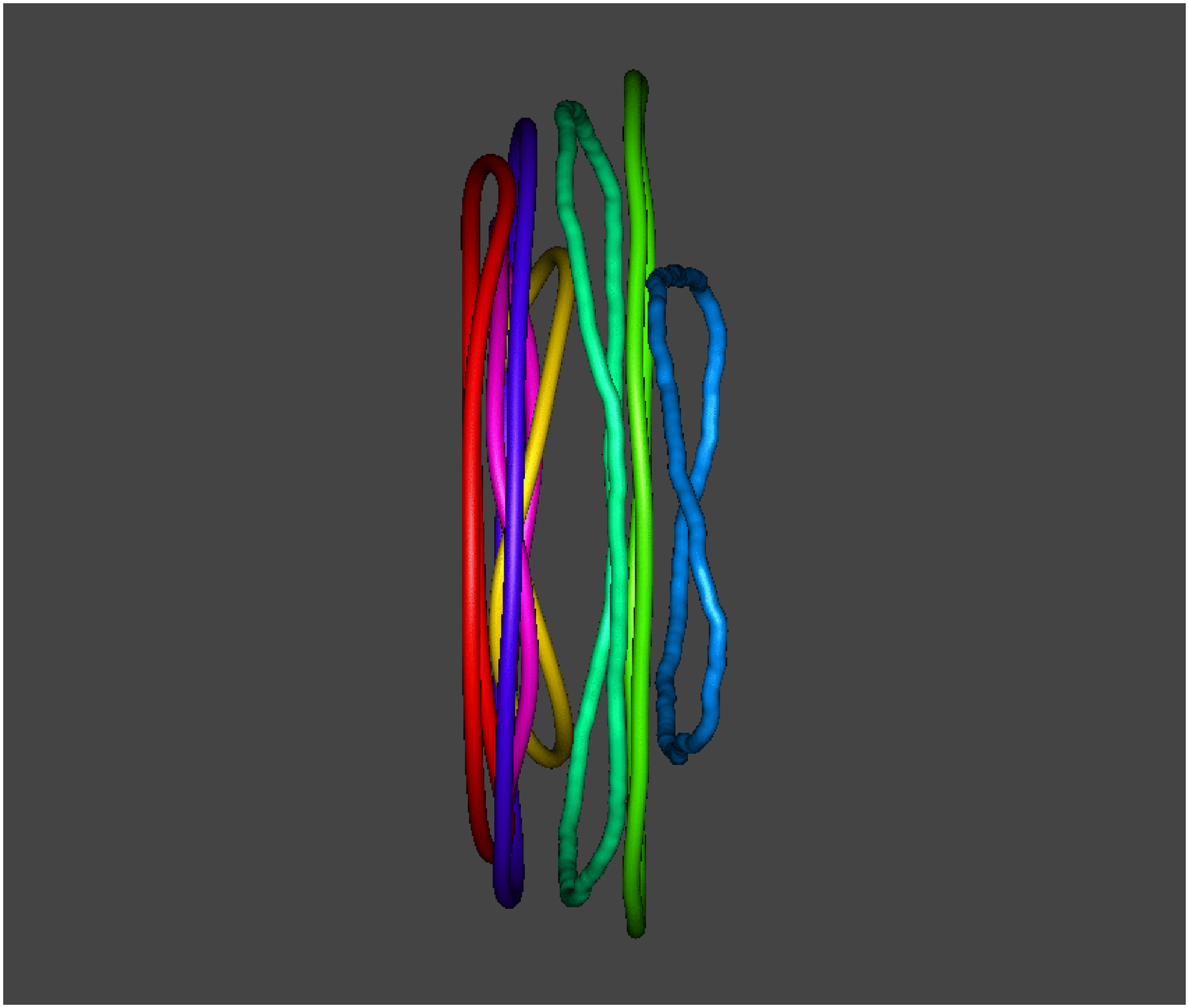}
}
\hspace{1em}
\subfloat[$t=67.50~\rm s$] {
  \includegraphics[scale=0.16]{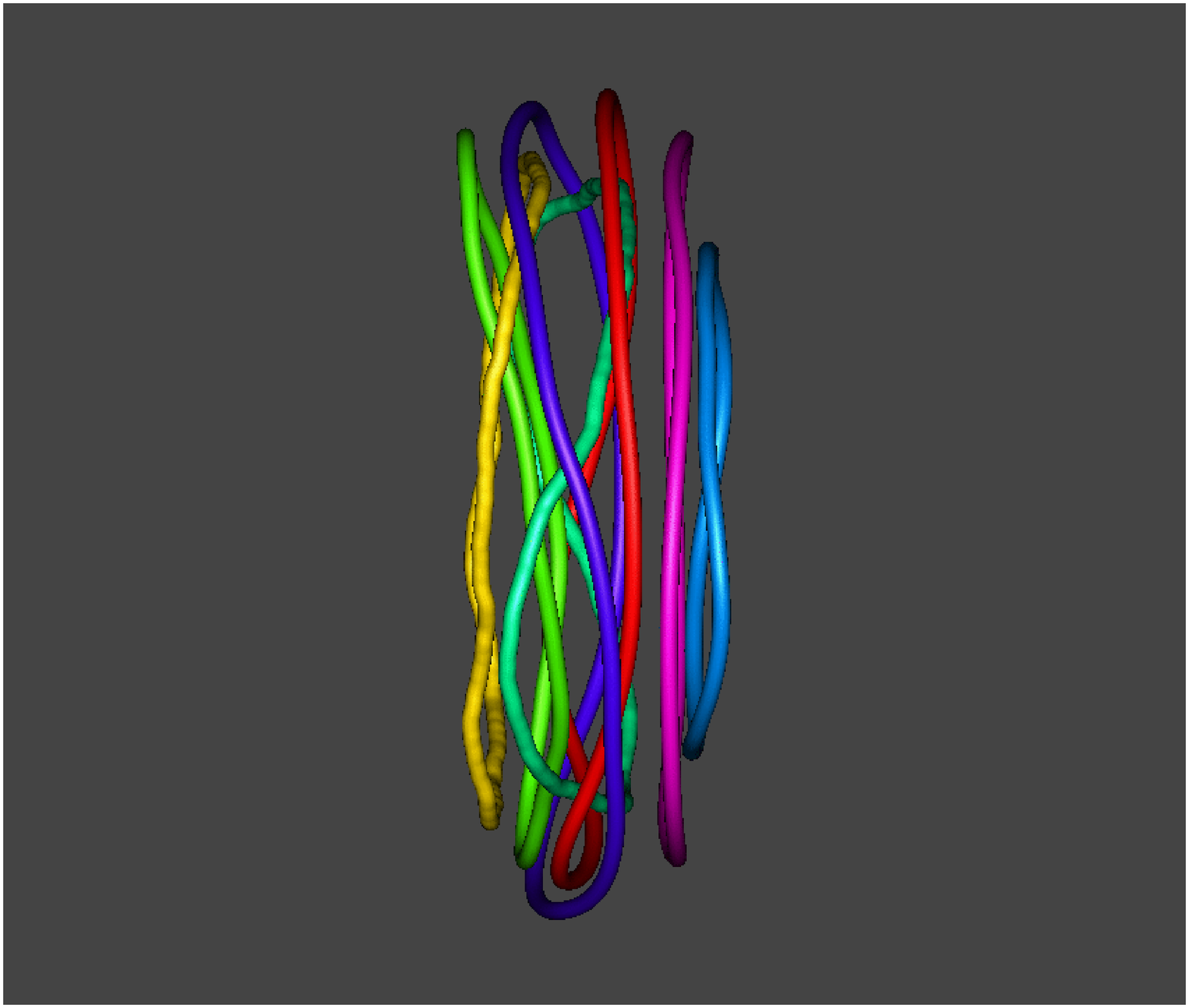}
}\\
\subfloat[$t=71.25~\rm s$] {
  \includegraphics[scale=0.16]{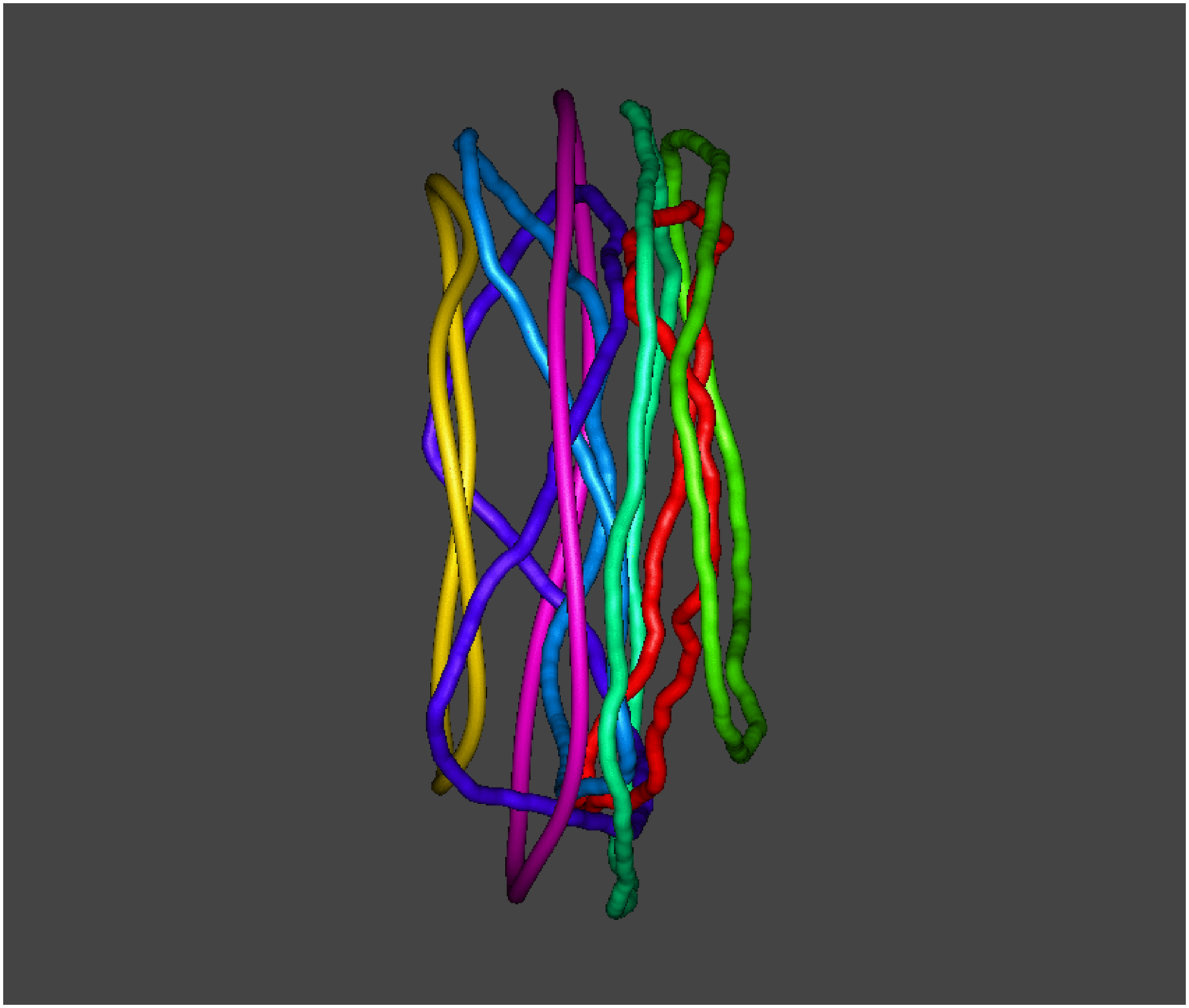}
}
\hspace{1em}
\subfloat[$t=75.0075~\rm s$] {
  \includegraphics[scale=0.16]{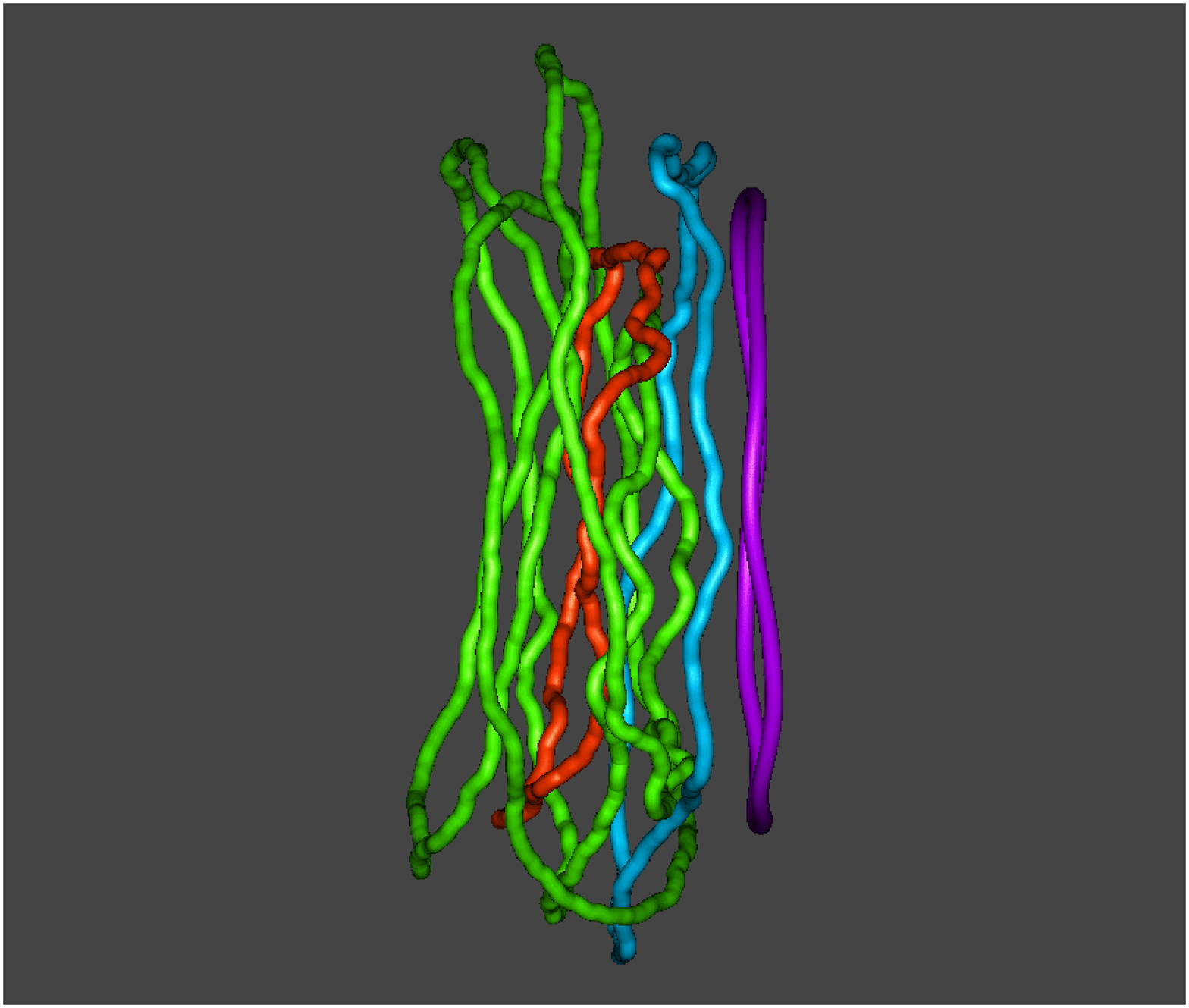}
}
\hspace{1em}
\subfloat[$t=78.75~\rm s$] {
  \includegraphics[scale=0.16]{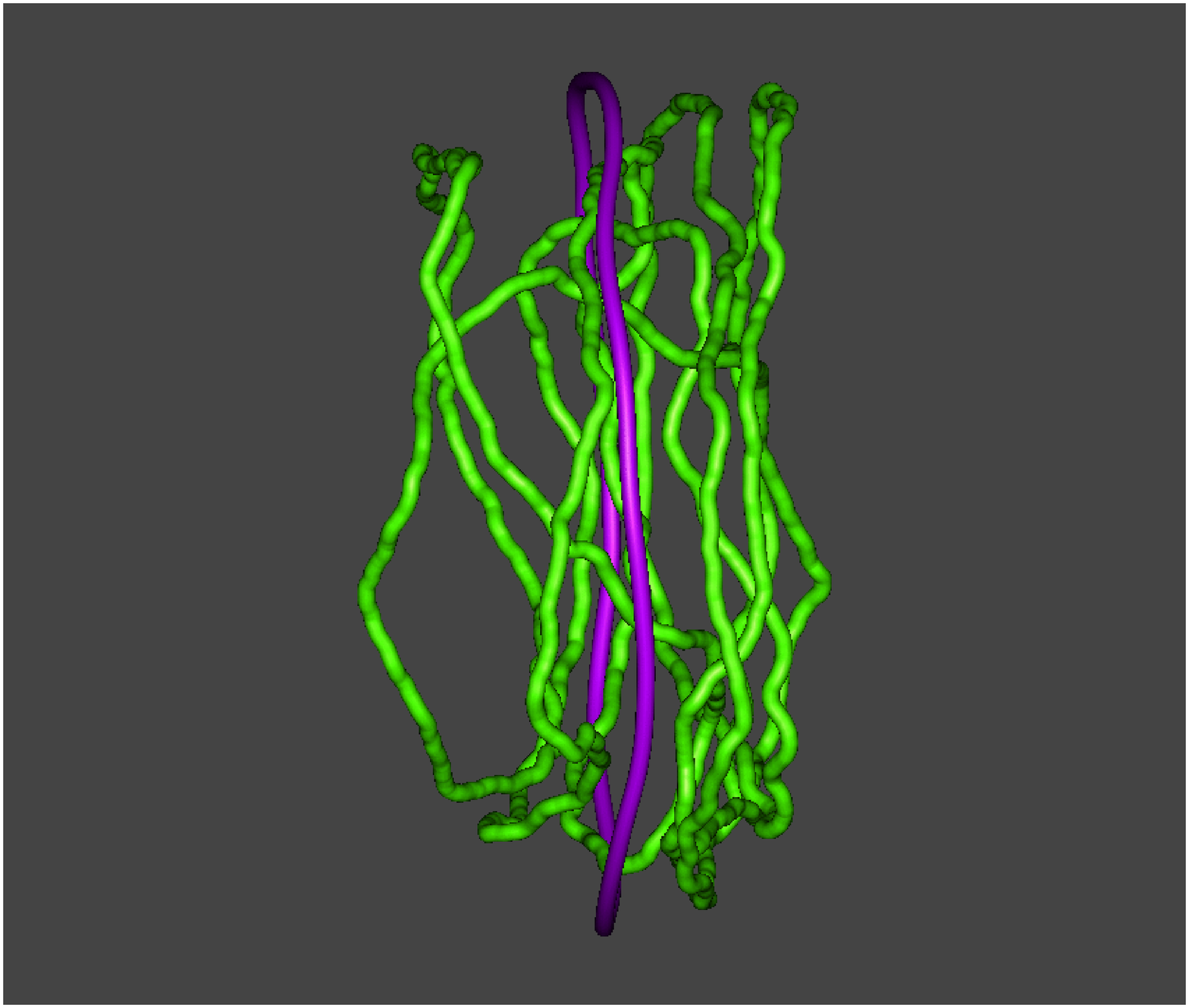}
}
\caption{Development of instabilities on $N=7$ vortex bundle, side views
at different times
(parameters: $R=0.0896~\rm cm$, $a=\ell=0.0223~\rm cm$,
$R/a=4$, $\Delta \xi=0.00149~\rm cm$, $\ell / \Delta \xi=15$, 
$\Delta t=5 \times 10^{-5}~\rm s$).
The rings are coloured arbitrarily (rings of the same colour in any two
images are not necessarily the same ring). For the sake of visibility,
the scale of the images and the
thickness of the vortex lines are also arbitrary.
}
\label{fig:11}
\end{figure}

\begin{figure}
\centering
\subfloat[$t=55.875~\rm s$] {
  \includegraphics[scale=0.16]{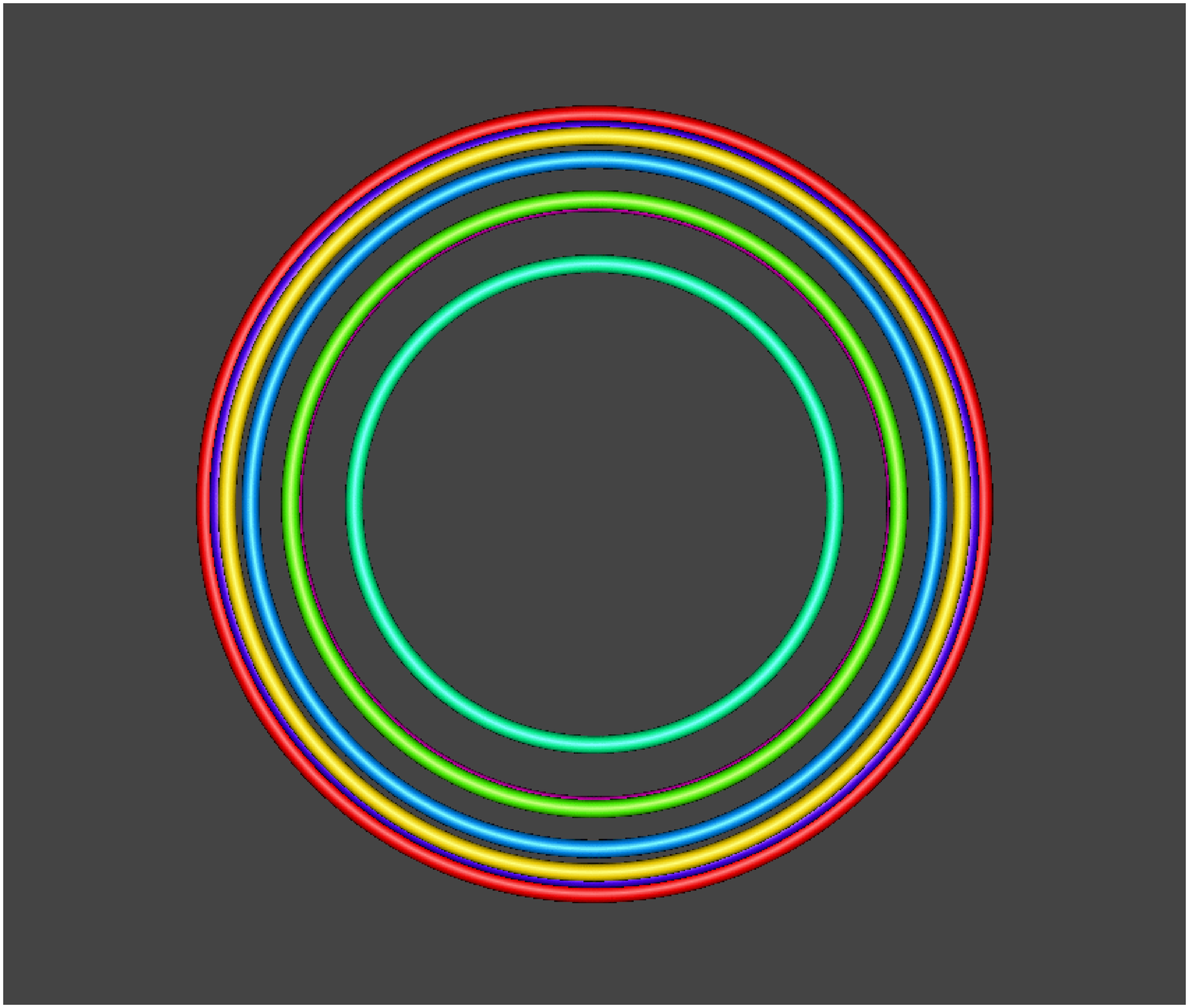}
}
\hspace{1em}
\subfloat[$t=55.75~\rm s$] {
  \includegraphics[scale=0.16]{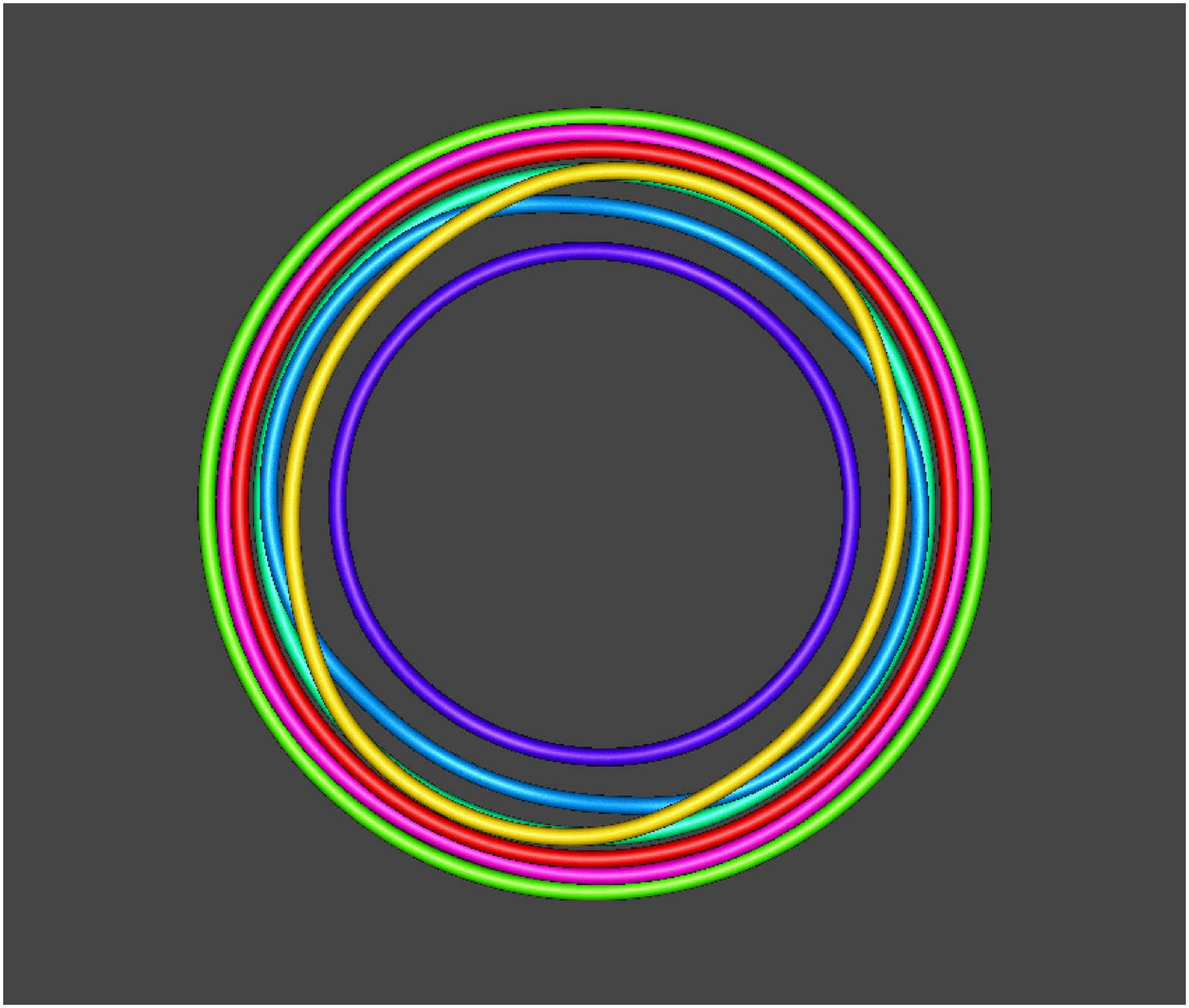}
}
\hspace{1em}
\subfloat[$t=58.50~\rm s$] {
  \includegraphics[scale=0.16]{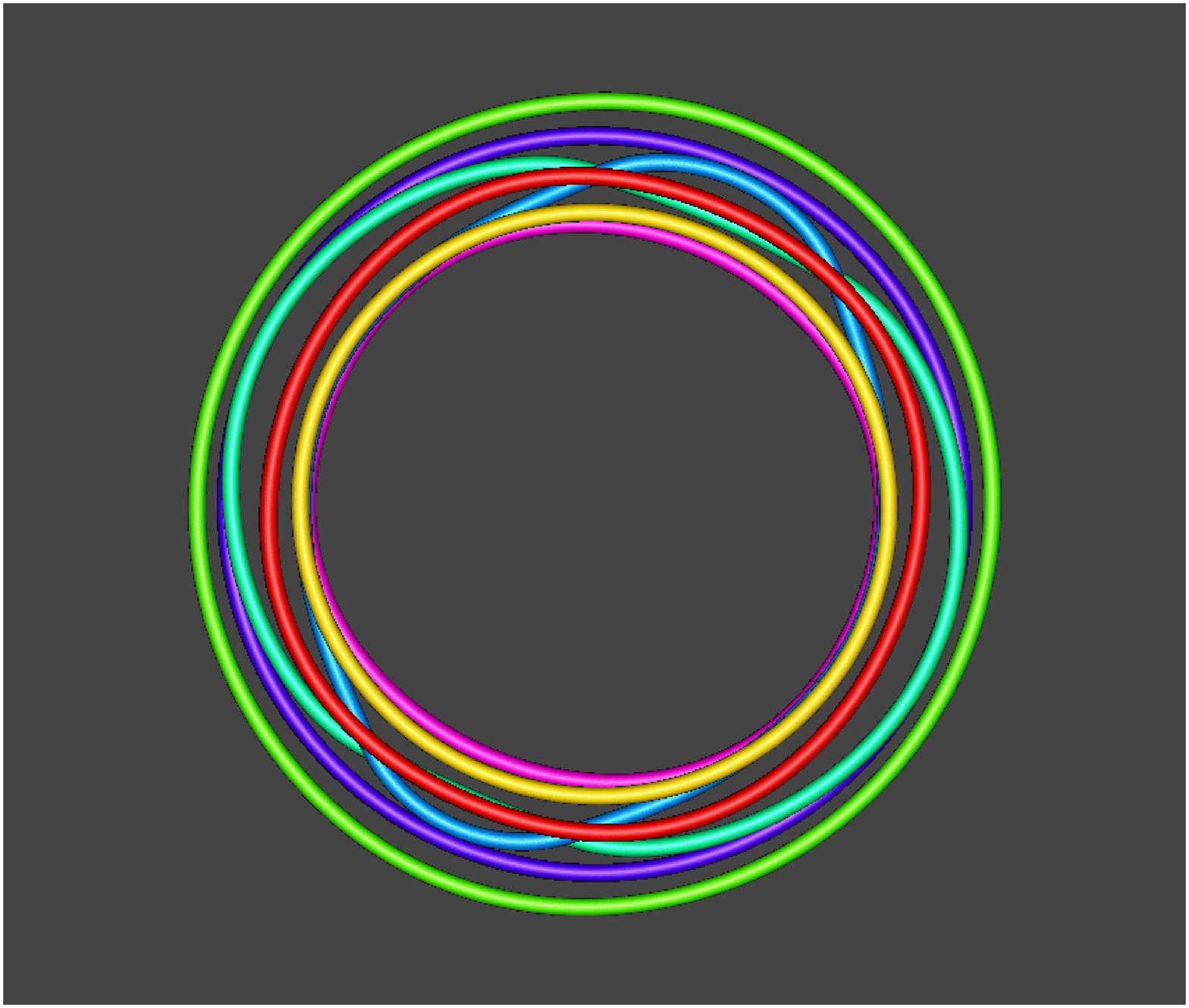}
}\\
\subfloat[$t=60.0~\rm s$] {
  \includegraphics[scale=0.16]{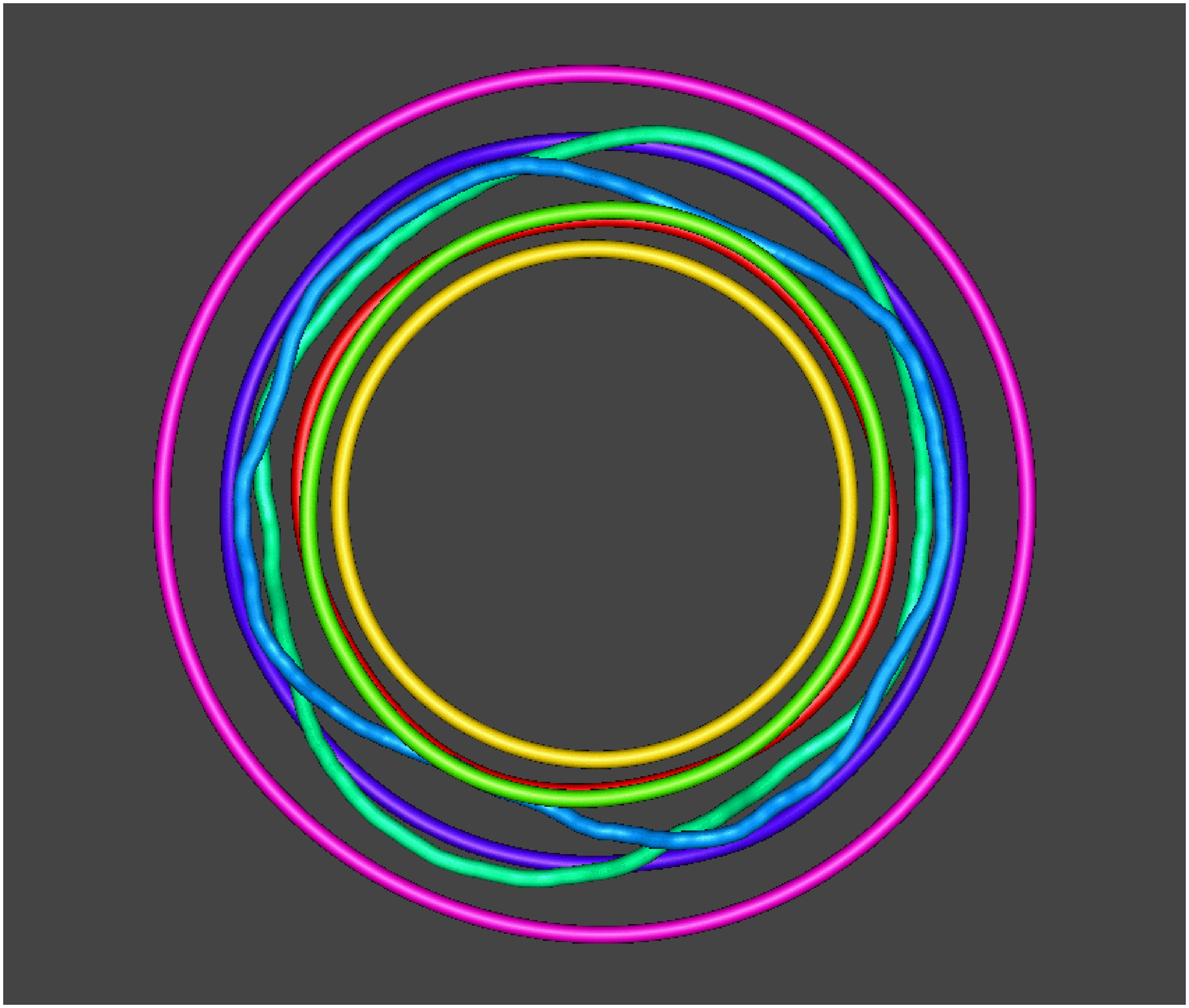}
}
\hspace{1em}
\subfloat[$t=63.75~\rm s$] {
  \includegraphics[scale=0.16]{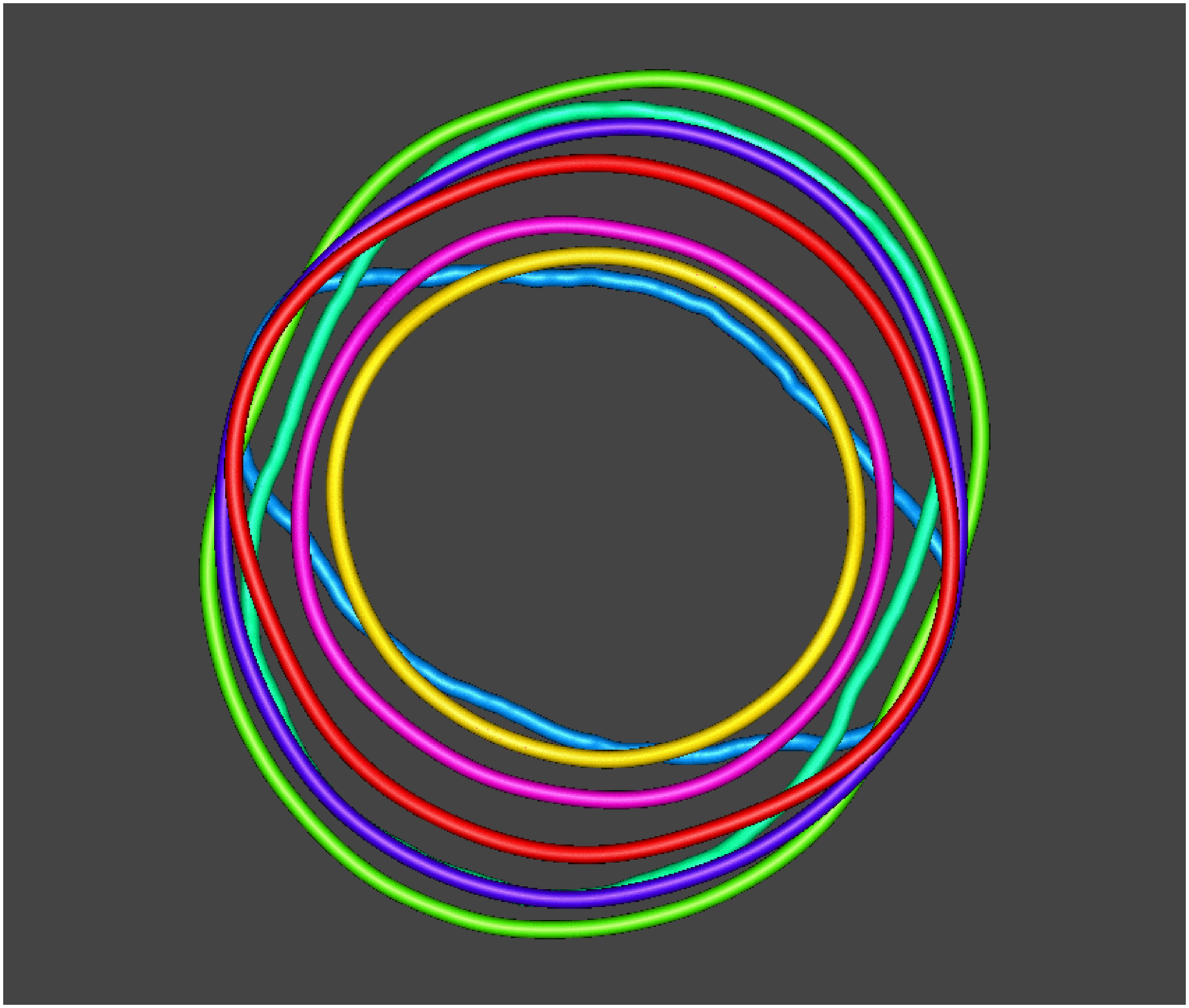}
}
\hspace{1em}
\subfloat[$t=67.50~\rm s$] {
  \includegraphics[scale=0.16]{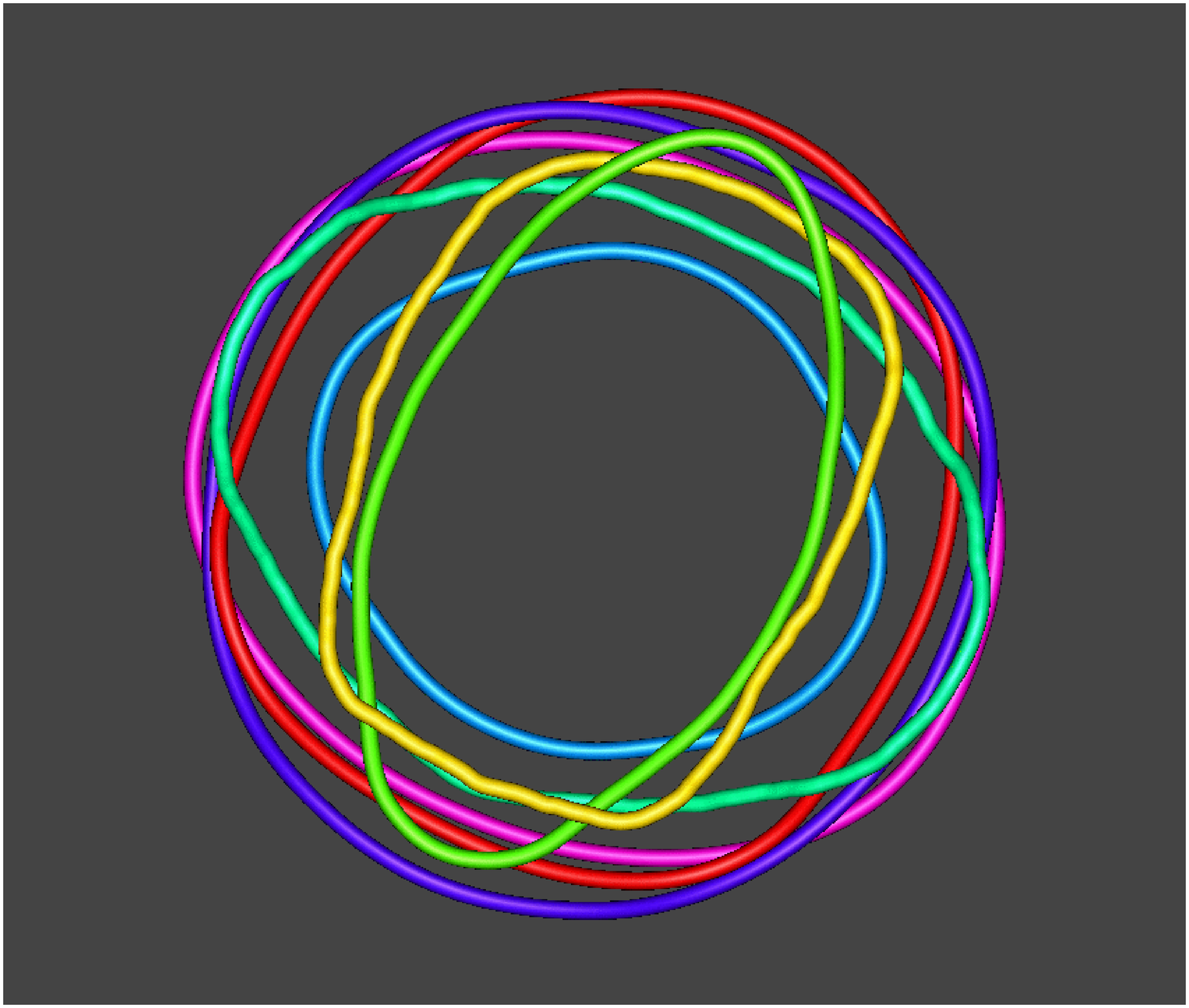}
}\\
\subfloat[$t=71.25~\rm s$] {
  \includegraphics[scale=0.16]{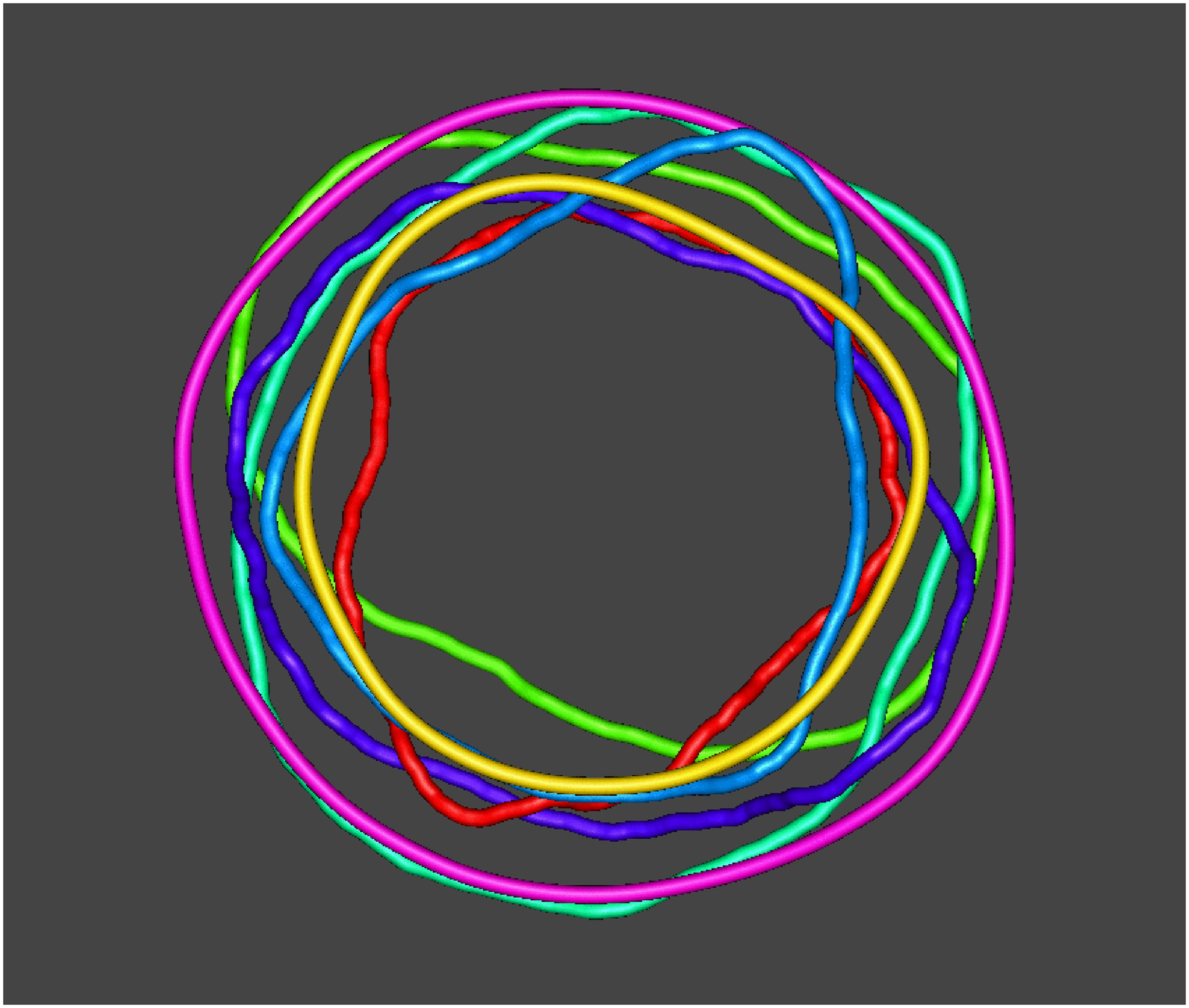}
}
\hspace{1em}
\subfloat[$t=75.0075~\rm s$] {
  \includegraphics[scale=0.16]{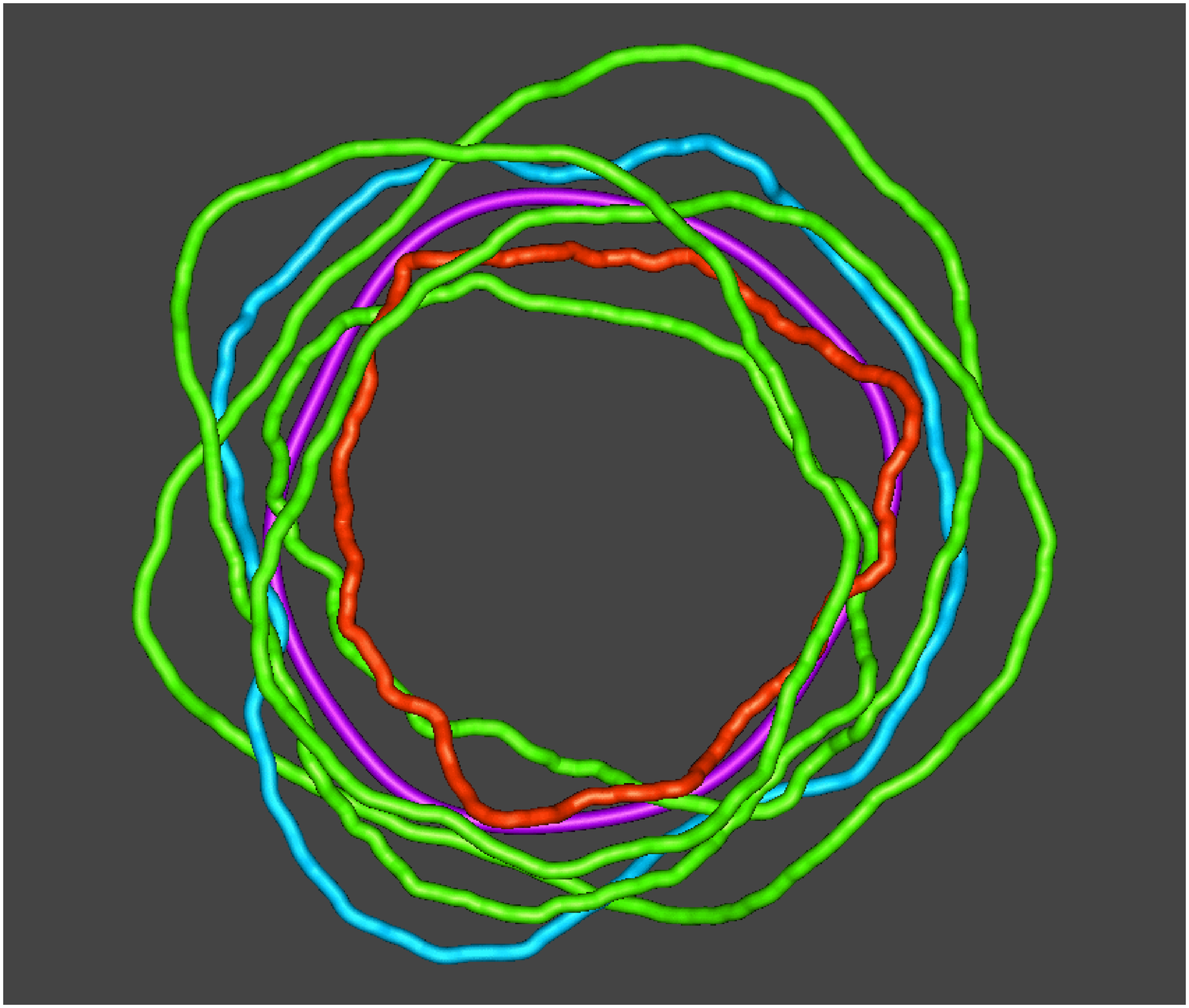}
}
\hspace{1em}
\subfloat[$t=78.75~\rm s$] {
  \includegraphics[scale=0.16]{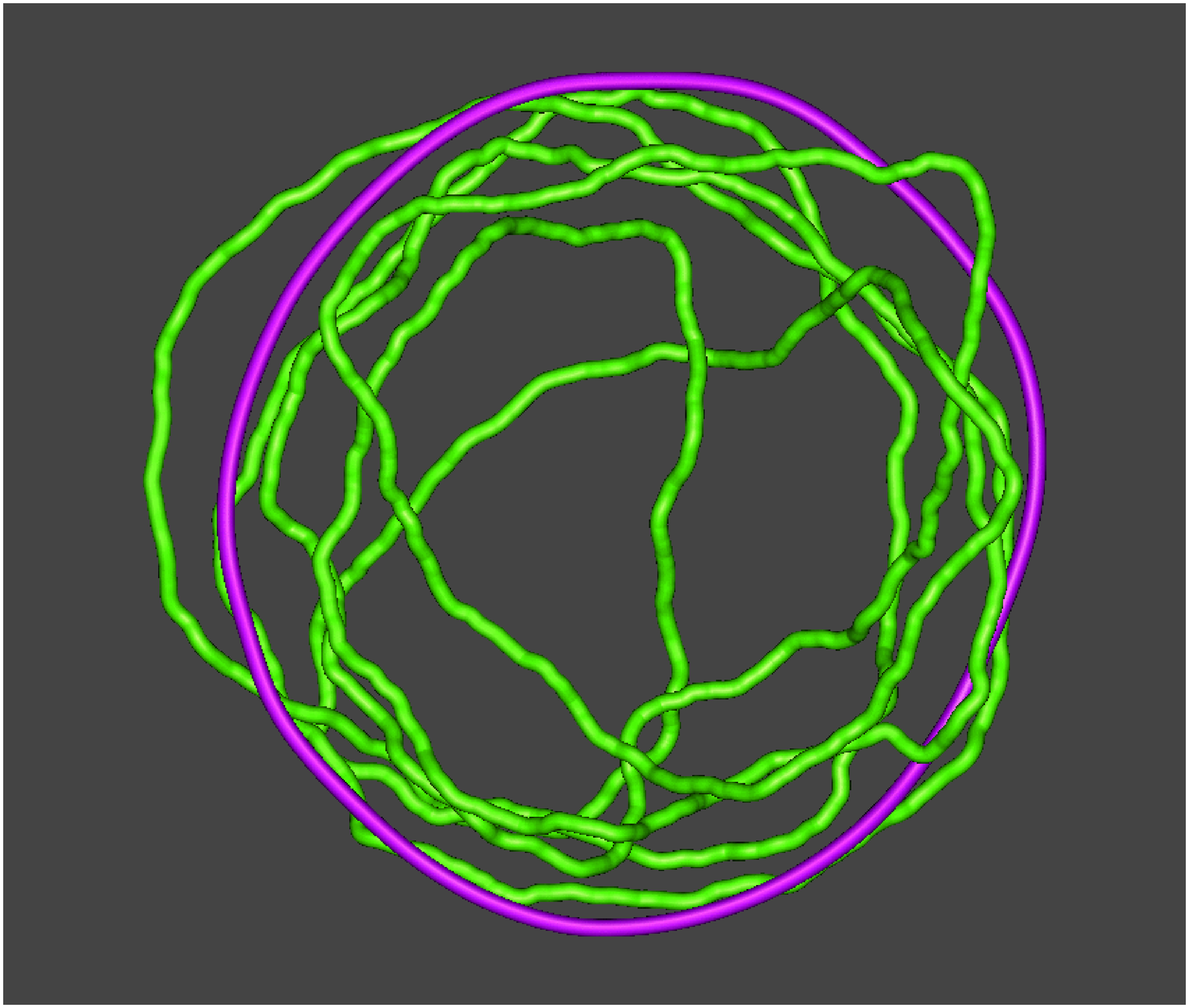}
}
\caption{As in Fig.~\ref{fig:11} but rear views.
}
\label{fig:12}
\end{figure}

Figs.~\ref{fig:11} and ~\ref{fig:12} show side and rear views of
what happens at large times; 
Table~\ref{tab:3} provides a quantitative description,
listing, at each selected time, the number of discretization points, $N_p$, 
the number of reconnections which have taken place, 
the number of rings, $N$, left in the bundle,
the vortex length, $\Lambda$,
the average curvature, $\bar c$, 
and distance travelled by the centre of vorticity $\Delta x$ in units of the 
initial diameter $D$.

We find that the vortex rings retain their circular coaxial shapes up to
about $57~\rm s$ when a long-wavelength Kelvin wave starts to grow and
eventually causes a vortex reconnection between vortex lines which are 
almost parallel to each other (in other words,
the angle $\theta$ between the
unit tangent vectors ${\bf s}=d{\bf s}/d\xi$ to the two reconnecting
vortex lines is small).
Despite the reconnection, the circular symmetry of the vortex
configuration is not much altered.  
An observer tracking the position and velocity of the vortex structure would
not notice any change.

In the absence of damping, 
the short-wavelength perturbations induced by the reconnection cusp 
travel around the vortex lines which were affected by the reconnection,
and induce similar waves on neighboring vortex lines (an effect
which we have described in a previous paper~\cite{Baggaley-cascade}).
The amount of Kelvin waves is quantified by the average curvature, 
which grows rapidly. The numerical algorithm has therefore to increase 
the number of discretization points (as shown in Table~\ref{tab:3}). 
More reconnections
take place, altering the total number of vortex rings in the bundle. 
At around $t \approx 70~\rm s$
short-wavelength perturbations are visible on the entire vortex bundle,
becoming the dominant feature, but still the velocity has not changed 
significantly yet.  Because of the Kelvin waves,
the vortex rings are not coplanar any longer, and the vortex length
grows significantly. 

We know from previous work that large-amplitude
Kelvin waves on individual vortex rings slow down the ring's
translational motion \cite{Hanninen,Helm}, and a similar effect seems to
take place here: when the perturbations are large enough,
the vortex bundle slows down significantly.
At this late stage of the evolution
the vortex bundle has lost most of its symmetry, the
number of vortex rings is less than the initial $N=7$, and the structure
consists of wiggly, disordered toroidal coils. During the last 
stage of the evolution, the
angle between reconnecting filaments tends to be larger 
(note Fig.~\ref{fig:12}(i), where some vortex strands are almost
perpendicualr to each other),
further scrambling the vortex structure. 

This scenario seems generic: we find it at other values of parameters and for
vortex bundles with a different initial number of rings.

\section{Conclusion}
\label{sec:conclusion}

We have computed numerically the evolution of bundles of thin-cored,
inviscid vortex filaments, using parameters motivated by the experiments 
in superfluid helium by Borner and Murakami.
We have found that these toroidal vortex structures are relatively robust.
They travel at constant speed (which we have modelled in a simple
way) and remain coherent for distances which are larger than their own size, 
confirming the interpretation of the experiments of Borner \cite{Borner1981}
and Murakami \cite{Murakami}. The individual vortex rings which make up the
bundles execute a collective form of leapfrogging motion, moving inside
and then outside the other rings, which generalizes the well-known 
leapfrogging of two vortex rings.

The evolution of these vortex structures is not much affected by
the presence of Kelvin waves on individual vortex strands and by
vortex reconnections - the translational velocity remains
approximately the same. Only when the disturbances are large 
and these large-scale vortex rings have been reduced to a few disordered
toroidal vortex coils the velocity and shape are affected.

Moving away from the parameters of the experiments, we have also
found that increasing $R/a$ leads to increased regularity of the
generalised leapfrogging motion, and increased stability of the
whole vortex structure.

We hope that these results will stimulate further work on the motion 
and the coherence of vortex filaments.
For example, it would be interesting to investigate 
the stability of two and three-dimensional bundles analytically, to find
whether one should expect stable behaviour for $t \to \infty$ 
(the results which we have presented only refer to length scales and
time scales of the experiments). 

It would also be interesting to attempt numerical investigations
with a very large number of vortex rings keeping the cross-section
of the torus constant (this computational may require
the use of the vortex tree algorithm \cite{Baggaley-tree} 
recently developed for superfluid turbulence).
The natural question is whether the continuous
limit of the classical Euler equation is achieved by superposing
vortex singularities. For example, one should compare the
displacement of vortex lines away from regular trajectories to the
distortions of the cross-section of a finite-size core with continous
vorticity distribution found in the context
of the Euler equation \cite{Fukumoto}.

\begin{acknowledgments}
\noindent
We thank the EPSRC for financial support and Dr A.J. Youd for help
with the graphics.
\end{acknowledgments}

\newpage

\begin{table}[h!!!]
\begin{center}
\begin{tabular}{ccccc}
\hline
$N$    & $\ell$      & $R$      & $a$         & $R/a$   \\
       & ($\rm cm$)    & ($\rm cm$) & ($\rm cm$)    &         \\
\hline
$1$    & -           & $0.06$   & -           & -     \\
$2$    & $0.015$     & $0.06$   & $0.0075$    & $8$     \\
$3$    & $0.015$     & $0.06$   & $0.00866$   & $6.92$     \\
$7$    & $0.015$     & $0.06$   & $0.015$     & $4$     \\
$19$   & $0.015$     & $0.12$   & $0.03$      & $4$     \\
\hline
\end{tabular}
\end{center}
\caption{Initial conditions used in the simulations described in Sect.~\ref{subsec:coherent}}
\label{tab:1}
\end{table}


\begin{table}[h!!!]
\begin{center}
\begin{tabular}{ccccccc}
\hline
$N$    & Event      & $t$        & $\Lambda$   &  $\bar c$       & $\Delta x/D$ & $v$       \\
       &            & $\rm (s)$  & $\rm (cm)$  &  $\rm cm^{-1}$  &              & $\rm cm/s$  \\
\hline
$1$  & start        & $0$        & $0.377$      & $16.7$          &    $0$       & $0.023$     \\
     & finish       & $60$       & $0.377$      & $16.7$          &    $11.48$   & $0.023$     \\
$2$  & start        & $0$        & $0.754$      & $17.2$          &    $0$       & $0.028$     \\
     & finish       & $50$       & $0.760$      & $16.7$          &    $11.11$   & $0.026$     \\ 
$3$  & start        & $0$        & $1.131$      & $17.1$          &    $0$       & $0.031$     \\
     & finish       & $40$       & $1.131$      & $17.2$          &    $10.24$   & $0.031$     \\ 
$7$  & start        & $0$        & $2.639$      & $18.1$          &    $0$       & $0.045$     \\
     & reconnection & $26.98$    & $2.712$      & $21.0$          &    $9.86$    & $0.040$     \\ 
     & finish       & $30$       & $2.841$      & $69.1$          &    $10.91$   & $0.038$     \\ 
$19$ & start        & $0$        & $14.326$     & $10.9$          &    $0$       & $0.047$     \\
     & reconnection & $19.89$    & $14.482$     & $11.9$          &    $3.83$    & $0.046$     \\ 
     & finish       & $60$       & $22.451$      & $199.5$        &    $10.38$   & $0.028$     \\ 
\hline
\end{tabular}
\end{center}
\caption{Outcomes of the simulations described in Sect.~\ref{subsec:coherent}.}
\label{tab:2}
\end{table}

\newpage

\begin{table}[h!!!]
\begin{center}
\begin{tabular}{cccccccc}
\hline
$t$       & $N_p$   & reconnections & $N$    &  $\Lambda$   &  $\bar c$       &   $\Delta x/D$ & $v$         \\
($\rm s$) &         &               &        &  ($\rm cm$)  &  ($\rm cm^{-1}$)&                &($\rm cm/s$)  \\
\hline
55.875    & 3528    & 0             & 7      &  3.95        &  12.1           &  9.30       &  0.029       \\
57.75     & 3528    & 0             & 7      & 3.976        &  11.9           &  9.60       &  0.029       \\
58.5      & 3593    & 0             & 7      & 3.979        &  11.8           &  9.72       &  0.029       \\
60.0      & 3747    & 5             & 7      & 4.017        &  27.0           &  9.97       &  0.029       \\
63.75     & 3714    &  5            & 7      & 4.026        &  25.8           &  10.57      &  0.028       \\
67.5      & 3925    & 5             & 7      & 4.087        &  25.7           &  11.18      &  0.027       \\
71.25     & 3944    & 17            & 7      & 4.207        &  57.3           &  11.76      &  0.025       \\
75.0075   & 4284    & 34            & 4      & 4.327        &  93.5           &  12.32      &  0.024       \\
78.75     & 4347    & 47            & 2      & 4.516        & 101.7           &  12.84      &  0.023       \\
82.5      & 4509    & 68            & 3      & 4.68         & 120.1           &  13.33      &  0.021       \\
86.25     & 4625    & 86            & 5      & 4.847        & 122.6           &  13.76      &  0.018       \\
90.0      & 4766    & 102           & 5      & 4.976        & 125.4           &  14.12      &  0.015       \\
93.75     & 4900    & 124           & 3      & 5.105        & 133.6           &  14.45      &  0.014       \\
97.5      & 4989    & 142           & 2      & 5.213        & 135.5           &  14.77      &  0.013       \\
\hline
\end{tabular}
\end{center}
\caption{Development of instabilities for vortex bundle of $N=7$ rings, 
see also Figs.~\ref{fig:11} and \ref{fig:12}.
At each time, the table lists the number of discretization points, $N_p$, 
the number of rings in the bundle, $N$, 
the number of reconnections which have taken place up to that time, 
the vortex length $\Lambda$, the average curvature $\bar c$, 
the distance travelled, $\Delta x$,  
in units of the initial
diameter, $D$, and the velocity $v$.
}
\label{tab:3}
\end{table}

\clearpage
\newpage

\clearpage
\newpage

\clearpage
\newpage

\clearpage
\newpage

\clearpage
\newpage

\clearpage
\newpage

%

\clearpage
\newpage

%

\clearpage
\newpage

%
\clearpage
\newpage

%
\clearpage
\newpage

%

\clearpage
\newpage

%

\clearpage
\newpage

\begin{thebibliography}{10}

\bibitem{Borner1981}
H. Borner, T. Schmeling and D. Schmidt,
Experimental investigation of the circulation of large-scale vortex
rings in HeII,
Physica B+C {\bf 108}, 1123-1125 (1981).

\bibitem{Borner1983}
H. Borner, T. Schmeling and D. Schmidt,
Experiments on the circulation and propagation of large-scale vortex
rings in HeII,
Phys. Fluids {\bf 26}, 1410-1416 (1983).

\bibitem{Borner1985}
H. Borner and T. Schmeling,
Investigation of large-scale vortex rings in HeII  by acoustic measurements
of circulation, in {\it Flow of real fluids}, ed. by G. Meier and F. Obermeier,
Lecture Notes in Physics, vol. 235, pp. 135-146, Springer, 
Berlin/Heidelberg (1985).

\bibitem{Donnelly}
R. J. Donnelly {\it Quantized vortices in helium II}, Cambridge U. Press
(1991).

\bibitem{BD2009}
CF Barenghi and R.J. Donnelly,
Vortex rings in classical and quantum systems,
Fluid Dyn. Res. {\bf 41}, 051401 (2009).

\bibitem{Golov}
P.M. Walmsley and A.I. Golov,
Quantum and quasi classical types of superfluid turbulence,
Phys. Rev. Letters {\bf 100}, 245301 (2008).

\bibitem{Bewley-rings}
G.P. Bewley, and K.R. Sreenivasan, 
The decay of a quantized vortex ring and the influence of tracer particles,`
J. Low Temp. Phys. {\bf 156} 84-94  (2009).

\bibitem{Murakami}
M. Murakami, M. Hanada, and T. Yamazaki,
Visualization study of large-scale vortex rings in HeII.
Jap. J. Applied Phys. Suppl. {\bf 26}, 107-108 (1987).

\bibitem{Stamm1}
G. Stamm, F. Bielert,  W. Fiszdon, and J. Piechna,
Counterflow induced macroscopic vortex rings in superfluid helium:
visualization and numerical simulation,
Physica B {\bf 193}, 188-194 (1994).

\bibitem{Stamm2}
G. Stamm, F. Bielert,  W. Fiszdon, and J. Piechna,
On the existence  of counterflow-induced macroscopic vortex rings in he II,
Physica B {\bf 194-196} 589-590 (1994).


\bibitem{Salort2010}
J. Salort J, et al.,
Turbulent velocity spectra in superfluid flows,
Phys. Fluids {\bf 22}, 125102 (2010).

\bibitem{Salort2012}
J. Salort, B. Chabaud, L{\'e}v{\^e}que E. and P.-E. Roche,
Energy cascade and the four-fifths law in superfluid turbulence,
Europhys. Lett. {\bf 97},  34006 (2012).


\bibitem{LNS} V.~S. L'vov, S.~V. Nazarenko, L.~Skrbek,
Energy Spectra of
Developed Turbulence in Helium Superfluids,
J. Low Temp. Phys. {\bf 145} 125 (2006).

\bibitem{Sasa2011}
N. Sasa, T. Kano, M. Machida, V. S. L'vov, O. Rudenko and M. Tsubota,
Energy spectra of quantum turbulence: Large-scale simulation and modelling,
Phys. Rev. B {\bf 84} 054525 (2011).

\bibitem{BLR}
C.F. Barenghi, V. L'vov and P.-E. Roche,
Turbulent velocity spectra in a quantum fluid:
experiments, numerics and models,
erXiv:1306.6248v1 (2013).

\bibitem{Vinen}
W.~F. Vinen and J.~J. Niemela (2002),
Quantum turbulence,
J. Low Temp. Phys. {\bf 128}, 167 (2002).

\bibitem{SS2012}
L. Skrebk and K.R. Sreenivasan,
Developed quantum turbulence and its decay,
Physics of Fluids {\bf 24}, 011301 (2012).
 
\bibitem{Sherwin2012}
A.W. Baggaley, L.K. Sherwin, C.F. Barenghi, and Y.A. Sergeev,
Thermally and mechanically driven quantum turbulence in helium II,
Phys. Rev. B {\bf 86}, 104501 (2012).


\bibitem{Baggaley-structures}
A.W. Baggaley, C.F. Barenghi, A. Shukurov, and Y.A. Sergeev,
Coherent vortex structures in quantum turbulence,
Europhys. Lett. {\bf 98}, 26002 (2012).

\bibitem{Frisch}
U. Frisch (1995),
{\it Turbulence. The legacy of A.N. Kolmogorov}
(Cambridge University press, Cambridge, UK).

\bibitem{Oshima}
Y. Oshima, T. Kambe, and S. Asaka,
Interaction of two vortex rings moving along a common axis
of symmetry,
J. Phys. Soc. Japan {\bf 38}, 1159-1166 (1975).

\bibitem{Riley}
N. Riley and D.P. Stevens,
A note on leapfrogging vortex rings,
Fluid Dyn. Res. {\bf 11}, 235-244 (1991).

\bibitem{Satti}
J. Satti and J. Peng,
Leapfrogging of two thick-cored vortex rings,
Fluid Dyn. Res. {\bf 45}, 035503 (2013).

\bibitem{Niemi}
A.J. Niemi,
Exotic statistics of leapfrogging vortex rings,
Phys. Rev. Lett. {\bf 94}, 124502 (2005).

\bibitem{note-core}
In our context, the small difference between hollow core model, 
solid body uniformly rotating core (Rankine) model, 
Gross-Pitaevskii core model and actual vortex core
in superfluid helium is not significant.

\bibitem{DB1998}
R.J. Donnelly and C.F. Barenghi,
The observed properties of liquid helium at the saturated vapor pressure,
J. Phys. Chem. Ref. Data {\bf 27}, 1217-1274 (1998).

\bibitem{Saffman}
P.G. Saffman, {\it Vortex Dynamics},
Cambridge University Press, Cambridge (1992).

\bibitem{Bewley-reconnections}
G.P. Bewley, M.S. Paoletti, K.R. Sreenivasan, and D.P. Lathrop,
Characterization of reconnecting vortices in superfluid helium,
PNAS {\bf 105}, 13707 (2008).

\bibitem{Zuccher}
S. Zuccher, M. Caliari, and C.F. Barenghi,
Quantum vortex reconnections,
Phys. Fluids {\bf 24}, 125108 (2012).

\bibitem{Paoletti}
M.~S. Paoletti, M.~E. Fisher, and D.~P. Lathrop,
Reconnection dynamics for quantized vortices,
Physica D {\bf 239}, 1367-1377 (2010).

\bibitem{Schwarz}
K.~W. Schwarz,
Three-dimensional vortex dynamics in superfluid $^4${H}e: homogeneous
superfluid turbulence,
Phys. Lett. B {\bf 38} 2398 (1988).

\bibitem{Adachi}
M. Tsubota \& H. Adachi,
Simulation of counterflow turbulence by vortex filament,
J. Low Temp. Phys. {\bf 162} 367-374 (2011).

\bibitem{Baggaley-cascade}
A.W. Baggaley and C.F. Barenghi,
Spectrum of turbulent Kelvin-waves cascade in superfluid helium,
Phys. Rev. B {\bf 83}, 134509 (2011).

\bibitem{Baggaley-fluctuations}
A.W. Baggaley and C.F. Barenghi,
Vortex-density fluctuations in quantum turbulence,
Phys. Rev. B {\bf 84 R}, 020504 (2011)

\bibitem{Baggaley-reconnections}
A.W. Baggaley,
The sensitivity of the vortex filament method
to different reconnection models,
J. Low Temp. Physics, {\bf 168}, 18 (2012)

\bibitem{Leadbeater}
M Leadbeater, T. Winiecki, D.C. Samuels, C.F. Barenghi and C.S. Adams,
Sound emission due to superfluid vortex reconnections,
Phys. Rev. Lett. {\bf 86}, 1410 (2001).

\bibitem{Didden}
N. Didden,
On the formation of vortex rings:  rolling-up and production of circulation,
Z. Angew. Math. Phys. {\bf 30}, 101-116 (1979).

\bibitem{Shariff}
K. Shariff and A. Leonard,
Vortex rings,
Ann. Rev. Fluid Mech.{\bf 24}, 235279 (1992).

\bibitem{knotplot}
Bob Scharein's knot theory computer package {\it KnotPlot} is available
at http://www.KotPlot.com.

\bibitem{Love}
A. Love, The motion of paired vortices with a common axis,
Proc. Lond. Math. Soc. {\bf 25}, 185194 (1894).

\bibitem{Acheson}
D. Acheson, Instability of vortex leapfrogging,
European J. Phys. {\bf 21}, 269273 (2000).


\bibitem{Hanninen}
C.F. Barenghi, R. Hanninen, and M. Tsubota,
Anomalous translational velocity of vortex ring with 
finite-amplitude Kelvin waves,
Phys. Rev. E {\bf 74}, 046303 (2006).

\bibitem{Helm}
J.L. Helm, C.F. Barenghi, and A.J. Youd
Slowing down of vortex rings in Bose-Einstein condensates
Phys. Rev. A {\bf 83}, 045601 (2011).

\bibitem{Baggaley-tree}
A.W. Baggaley and C.F. Barenghi,
Tree Method for quantum vortex dynamics,
J. Low Temp. Physics {\bf 166}, 3-20 (2012)


\bibitem{Fukumoto}
Y. Fukumoto and Y. Hattori,
Curvature instability of a vortex ring,
J. Fluid Mech. {\bf 526}, 77-115 (2005).

\end{thebibliography}
\end{document}